\documentclass[fleqn,usenatbib]{mnras}

\usepackage{newtxtext,newtxmath}
\usepackage[T1]{fontenc}
\usepackage{graphicx}
\usepackage{amsmath}
\usepackage{siunitx}
\usepackage{bm}
\usepackage{enumitem}
\usepackage[dvipsnames]{xcolor}

\usepackage{caption}
\DeclareCaptionFormat{cont}{#1 (cont.)#2#3\par}

\newcommand{\kmsMID}{\ensuremath{\text{km s}^{-1}} }
\newcommand{\kmsEND}{\ensuremath{\text{km s}^{-1}}}

\DeclareUnicodeCharacter{0308}{}

\DeclareRobustCommand{\VAN}[3]{#2}
\let\VANthebibliography\thebibliography
\def\thebibliography{\DeclareRobustCommand{\VAN}[3]{##3}\VANthebibliography}

\title[ALMACAL XIV]{ALMACAL XIV: X-Shooter Spectroscopy, Infrared Properties and Radio SEDs of Calibrators}

\author[Simon Weng et al.]{
Simon Weng,$^{1,2,3,4}$\thanks{E-mail: simonw358@gmail.com}
Elaine M. Sadler,$^{1,2,3}$
Emily Kerrison, $^{1,2,3}$
Victoria Bollo,$^{5}$
{C\'eline P\'eroux},$^{4,5}$
\newauthor
Martin Zwaan,$^{5}$ 
Elizabeth K. Mahony,$^{3}$ 
James R. Allison,$^{6}$
Jianhang Chen,${^7}$
Roland Szakacs,$^{5}$
\newauthor
Hyein Yoon,$^{8,9}$ 
\\
$^1$ Sydney Institute for Astronomy, School of Physics A28, University of Sydney, NSW 2006, Australia\\
$^2$ ARC Centre of Excellence for All Sky Astrophysics in 3 Dimensions (ASTRO 3D)\\ 
$^3$ ATNF, CSIRO Space and Astronomy,  PO Box 76, Epping, NSW 1710, Australia \\
$^4$ Aix Marseille Universit\'e, CNRS, LAM (Laboratoire d'Astrophysique de Marseille) UMR 7326, 13388, Marseille, France \\
$^5$ European Southern Observatory, Karl-Schwarzschildstrasse 2, D-85748 Garching bei M{\"u}nchen, Germany\\
$^6$ First Light Fusion Ltd., Unit 9/10 Oxford Pioneer Park, Mead Road, Yarnton, Kidlington OX5 1QU, UK\\
$^7$ Max-Planck-Institut für Extraterrestrische Physik (MPE), Giessenbachstrasse 1, D–85748 Garching, Germany\\
$^8$ Institute for Data Innovation in Science, Seoul National University, 1 Gwanak-ro, Gwanak-gu, Seoul 08826, Republic of Korea\\
$^9$ Astronomy Program, Department of Physics and Astronomy, Seoul National University, 1 Gwanak-ro, Gwanak-gu, Seoul 08826, Republic of Korea
}

\date{Accepted XXX. Received YYY; in original form ZZZ}

\pubyear{2025}

\begin{document}
\label{firstpage}
\pagerange{\pageref{firstpage}--\pageref{lastpage}}
\maketitle

\begin{abstract}
The ALMACAL$-$22 survey includes over 2700 hours of observations of ALMA phase and amplitude calibrators, spanning frequencies from 84 to 950 GHz across bands 3 to 10. 
In total, 687 out of the 1,047 calibrators have redshifts confirmed with spectroscopy and we find an additional 50 featureless blazars.
The redshift distribution of the ALMACAL-22 sample peaks at $z \approx 1$ and spans a wide range, from the nuclei of nearby galaxies at $z \ll 0.01$ to quasars at $z = 3.742$. 
70 new VLT/X-Shooter spectra of these sources covering UV to NIR wavelengths are also presented, which will be used in future stacking experiments to search for cold gas in the circumgalactic medium. 
Infrared magnitudes from WISE indicate that the majority of the sources are consistent with being quasars or blazars. 
{After fitting the radio spectral energy distributions of the calibrators, we find that most ALMA calibrators exhibit peaked spectra or are re-triggered which is surprising given the large number of blazars in the sample. 
The peak frequencies span three orders of magnitude from 100 MHz to 170 GHz, corresponding to linear sizes ranging from sub-pc to $>$ 10 kpc.}
In the future, when combined with high-resolution radio imaging, these results will offer valuable constraints on the molecular gas content of the CGM, as well as the ages and duty cycles of AGN jets. 
The ever-growing ALMACAL data set will remain an indispensable resource for studying the various aspects of galaxy formation and evolution.
\end{abstract}




\section{Introduction}
The Atacama Large Millimetre Array (ALMA) is a powerful radio interferometer operating at frequencies of 35 to 950 GHz. 
Each science project using ALMA requires observations of bright ($\approx$1 Jy), compact calibrators near the target. 
The ALMACAL survey compiles ALMA observations of these calibrator sources used for flux, phase, and bandpass calibration \citep{Zwaan2022}. 
These calibrator observations are taken at different times, in various ALMA bands and array configurations, depending on the science project, thus providing extensive coverage across the sub-mm spectrum.
Additionally, the bright, compact nature of the calibrators allows for precise self-calibration. 

The total on-source time for calibrators accumulated until May 2022 exceeds $2700$ hours and covers a sky area greater than 1000 arcmin$^2$ \citep{Bollo2024}. 
We will refer to this dataset as the ALMACAL$-$22 sample. 
So far, various scientific investigations have used fields containing ALMA calibrator sources. 
The calibrator data have been employed to search for dusty star-forming galaxies \citep[DSFGs;][]{Oteo2016, Klitsch2020, Chen2023, Chen2023b} and study their properties at higher resolution when data permits \citep{Oteo2017}. 
{Importantly, these calibrator fields are distributed across the sky, making the measured number counts of DSFGs largely immune to cosmic variance, the variation of galaxy clustering properties in different patches of sky. }
Furthermore, searching blindly for carbon monoxide (CO) emission that traces molecular gas helps constrain the evolution of the cosmic molecular gas density \citep{Klitsch2019b, Hamanowicz2023, Bollo2025}. 
Serendipitous observations of galaxies associated with quasar absorption lines have also enabled case studies of ionized, neutral, and molecular gas in galaxies \citep{Klitsch2018, Klitsch2019a}. 
Additionally, stacking the sub-mm spectra of calibrator sources at the redshift of nearby galaxies has been used to place constraints on the amount of extended molecular gas in the circumgalactic medium \citep[CGM;][]{Klitsch2023}. 
As the volume of calibrator data continues to grow, these various scientific endeavors will be pursued in greater depth in the future.

To fully exploit the ALMACAL data, understanding the properties of the calibrators is crucial. 
For instance, obtaining accurate redshifts is necessary to ensure that the calibrator is at a higher redshift than foreground galaxies in stacking experiments \citep{Klitsch2023}. 
Moreover, verifying that CO emitters are at different redshifts from the calibrators is important to ensure that measured CO luminosity functions are not affected by clustering \citep{Hamanowicz2023, Chen2023b}. 
\citet{Bonato2018} presented the first catalogue of 754 ALMA calibrators based on observations up until September 2017, focusing on radio classification, variability, and global spectral energy distributions of the calibrators.

{In this work, we present an updated catalogue of 1,047 calibration sources in the ALMACAL$-$22 sample, including new redshifts obtained from recent Very Large Telescope (VLT) observations (Section \ref{sec:Methods}).}
We examine the redshift properties of the calibrators, as well as their infrared classifications based on WISE data (Section \ref{sec:optical}). 
Finally, we re-evaluate the radio spectral energy distributions of the calibrators (Section \ref{sec:radio}) and discuss future research directions using the expanding ALMACAL sample (Section \ref{sec:science}). 




\section{Methods}
\label{sec:Methods}
\subsection{The ALMACAL$-22$ sample}
\begin{figure*}
    \includegraphics[width=1.0\textwidth]{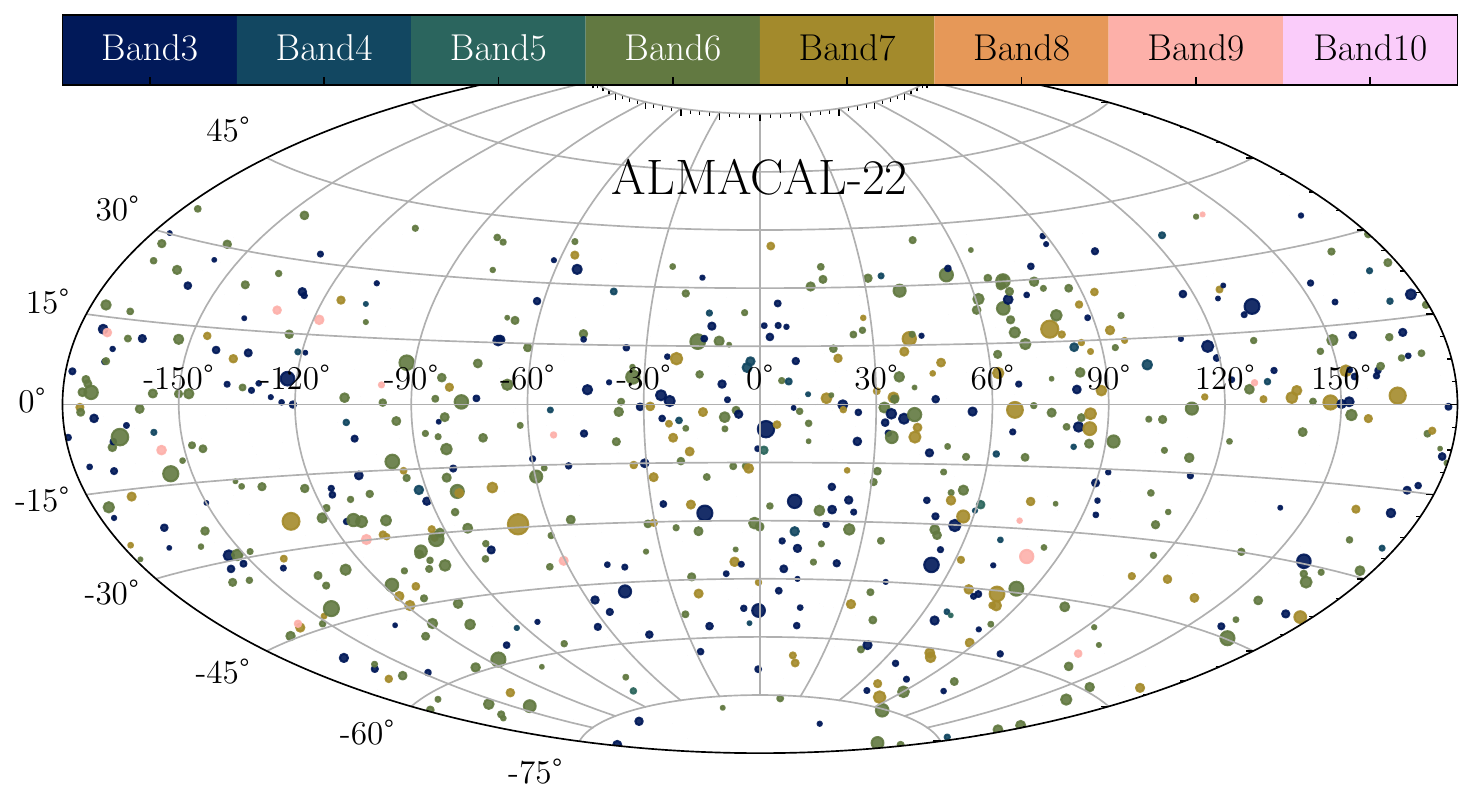}
    \caption{The distribution of 1,047 ALMACAL sources on the sky using an Aitoff projection. 
    Each source is coloured by the band with the longest on-source time, while the size of each source is proportional to the square root of the total integration time. 
    We refer the reader to \citet{Chen2023} for the distribution of on-source times across each band. }
    \label{fig:skyplot}
\end{figure*}

The ALMACAL$-22$ sample presented in this work includes 1,047 sources observed between 2012 August and 2022 May for approximately 2750 hours. 
Observations span bands 3 to 10 and a distribution of integration times across the eight bands can be found in \citet{Bollo2024}. 
In \autoref{fig:skyplot}, we depict the distribution of sources on the sky with colour representing the band dominating observations and marker size proportional to the square root of the total on-source time ($\propto \sqrt{T_{\rm tot}}$). 

\subsection{Optical cross-matching}
\begin{figure}
    \includegraphics[width=1.0\linewidth]{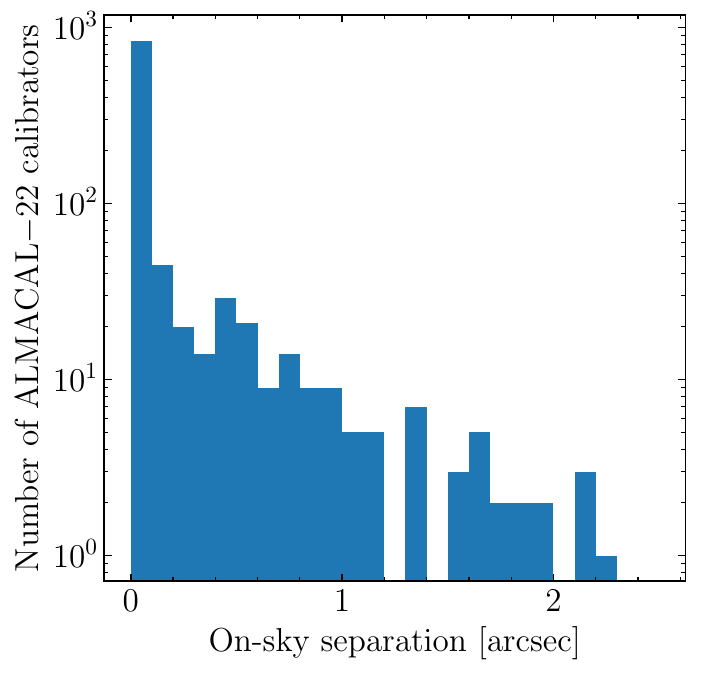}
    \caption{A histogram of the separations between the ALMACAL radio source and cross-matched optical source. 
    90\% (97\%) of radio sources have an optical counterpart within 0.5 (1.0) arcseconds. 
    Only 6 out of 1,047 ALMACAL$-22$ sources do not have an known optical counterpart within 3 arcseconds.}
    \label{fig:Separation}
\end{figure}

We use the NASA Extragalactic Database (NED\footnote{\url{https://ned.ipac.caltech.edu/}{}}) and SIMBAD\footnote{\url{https://simbad.cds.unistra.fr/simbad/}} databases to search for optical counterparts to the ALMACAL sources. 
Additionally, we perform a cross-match with an optical catalogue of sources in the Australia Telescope
20 GHz (AT20G) Survey \citep{Mahony2011}. 
There is substantial overlap between the two samples, as many of the southern ALMA calibrators were selected from bright sources in \citet{Mahony2011}. 
As ALMACAL sources are typically bright and compact to ensure consistent flux, phase and bandpass calibration, much of the difficulty and uncertainty in identifying optical counterparts to radio sources is mitigated. 
In \autoref{fig:Separation}, we show a histogram of on-sky separations after cross-matching ALMACAL sources with nearby optical sources. 
97 per cent of optical counterparts are found within 1 arcsecond of the radio position, while only 6 radio sources do not have any counterpart within 3 arcseconds. 

Sources without optical counterparts are tabulated in \autoref{tab:NoCounterpart}. 
We give the names of and separations to the nearest optical source. 
Three sources (J1032-5917, J1220-5604 and J1555-4150) are found at low Galactic latitudes in crowded stellar fields, while the remainder are likely too faint at visible wavelengths. 
{All six sources without counterparts are within the current footprint observed by the Vera Rubin Observatory \citep{RubinObservatory}, and an optical counterpart will likely be identifiable in these deeper images.}

{
Our choice of a 3-arcsecond threshold to determine whether an ALMA calibrator is matched to an optical source is somewhat arbitrary. 
While ALMA calibrator positions are typically accurate to within 3 mas due to their very-long-baseline interferometric (VLBI) observations \citep{Petrov2025}, the astrometric accuracy of optical sources is more variable, as they are drawn from large databases.
To assess the likelihood of mismatches between radio and optical sources, we cross-matched the ALMACAL catalog with NED after shifting the radio coordinates by one arcminute in both RA and Dec. 
We find that [0.5, 1.3, 3.5, 6.7] per cent of the radio sample with scrambled positions matches an optical source within [0.5, 1, 2, 3] arcseconds. 
This suggests that at a 3-arcsecond separation, the probability of a false positive in the cross-match is approximately 7 per cent, while the error rate drops to just 0.5 percent within 0.5 arcseconds (941 out of 1047 calibrators have separations smaller than 0.5 arcseconds).}

\begin{table*}
    \centering
    \begin{tabular}{l c c l c}
        \hline\hline
            ALMACAL ID & RA & Dec & Nearest Source & On-sky Separation\\
            & (hh:mm:ss) & (dd:mm:ss) &  & (arcseconds) \\
        \hline
            J0304+3348 & 00:32:51 & 02:47:41 & 2MASS J08123858+0247496 & 15.1 \\
            J0501-0159 & 04:51:57 & 00:06:11 & MACS J0451.9+0006 & 10.1 \\
            J1032-5917 & 10:32:43 & -59:17:57 & 2MASS J10324189-5917488 & 9.5  \\
            J1220-5604 & 12:20:26 & -56:04:43 & WISEA J122026.74-560446.0 & 4.8 \\
            J1555-4150 & 15:55:18 & -41:50:32 & WISEA J155517.78-415028.4 & 4.9  \\
            J1717-3948 & 12:41:57 & -02:10:53 & 2MASS J12415707-0211059 & 13.1 \\
         \hline
    \end{tabular}
    \caption{
    The ALMACAL sources without an optical counterpart within 3 arcesconds. 
    Along with the ALMACAL name and coordinates, we provide the nearest optical source from NED, including the separation in arcseconds to that optical source. 
    The location of these sources near the Galactic plane or at very low declinations are some of the reasons why an optical counterpart is not easily identifiable. }
    \label{tab:NoCounterpart}
\end{table*}

\subsection{VLT X-Shooter UV--NIR observations}
We obtained observations of cross-matched optical sources without redshifts or spectra using VLT/X-Shooter \citep{Vernet2011} in programmes 0101.A-0528 (PI. Mahony) and 111.253L.001 (PI. Weng). 
We observed 70 southern ALMACAL calibrators across both programmes, including repeat observations of P101 targets with low signal-to-noise ratio (SNR $< 3$) spectra. 
Optically bright ($R < 21$) quasars were targeted to facilitate observations under `any-weather' conditions. 
We chose slit widths of 1.3, 1.2 and 1.2 arcseconds for the UVB, VIS and NIR arms (covering 300$-$2500 nm) to account for poor seeing conditions. 
These widths correspond to a resolving power of $R = 4100, 6500, 4300$, respectively. 
Each target was observed in the NODDING mode to improve the removal of sky emission. 
Exposure times range from 900 to 2,700 seconds, depending on the magnitude of the target. 

We processed the X-shooter data from the two programs using version 3.6.3 of the ESO X-shooter pipeline \citep{Modigliani2010} within the ESO Reflex environment \citep{Freudling2013}. 
The data from each arm were independently reduced, using the nodding mode setting in the pipeline. 
This automated workflow involves subtracting detector bias and dark current, followed by wavelength and flux calibration, to produce flux-calibrated 1D spectra. 
In nodding mode, science frames at different nodding positions along the slit are combined into an image sequence, with pairs of combined frames subtracted to achieve sky subtraction by differencing the images from nodding positions A and B. 
After visually inspecting the two-dimensional spectra, we extract the one-dimensional spectra using the pipeline. 
For the P101 data, we reach median SNR values of [7.4, 12.7, 3.0], while for the P113 data, the median SNR values are [9.4, 13.0, 3.1] for the UV, VIS, and NIR arms, respectively. 
{The signal-to-noise ratio for the visible and infrared arms is calculated in regions away from significant telluric bands, namely at $\lambda \simeq 7590--7700$, $13450-14500$ \AA\ and $18000-20950$ \AA}. 
We use \textsc{marz}\footnote{https://samreay.github.io/Marz/\#/overview} \citep{Hinton2016} to determine redshifts. 
\textcolor{blue}{Examples of a featureless blazar and QSO are shown in Figures \ref{fig:XSHOOexampleBL} and \ref{fig:XSHOOexampleQSO}, with the remainder provided in Appendix \ref{sec:XSHOOSpec}. }

\begin{figure*}
    \includegraphics[width=0.95\textwidth]{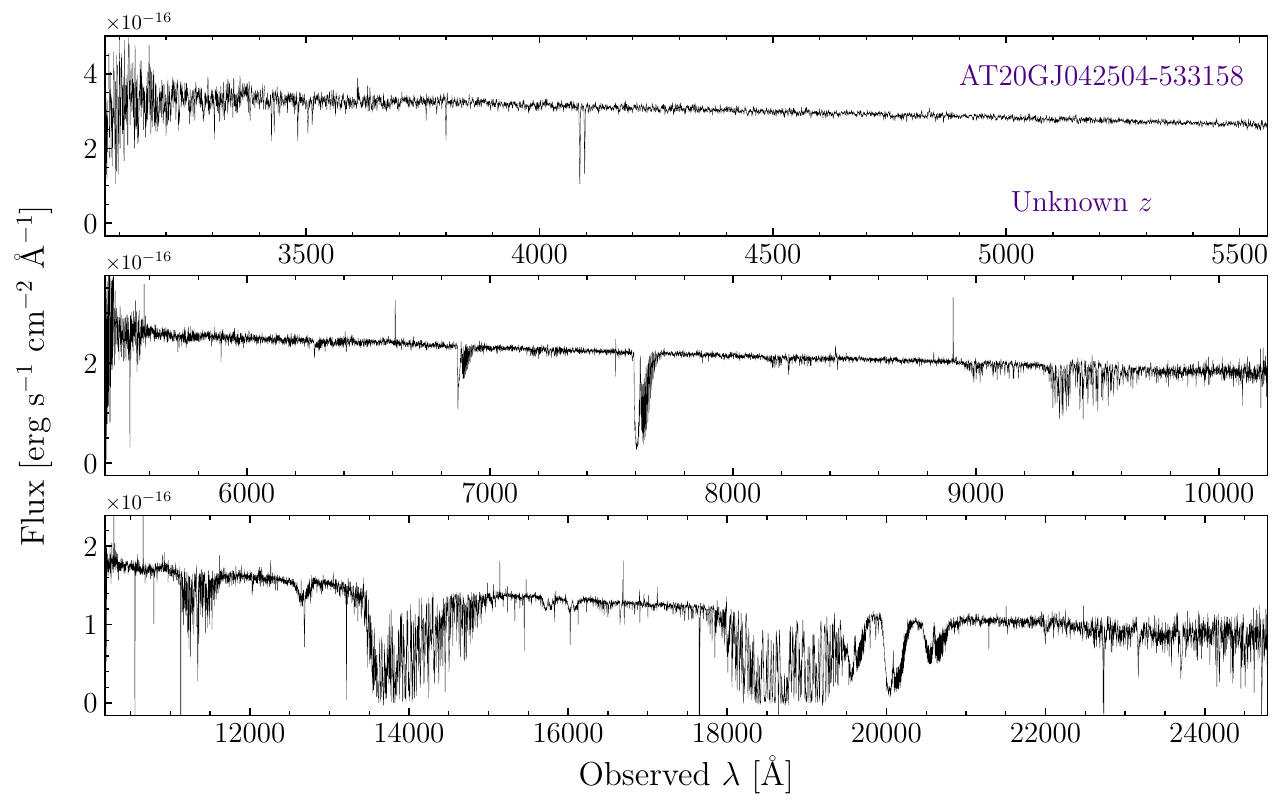}
    \caption{Example of an X-Shooter spectrum of a featureless blazar with \ion{Mg}{ii} $\uplambda\uplambda$ 2796, 2803 absorption lines at $z_{\rm abs} = 0.462$ highlighted in green. 
    We include a simulated atmospheric transmission curve in red for reference only for the optical and near-infrared wavelengths where there is telluric absorption \citep{Bertaux2014}.} 
    \label{fig:XSHOOexampleBL}
\end{figure*}

\begin{figure*}
    \captionsetup{list=off,format=cont}
    \includegraphics[width=0.95\textwidth]{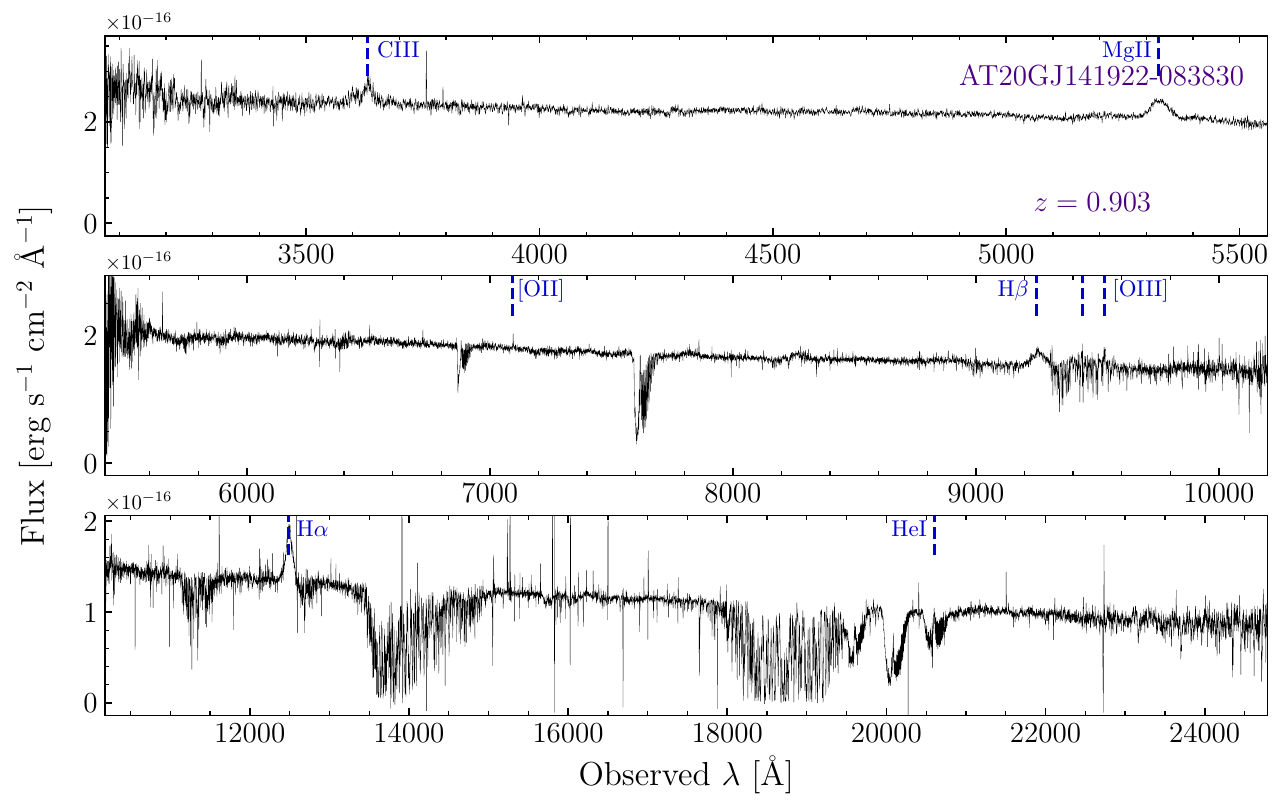}
    \caption{Example of an X-Shooter spectrum of a QSO with broad emission lines. 
    The identical simulated transmission curve in \autoref{fig:XSHOOexampleBL} is shown here in red.}
    \label{fig:XSHOOexampleQSO}
\end{figure*}

\subsection{ALMACAL redshift catalogue construction}
To refine the redshift determination method, the co-authors (CP, EK, EM, ES, HY, JA, JC, RS, SW) studied an initial sample of the same 20 calibrator sources. 
We noted any discrepancies and discussed ways to standardise and optimise the process. 
{In addition to large surveys such as the Sloan Digital Sky Survey \citep[SDSS;][]{York2000} and Dark Energy Spectroscopic Instrument (DESI) survey}\citep{DESI2016, DESIDR1}, we found that several works following up bright radio sources contributed multiple redshifts to the ALMACAL catalogue and contained corresponding spectra \citep[e.g.][]{Drinkwater1997, Hook2003, Shaw2013, Titov2013} or a list of emission lines \citep{Hewitt1987}. 
These works were noted by the team and increased the efficiency of determining redshifts and their confidence. 
Spectroscopic redshift measurements for objects not belonging to large surveys were each checked independently by two members of the collaboration. 
An important part of this process was establishing the provenance of each redshift by tracking back through the literature to identify and record the source of the original measurement. 
In completing the redshift determinations, the substantial dedication and meticulous effort invested by the team were paramount. 
This labour-intensive and detail-oriented process was crucial in enhancing both the precision and dependability of the redshift values in the final ALMACAL catalogue. 

The cross-matched ALMACAL sources were initially split into two distinct catalogues based on whether the optical counterpart belonged to large surveys such as the Two-degree-Field Galaxy Redshift Survey \citep[2dFGRS;][]{Colless2001}, 6dF Galaxy Redshift Survey \citep[6dFGRS;][]{Jones2004}, SDSS and {DESI DR1} \citep{DESIDR1}.  
Redshifts of sources with counterparts in these surveys could more easily be determined by directly accessing the spectra. 
Sources observed by X-Shooter were differentiated from the remaining sources that required more extensive examination using the NED and SIMBAD databases. 

\subsubsection{Format}
The final ALMACAL redshift catalogue is publicly available online on GitHub\footnote{\url{https://github.com/simonw358/ALMACAL}} and VizieR. 
The column numbers, names and descriptions of the ALMACAL redshift catalogue are described in \autoref{tab:ALMACALcatlabel}. 

\begin{table*}
    \centering
    \begin{tabular}{l l l}
        \hline\hline
            Number & Name & Description\\
        \hline
            1 &  Name\_ALMACAL  & the name of the source in ALMACAL that is consistent with previous publications \\
            2 &  RA\_ALMACAL & the right ascension of the radio source in ICRS co-ordinates (decimal degrees)  \\
            3 &  Dec\_ALMACAL & the declination of the radio source in ICRS co-ordinates (decimal degrees)  \\
            4 &  Redshift & the redshift of the source  \\
            5 &  Spectrum & a boolean 1/0 (TRUE/FALSE) flag indicating whether the optical spectrum for the source is publicly available  \\
            6 &  Provenance & the original publication containing the redshift and spectrum, if available \\
            7 &  Confidence & the confidence of the redshift measurement  \\
            8 &  Name\_IAU & an IAU name of the source obtained from NED or SIMBAD  \\
            9 &  RA\_Optical & the right ascension of the optical source in ICRS co-ordinates (decimal degrees)  \\
            10 & Dec\_Optical & the declination of the optical source in ICRS co-ordinates (decimal degrees) \\
            11 & Separation & the separation between the radio and optical source (arcseconds)  \\
         \hline
    \end{tabular}
    \caption[The ALMACAL catalogue header description]{
    The numbers, names, and descriptions of the columns in the ALMACAL catalogue in this table. 
    Magnitudes can be retrieved using the optical coordinates on NED or SIMBAD; magnitudes are not included in the table due to the large variety of instruments and filters.}
    \label{tab:ALMACALcatlabel}
\end{table*}

In the case where multiple sources cite the same redshift value, precedence is given to papers with a publicly accessible spectrum. 
In the `Confidence' column, we assign a quality value from 0 to 4, or 6 or 7, classifying each source. 
This notation is adapted from the 6dF survey \citep{Jones6dFDR3}, with an additional quality flag 7 included to indicate a featureless BL Lac spectrum. 
They are as follows: 
    (0) sources lacking both spectrum and redshift measurement, 
    (1) sources with a spectrum but no viable redshift estimate (excluding featureless BL Lac sources), 
    (2) sources with or without a spectrum where a possible redshift is derived from weak absorption or emission lines, 
    (3) sources with a spectroscopic redshift from a published paper but without a publicly available spectrum, or 
    sources with a spectrum showing a single strong absorption or emission feature and weaker lines, 
    (4) sources with a spectrum featuring multiple strong absorption or emission lines, 
    (6) {sources that are the nuclei of galaxies at redshift $z \ll 0.01$ such as M87}, and 
    (7) BL Lac objects without detectable emission lines. 
Hence, objects without a publicly available spectrum can only be assigned a confidence of 3 or lower, and this includes older works where only the emission lines are tabulated \citep[e.g.][]{Hewitt1987}. 
Quantitatively, sources with redshift errors $\lesssim 300$ \kmsMID are classified with confidence 4, whereas a confidence of 3 means that no error measurement is possible or errors are $\gtrsim 300$ \kmsEND. 

\begin{figure}
    \includegraphics[width=1.0\linewidth]{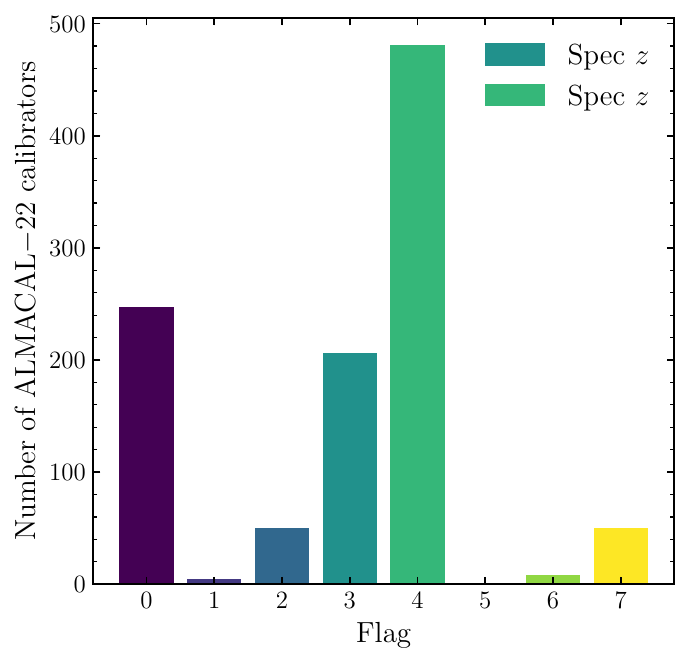}
    \caption{Bar chart illustrating the number of sources associated with various flag values. 
    The $x$-axis represents the flag values, and the $y$-axis denotes the number of sources. 
    The flags represent specific criteria: 
    (0) sources without a spectrum or redshift measurement,
    (1) sources with a spectrum but no possible redshift estimate (excluding BL Lac objects), 
    (2) sources with or without a spectrum where a possible redshift is indicated by weak absorption or emission lines, 
    (3) sources with a spectroscopic redshift reported in a published paper but lacking a publicly available spectrum, or 
    sources with a spectrum showing a single strong absorption or emission feature, 
    (4) sources with a spectrum showing multiple strong absorption or emission features, 
    (6) {active nuclei of galaxies such as M87 at redshift $z \ll 0.01$}, and 
    (7) BL Lac objects. 
    A total of 687 objects have flag values of 3 or 4, indicating that approximately 65 per cent of the sample possesses reliable redshift measurements.
    }
    \label{fig:zFlag}
\end{figure}

In \autoref{fig:zFlag}, the bar chart depicts the distribution of flags assigned to the 1,047 ALMACAL sources. 
$687$ objects are assigned flag values of $3$ or $4$, signifying that $\approx$65 per cent of the sample have reliable redshift measurements. 
Additionally, $50$ objects are spectroscopically-confirmed BL Lacs without detectable emission lines. 
{A further $8$ calibrators appear to be the active nuclei of nearby galaxies such as M87 used for Event Horizon Telescope observations \citep{EHT2019}.}


\section{Sample Properties}
\label{sec:optical}
The properties of the ALMACAL calibrators have been previously discussed in \citet{Bonato2018}, where flux, spectral index, and redshift distributions were presented using an earlier sample of 754 calibrators based on observations up to September 2017. 
In this work, we revisit the redshift distribution with an updated catalogue of 1,047 sources, incorporating revised redshifts. 
Additionally, we include classifications of the sources from the Wide-field Infrared Survey Explorer survey \citep{Wright2010}. 
A comprehensive study of the ALMACAL$-22$ data products, including properties such as sensitivity, resolution, and integration time for the same sample, can be found in \citet{Bollo2024}.

\subsection{Redshift distribution}
\begin{figure*}
    \includegraphics[width=\linewidth]{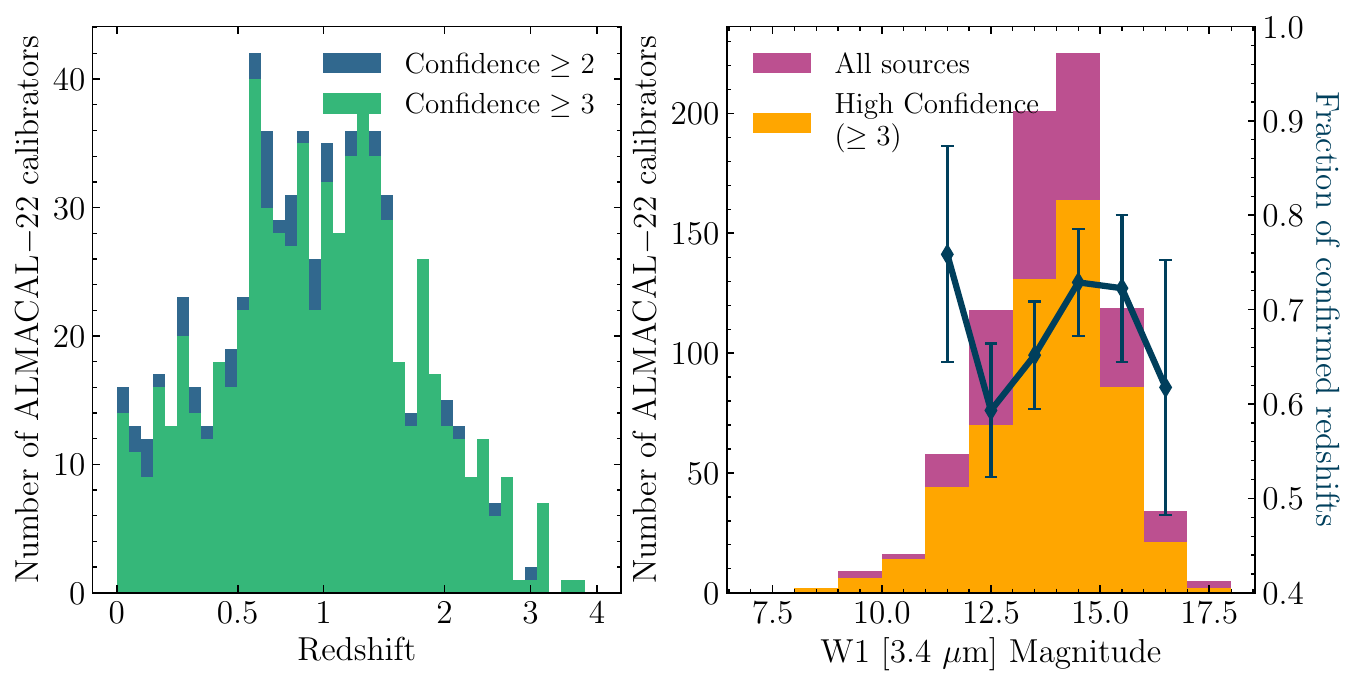}
    \caption{The left histogram shows the redshift distribution of ALMA calibrators. 
    Sources with redshift confidence $\geq 2$ are represented in blue, while sources with high confidence (3 or 4) are shown in green. 
    The distribution peaks at $z \approx 1$, with the highest redshift source at $z = 3.789$. 
    On the right, we display how the fraction of successfully redshifted sources varies with their WISE [3.4 $\mu$m] magnitude. 
    On the right, we display how the fraction of successfully redshifted sources varies with their WISE [3.4 $\mu$m] magnitude. 
    Sources with a measured W1 magnitude are shown in pink, and the subsample with a confident redshift is shown in orange.
    The navy blue line represents the fraction of sources with confident redshifts in each WISE magnitude bin, and we include this fraction only for bins with more than 50 sources to minimise the variance associated with small sample sizes. 
    We find evidence that the fraction increases (between W1 $= 12.5$ to 15 mag) or plateaus (after accounting for errors) as sources become fainter, which would be consistent with the observation that a significant portion of the ALMACAL sample consists of blazars.    
    }
    \label{fig:zDist}
\end{figure*}

We show the redshift distribution of the ALMA calibrators in the left panel of \autoref{fig:zDist}.  
The blue histogram represents the distribution of $735$ sources with redshift confidence $\geq 2$, whereas the green histogram represents sources with confidence 3 or 4. 
Similar to the distribution of band 3 and 6 ALMACAL sources in \citet{Bonato2018}, the distribution peaks at $z \approx 1$, and the highest-redshift source is at $z = 3.789$. 

Additionally, we show in \autoref{fig:zDist} (right panel) how the fraction of sources with a successful redshift varies with their W1 magnitude from WISE. 
We include error bars proportional to $\sqrt N$, where $N$ is the number of sources in a given bin. 
Unlike typical redshift surveys where completeness decreases with increasing magnitude, we find a more varied relationship where completeness appears to mildly increase (or plateau, considering the errors) between W1 $= 12.5$ and 15 mag. 
This is likely due to a significant fraction of the brighter sources in the sample being blazars, where the non-thermal continuum masks emission features, making redshift determination difficult. 
The precise classification of ALMACAL$-22$ sources will be discussed in the forthcoming section.

\subsection{WISE classification} 
\begin{figure}
    \includegraphics[width=1.0\linewidth]{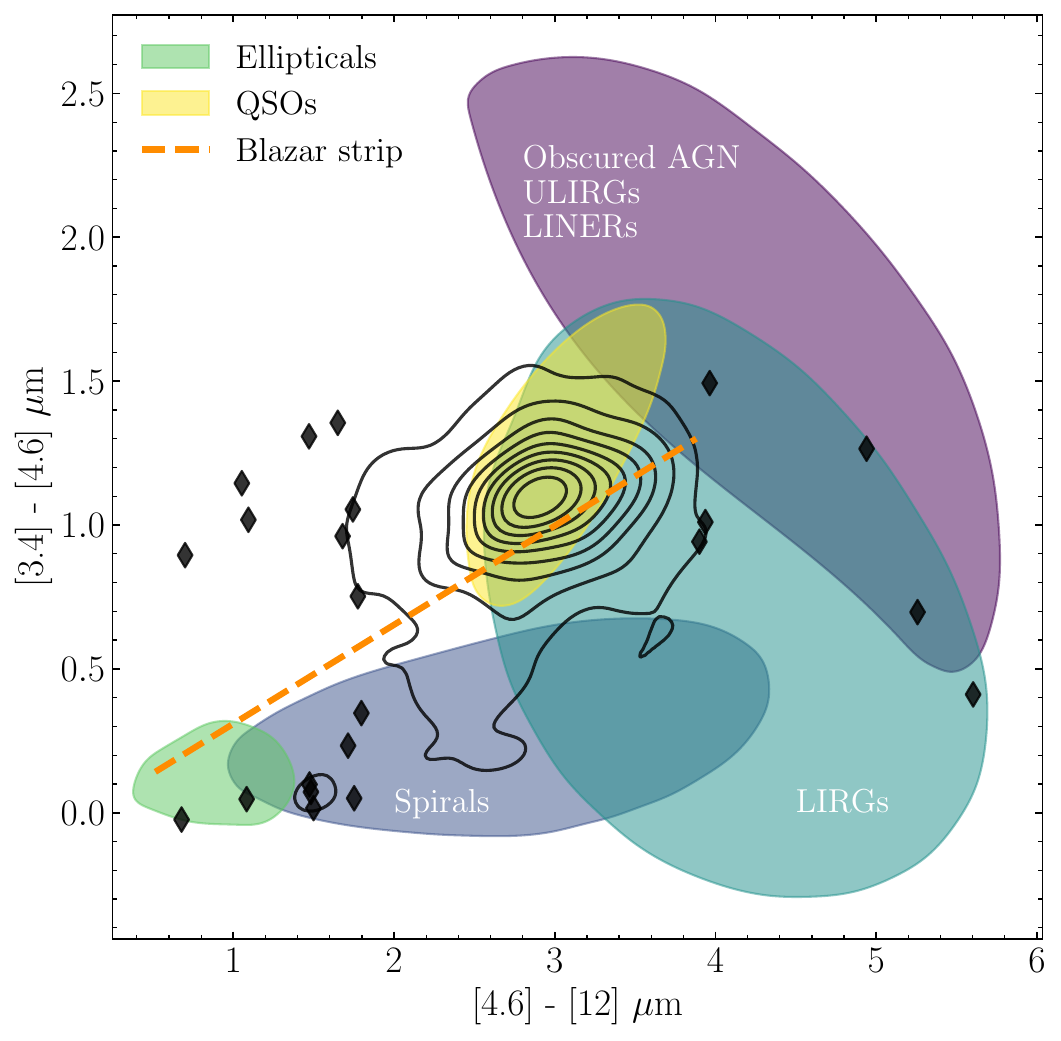}
    \caption{{The positions of the ALMACAL sources are shown with black contours and black diamonds, reflecting outliers, on a reconstructed WISE colour-colour plot \citep{Wright2010}. 
    The orange dashed line indicates the position of blazars, which are dominated by non-thermal emission \citep{Massaro2011}.}
    The inner contour encloses approximately 10 per cent of the sources, with each successive contour increasing the number of enclosed calibrators by an additional 10 per cent. 
    According to the WISE classifications, the majority of ALMACAL sources are quasars and blazars, aligning with previous findings \citep{Bonato2018}. 
    \label{fig:WISE}}
\end{figure}

We present classifications of the sources based on their infrared magnitudes from the WISE survey in \autoref{fig:WISE} \citep{Wright2010}.
Using \textsc{astroquery} \citep{Ginsburg2019}, we retrieve WISE magnitudes for the cross-matched optical positions discussed in \autoref{sec:Methods} from the NASA Extragalactic Database.
In total, 774 sources have measurements of the W1, W2, and W3 magnitudes, corresponding to 3.4, 4.6, and 12 $\mu$m, respectively, which are required to produce the WISE colour-colour plot \citep{Wright2010}.

Using the W1, W2, and W3 magnitudes, the WISE colour-colour plot highlights the regions typically occupied by different object types. 
As seen in the contours from a kernel density estimator, most ALMACAL sources fall within the QSO and Seyfert loci, consistent with these sources being active galactic nuclei (AGN), particularly quasars and blazars. 
Objects outside these contours (22 purple diamonds) are potentially dust-obscured AGN or the nuclei of nearby galaxies. 
Blazars have been found to delineate a specific region of the WISE colour-colour diagram, known as the WISE blazar strip \citep{Massaro2011, Massaro2012}, which corresponds to the green dashed line in \autoref{fig:WISE}. 
Separating blazars that are dominated by non-thermal emission from the larger population of quasars is challenging based on WISE colours alone, as they occupy similar regions in the diagram. 
{However, the alignment of the contours with the gradient of the blazar strip suggests that the ALMACAL$-22$ sample is predominantly composed of blazars. 
This result independently corroborates the findings of \citet{Bonato2018}, which, based on an earlier version of the ALMACAL catalogue (up to September 2017), determined that 97 per cent of the sources are consistent with being blazars, with approximately half classified as flat-spectrum radio quasars. 
Their classification was based on a cross-match with the Roma Multifrequency Catalogue of Blazars \citep{Massaro2009} and an analysis of the low-frequency spectral index ($\alpha_{\rm low}$, measured between 1 and 5 GHz), supplemented by variability studies and the presence of associated $\gamma$-ray emission.
}

\section{Radio Spectral Energy Distribution}
\label{sec:radio}
Fitting the radio spectral energy distributions (SEDs) of ALMACAL sources is particularly interesting because they represent the brightest (sub-)mm sources in the sky. 
Typically, SED fitting is limited to frequencies $\leq$ 20 GHz, with the upper frequency set by the AT20G survey \citep{Murphy2010}. 
This process comes with challenges; these sources are often highly variable, and dust emission may need to be accounted for at these high frequencies. 
Nonetheless, these challenges also provide opportunities to study the radio properties of these sources using SEDs.

In this section, we present radio SED fits using the recently developed \textsc{RadioSED}\footnote{\url{https://github.com/ekerrison/RadioSED/}} algorithm \citep{Kerrison2024}. 
\textsc{RadioSED} offers an automated framework to retrieve data from large-area surveys and fit the corresponding data using a Bayesian nested sampling approach. 
The SED is fitted using four analytical models: 
i) a power law for steep, flat, or inverted spectra, 
ii) a simple peaked spectrum source model \citep{Snellen1998}, 
iii) a peaked spectrum model with curvature \citep{Orienti2007} and 
iv) a re-triggered source modelled by a linear combination of the power law and simple peaked spectrum source models.  
For a more detailed explanation of the modelling framework and its applications, we refer readers to \citet{Kerrison2024}. 
Out of the 1,047 ALMACAL sources, we perform SED fitting for 897 objects that are observed in the Rapid ASKAP Continuum survey \citep[RACS-low;][]{Hale2021}. 

\begin{figure*}
    \includegraphics[width=\linewidth]{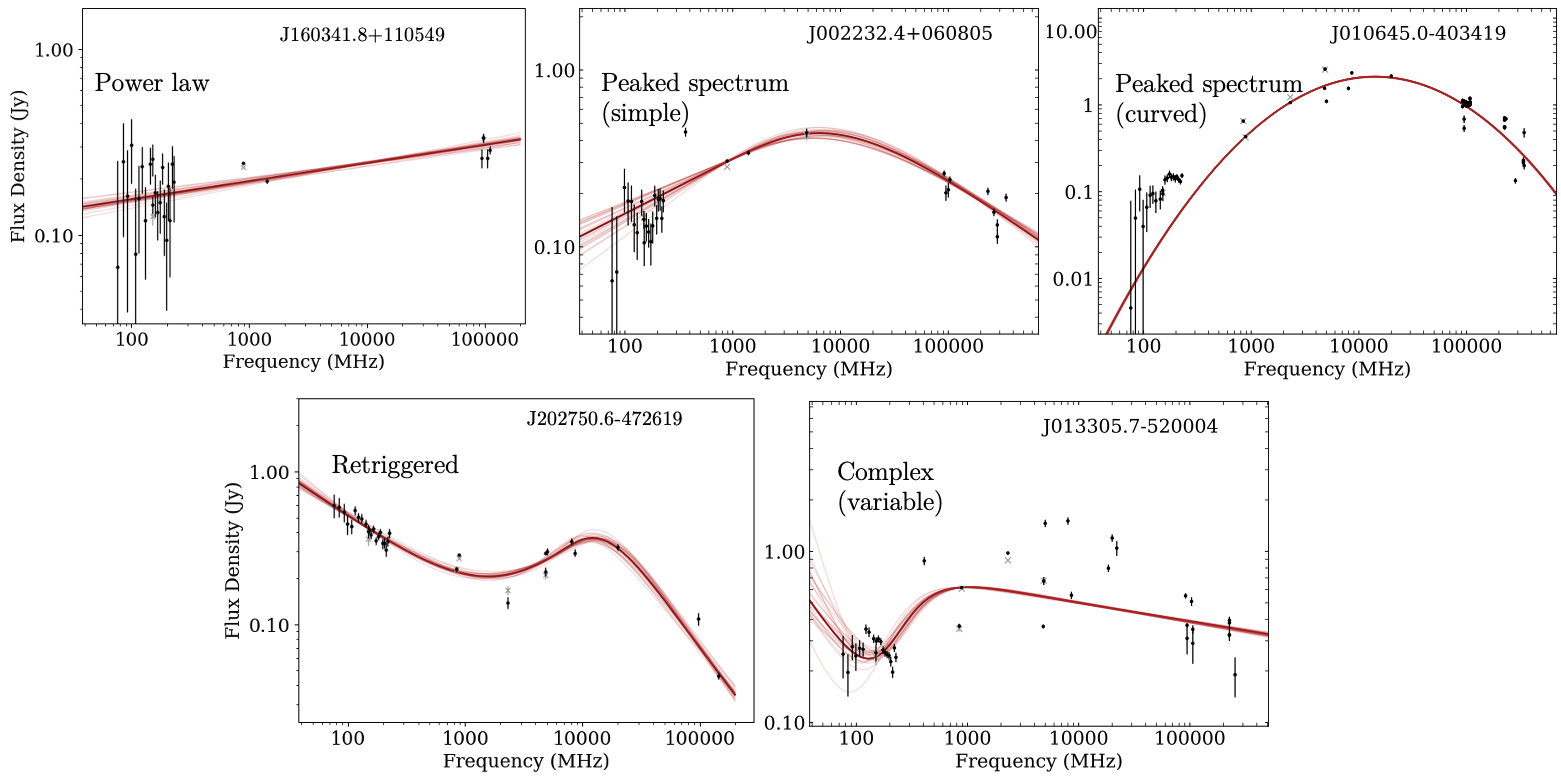}
    \caption{
    Example spectral energy distributions and the four analytic models fitted (power law, simple peaked spectrum, curved peaked spectrum and a re-triggered source). 
    A final source is included which has been labelled as `complex' because the fit is unreliable due to variability. 
    }
    \label{fig:SEDExample}
\end{figure*}

Five example SEDs are displayed in \autoref{fig:SEDExample}, with four labelled by their corresponding models: power law, peaked spectrum (simple and curved) or re-triggered. 
The fifth example has been fit with a re-triggered model that has significant residuals due to variability. 
While there are four models used to fit the data, we use a marginally different set of classifications by grouping both peaked spectrum models together. 
We also introduce a final class labelled `complex' to describe data that is poorly-fitted, has unusual model parameters or unnatural spectral indices. 
Variability is typically responsible for these fits as shown in the bottom-right SED of \autoref{fig:SEDExample}. 
The data span more than 20 epochs which can allow sources to exhibit variability. 
We note that after visually inspecting each SED, we find no evidence for dust emission that could not be explained by variability. 

\begin{figure*}
    \includegraphics[width=\linewidth]{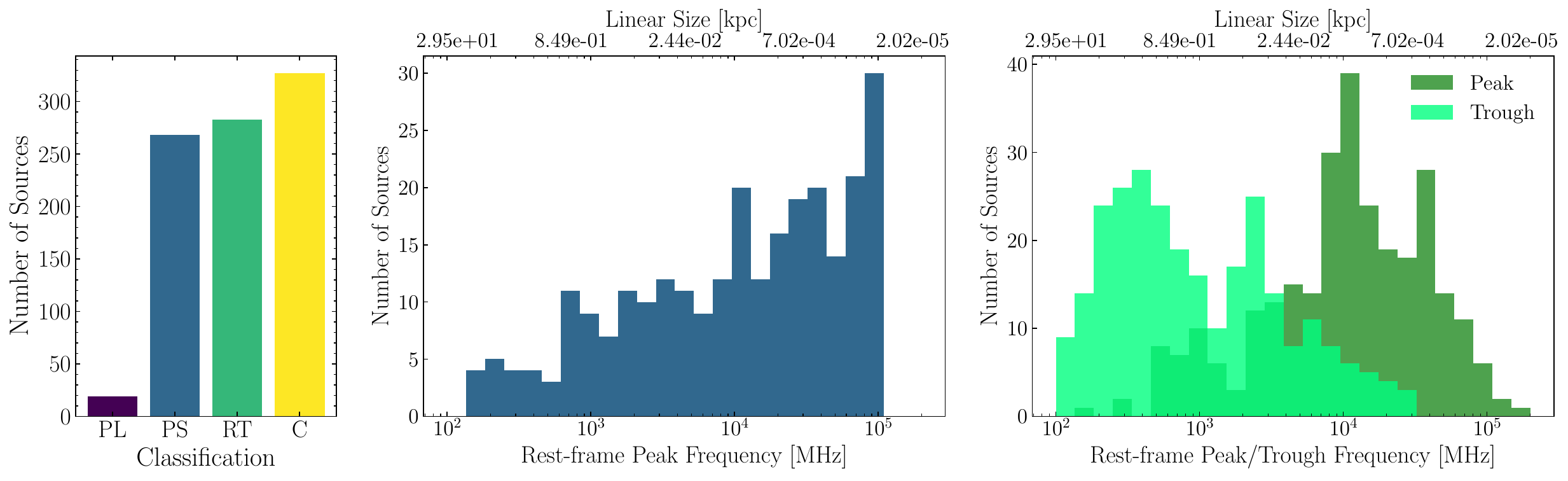}
    \caption{
    The distribution of radio spectral energy distribution classifications for 897 sources is shown in the left histogram. 
    The categories include power law (PL), peaked spectrum (PS), re-triggered (RT), and complex (C) SEDs. 
    For the peaked spectrum sources, we display their rest-frame peak frequencies, which approximately correspond to the linear sizes shown on the secondary $y$-axis (derived using the scaling relation from \citet{Jeyakumar2016}). 
    We repeat this for re-triggered sources in green, showing both the peak and trough frequencies in darker and lighter shades, respectively. 
    }
    \label{fig:SEDDist}
\end{figure*}

The resulting distribution of the final classifications are shown in the left panel of \autoref{fig:SEDDist}. 
We find that ALMA calibrator sources are typically peaked spectrum (PS) or re-triggered sources (RT). 
`Complex' (C) sources form the largest category, which is unsurprising given we have found that the majority of ALMACAL sources are blazars. 
In the right histograms of the same figure, we display the distribution of rest-frame peak frequencies for peaked spectrum sources in blue, and both the peak and trough frequencies for the re-triggered sources in green (the trough being the frequency at which the low-frequency upturn begins to dominate the spectrum). 
The sample is clearly biased towards high-frequency peaked spectrum sources which is consistent with the fact that the ALMACAL sample represents the brightest mm-sources on the sky. 
We also include an approximation of the linear source size on the secondary $y$-axis using a scaling relation \citep{Jeyakumar2016}. 
These highlight that the majority of sources are compact, with sub-kpc linear sizes. 

The 30 percent fraction of peaked spectrum sources in this sample is higher than in other samples with lower-frequency coverage \citep{Odea1998, Callingham2017}, and is surprising given the large proportion of blazars this sample contains (typically PS and blazar are considered mutually-exclusive categories, given blazars are highly variable, while peaked spectrum sources are largely quiescent; \citealt{OdeaSaikia2021}). 
Moreover, the peak frequencies are higher, representing a subclass of peaked spectrum sources known as high-frequency peaked (HFP) objects that peak at $\gtrsim$ 5 GHz \citep{Odea1998, Dallacasa2000}. 
While we expect a higher false detection rate for peaks at higher frequencies near the edge of the measurement range \citep{Kerrison2024}, HFP sources are still over-represented in this sample. 
The potential science with this selection of sources will be discussed in the forthcoming section. 

\section{Discussion}
\label{sec:science}
In this section, we discuss ongoing science cases that are being pursued using the ALMACAL survey. 
The unique combination of high-sensitivity ALMA observations and multi-wavelength coverage enables a range of investigations into how galaxies form and evolve. 
We focus on two key topics: the properties of radio-loud active galactic nuclei and the presence of cold molecular gas in the circumgalactic medium.

\subsection{A unique population of radio-loud AGN}
Peaked spectrum sources are often associated with young active galactic nuclei with compact radio emission \citep{Odea1998, Callingham2017, OdeaSaikia2021}. 
The ALMACAL sample consists of the brightest mm-sources on the sky and, hence, is a particularly interesting sample for studying radio-loud AGN with recent jet activity.  
Typically, these sources are considered distinct from blazars, which exhibit compact radio emission due to the alignment of the jet axis and the line of sight \citep{Massaro2011}. 
So the high fraction of both blazars and sources with a peaked SED in ALMACAL$-$22 makes it a useful sample for exploring the multiwavelength properties of each type of object. 
Multi-epoch, multi-band radio monitoring of these sources in the manner of \citet{Ross2022} would allow for a deeper understanding of their radio continuum properties, confirming the peaks identified here from inhomogeneous radio data, and potentially providing insights into the intersection between these two classes of AGN.
In addition, re-triggered sources may provide insights into the timescales at which radio jets switch on and off, placing an upper limit on the AGN duty cycle, or into the environments responsible for these multiple epochs of activity \citep{Hogan2015}.
Estimating an age from the SED requires a spectral break above the peak frequency \citep[see][for an example]{Allison2019}, which, after visual inspection for all sources, is not evident in this sample. 
An alternative is to track sources using imaging to obtain a dynamical age. 
This is more feasible for the ALMACAL sample, given that many sources have VLBI observations spanning more than 20 epochs \citep{Petrov2025}. 
The broad range of redshifts ($0 < z < 4$) also allows for studying jet ages and duty cycles across cosmic time.

Moreover, serendipitous CO emission, consistent with the calibrator redshift, has been detected in the ALMACAL fields \citep{Bollo2025}.
This presents a unique opportunity to study the impact of radio jets on molecular gas reservoirs in AGN and investigate potential jet-induced outflows.
In cases where galaxies at the calibrator redshift have been detected, analyzing their environment could offer insights into the mechanisms that may have re-triggered AGN activity.
The mechanisms driving AGN activation remain uncertain \citep{Fabian2012, Harrison2024}, but further analysis of the ALMACAL sample may yield valuable insights.

\subsection{Probing the molecular gas content of the CGM}
Theoretical and observational studies increasingly suggest that cold molecular gas can persist in the hot environment of the circumgalactic medium \citep{Ginolfi2017, Cicone2021}. 
Recent simulations indicate that sufficiently large molecular clouds in a hot medium ($T > 10^5$ K) can ``splinter'' into a mist of cold droplets that survive \citep{Farber2023}. 
$21$-cm observations of high-velocity clouds (HVCs) in the Milky Way reveal a cold neutral gas component \citep{Kalberla2006, Moss2013, Westmeier2018} and molecular gas has also been detected through the H$_2$ line at UV wavelengths \citep[][for a review]{Richter1999, Richter2001, Sembach2001, Putman2012}. 
The cold phase has also been found entrained within the nuclear wind of the Milky Way \citep{Teodoro2020, Cashman2021b}. 
While individual extragalactic absorbers have been detected \citep{Noterdaeme2017, Balashev2019, Klitsch2021}, absorption-line stacking experiments suggest a scarcity of dust in the CGM, implying a lack of molecular gas \citep{Menard2010, Zhu2013}. 

In a previous experiment, \citet{Klitsch2023} placed constraints on the molecular gas covering fraction of galaxies with a median redshift of $z \approx 0.05$. 
Using a sample of mm-bright quasars and foreground galaxies, the absence of any CO absorption after stacking rules out large, smooth molecular gas reservoirs like those seen at $z > 2$ in extreme overdensities \citep{Cicone2014, Cicone2021, Emonts2016} and around massive star-forming galaxies \citep{Ginolfi2017}. 
Specifically, the observations show that clouds with radii of approximately 30 pc and column densities $N(\rm{CO}) \gtrsim 10^{16}$ cm$^{-2}$ must have volume-filling factors of less than 0.2 per cent. 
Instead of stacking at the redshifts of foreground galaxies, one can stack CO spectra at the redshifts corresponding to intervening metal absorption lines such as \ion{Mg}{ii}.  
Ongoing efforts are being made to search for intervening absorbers in the X-Shooter spectra, as well as in publicly available SDSS \citep{York2000, Anand2021, Fresco2024} and DESI \citep{DESIDR1, Napolitano2023} spectra, to generate a sample of over 100 absorbers that can then be used to stack CO spectra. 
With future QSO surveys such as 4MOST ByCycle at resolution $R = 20,000$, the velocity precision of the stacking will further improve \citep{Peroux2023}. 

\section{Conclusions}
The ALMACAL sample is a continuously evolving data set with numerous applications in the field of galaxy evolution and formation. 
We present and release the catalogue of 1,047 calibrators observed up until May 2022, with updated, visually inspected redshifts. 
Our major results are summarised below:

\begin{itemize} 
\item Approximately 65 per cent of the sample have reliable redshift measurements, and an additional 50 objects are spectroscopically confirmed featureless BL Lac objects. 
We also present 70 X-Shooter observations of calibrator sources for the first time. 
More calibrators will be observed in ongoing and future surveys of QSOs \citep{Pieri2016, DESI2016, Peroux2023}.  
\item {The ALMACAL-22 sample has a redshift distribution that peaks at $z \approx 1$ and ranges from the nuclei of galaxies at $z \ll 0.01$, such as M87, to quasars at $z = 3.742$.} 
The fraction of sources with high-confidence redshifts increases or plateaus as sources become fainter in the WISE W1 band, between magnitudes 13 and 15 (\autoref{fig:zDist}). 
This suggests that a significant fraction of the ALMACAL sources are blazars, a result confirmed by plotting their positions on the WISE colour-colour plot (\autoref{fig:WISE}). 
\item Analysing the radio spectral energy distributions of 897 ALMA calibrators reveals that roughly 30 per cent of the objects have peaked spectra, with another third being re-triggered. 
{The fraction of sources with peaked spectra is higher than in other samples based on lower-frequency data \citep{Odea1998, Callingham2017}, which is unexpected given the large number of blazars in this sample.}
Sources with complex SEDs, largely caused by variability, comprise 36 per cent of the fitted calibrators, while only two per cent are consistent with a power-law spectrum. 
The peak frequencies of the peaked spectrum and re-triggered sources span three orders of magnitude, between 100 MHz and 170 GHz, with estimated linear sizes ranging from sub-pc to $> 10$ kpc scales. 
\end{itemize}

The number of ALMA calibrator observations will only grow with time, meaning that more calibrators will be observed, at greater depths, or across more bands. 
Establishing a reliable redshift catalogue is necessary for the many scientific cases possible with this data. 
For example, the search for dusty star-forming galaxies \citep{Chen2023b} and CO emitters \citep{Bollo2025} requires an estimation of the clustering bias introduced when searching around a calibrator source. 
Hence, understanding the properties of the calibrator itself is crucial for controlling this bias.

Moreover, the calibrators themselves are key in a variety of scientific cases that will be explored in the future. 
In an upcoming paper, we will stack CO spectra for the elusive cold gas seen in the circumgalactic medium at the redshift of metal absorption lines in the optical spectra of calibrators. 
Furthermore, combining radio spectral energy distributions with imaging of the calibrator sources may yield unconstrained properties, such as the AGN age and their duty cycle. 
The ALMACAL data set will be a continually valuable resource for the study of how galaxies form and evolve.

\section*{Acknowledgements}
We thank the anonymous referee for providing insightful comments. 
This research is supported by an Australian Government Research Training Program (RTP) Scholarship.
EK, EMS and SW acknowledge the financial support of the Australian Research Council through grant CE170100013 (ASTRO3D).
This research was supported by the International Space Science Institute (ISSI) in Bern, through ISSI International Team project \#564 (The Cosmic Baryon Cycle from Space). 

This research has made use of the NASA/IPAC Extragalactic Database (NED), which is funded by the National Aeronautics and Space Administration and operated by the California Institute of Technology, and the SIMBAD database, operated at CDS, Strasbourg, France. 
The scientific colour map batlow \citep{Crameri2018} is used in this study to prevent visual distortion of the data and exclusion of readers with colour-vision deficiencies \citep{Crameri2020}.

\section*{Data Availability}
The ALMACAL catalogue with redshift data is available online and the reduced X-Shooter spectra can be found in the appendix. 
Based on observations collected at the European Organisation for Astronomical Research in the Southern Hemisphere under ESO programmes 0101.A-0528 and 111.253L.001. 
All other data is available upon request to the corresponding author. 



\bibliographystyle{mnras}
\bibliography{References} 

\appendix
\section{X-Shooter Spectra}
\label{sec:XSHOOSpec}
In this section, we present the remaining 68 out of 70 total X-Shooter spectra from programmes 0101.A-0528 and 111.253L.001. 
Each spectrum has common quasar and galaxy emission lines marked along with the redshift. 
The SNR ranges from 1 to over 100.

\begin{figure*}
    \includegraphics[width=0.95\textwidth]{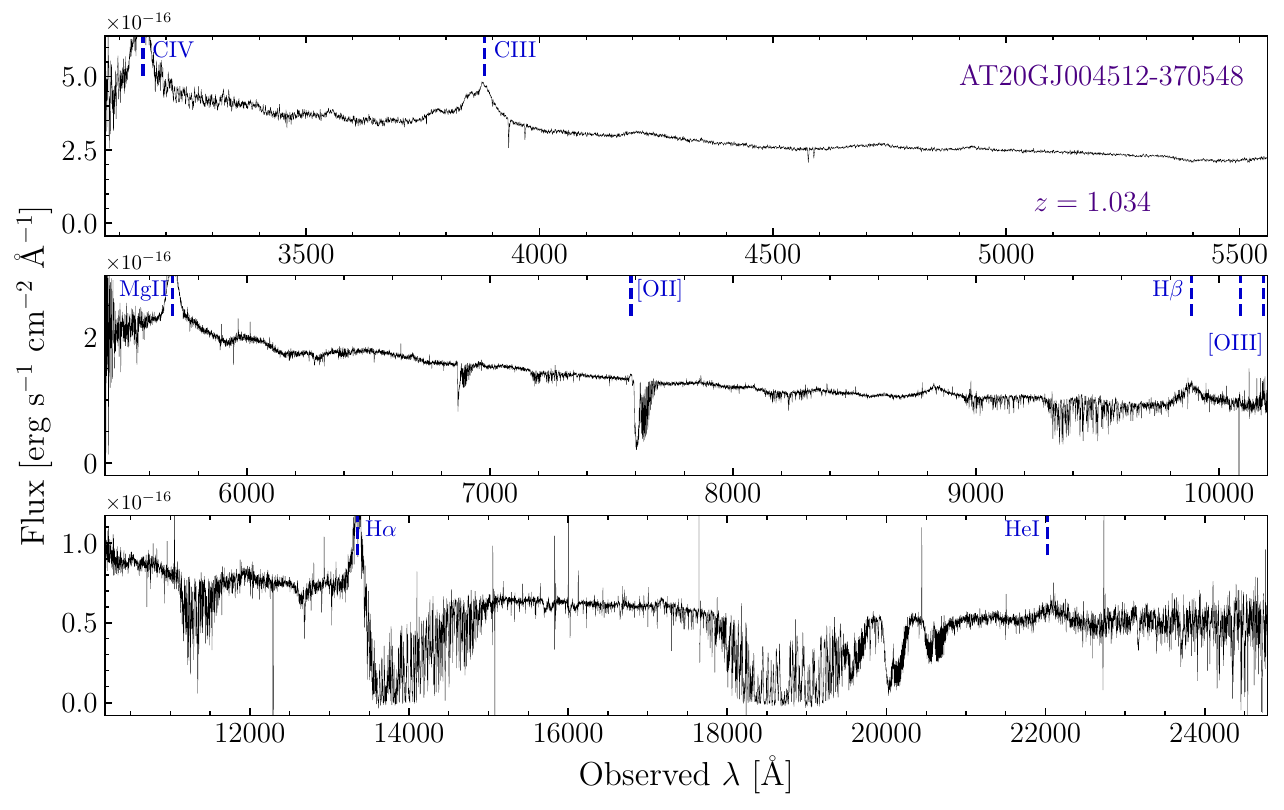}
    \caption{41 X-Shooter spectra from programme 0101.A-0528. 
    }
    \label{fig:XSHOOP101}
\end{figure*}

\begin{figure*}
    \ContinuedFloat
    \captionsetup{list=off,format=cont}
    \includegraphics[width=0.95\textwidth]{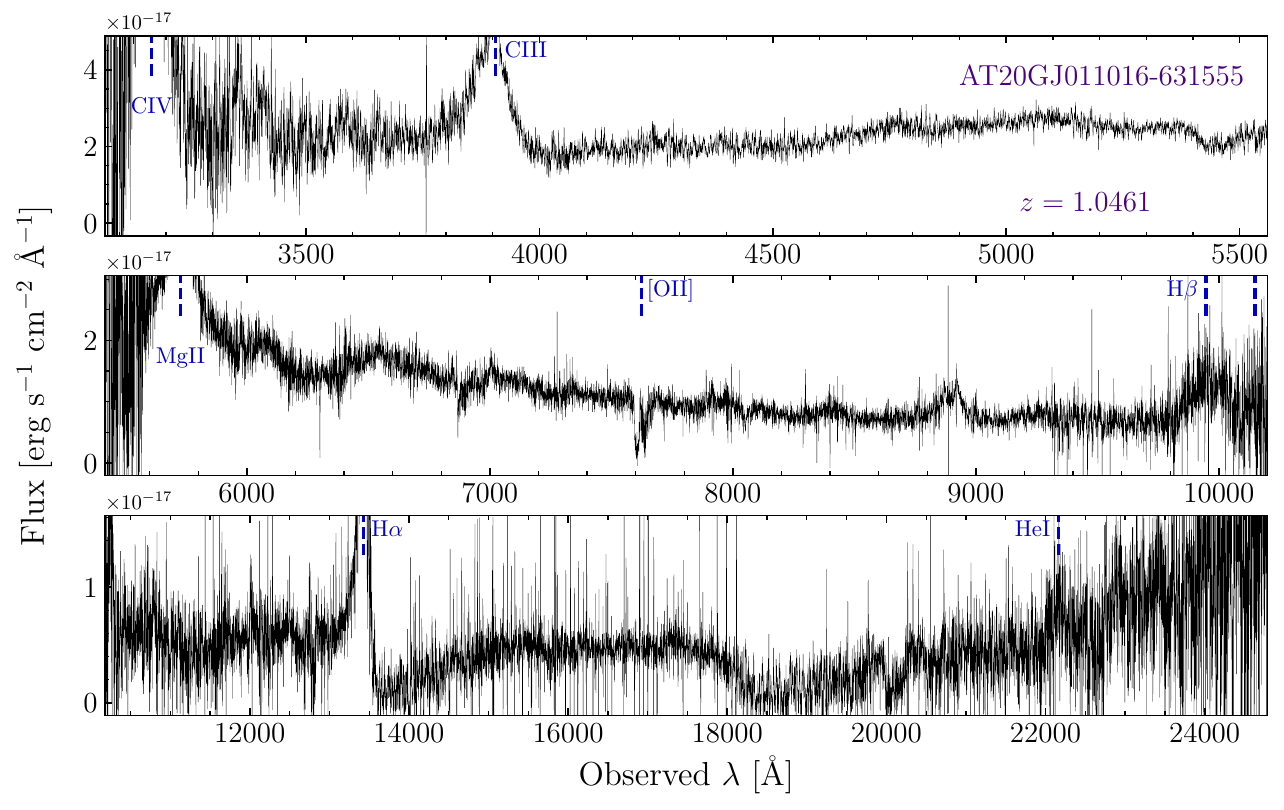}
    \caption{
    }
\end{figure*}

\begin{figure*}
    \ContinuedFloat
    \captionsetup{list=off,format=cont}
    \includegraphics[width=0.95\textwidth]{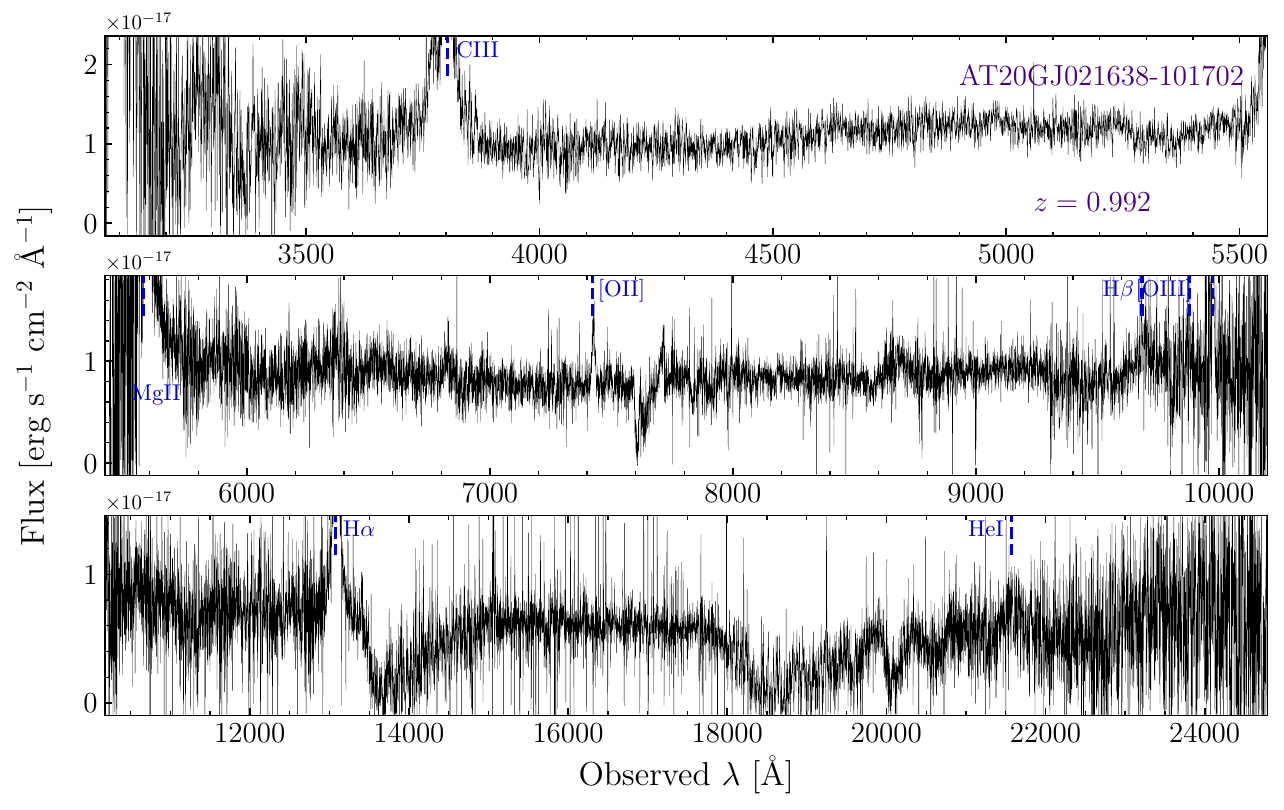}
    \caption{
    }
\end{figure*}

\begin{figure*}
    \ContinuedFloat
    \captionsetup{list=off,format=cont}
    \includegraphics[width=0.95\textwidth]{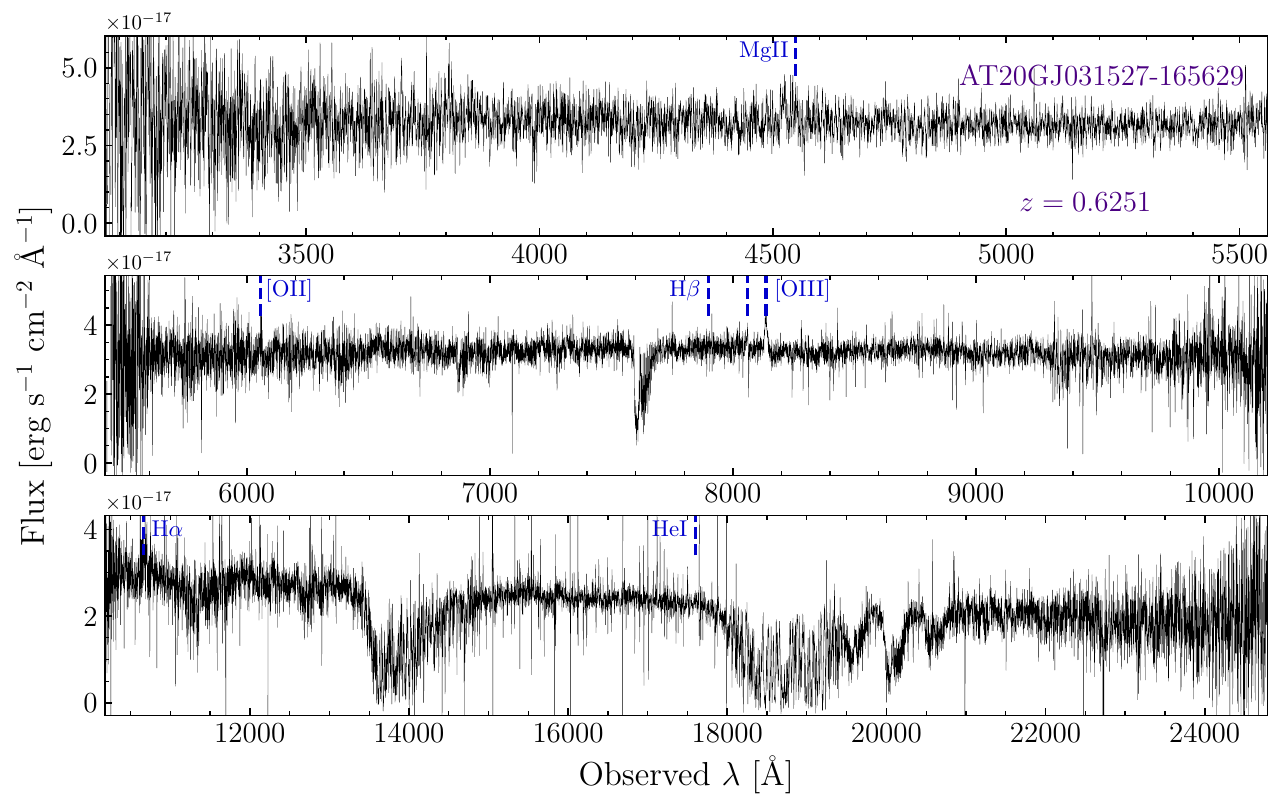}
    \caption{
    }
\end{figure*}

\begin{figure*}
    \ContinuedFloat
    \captionsetup{list=off,format=cont}
    \includegraphics[width=0.95\textwidth]{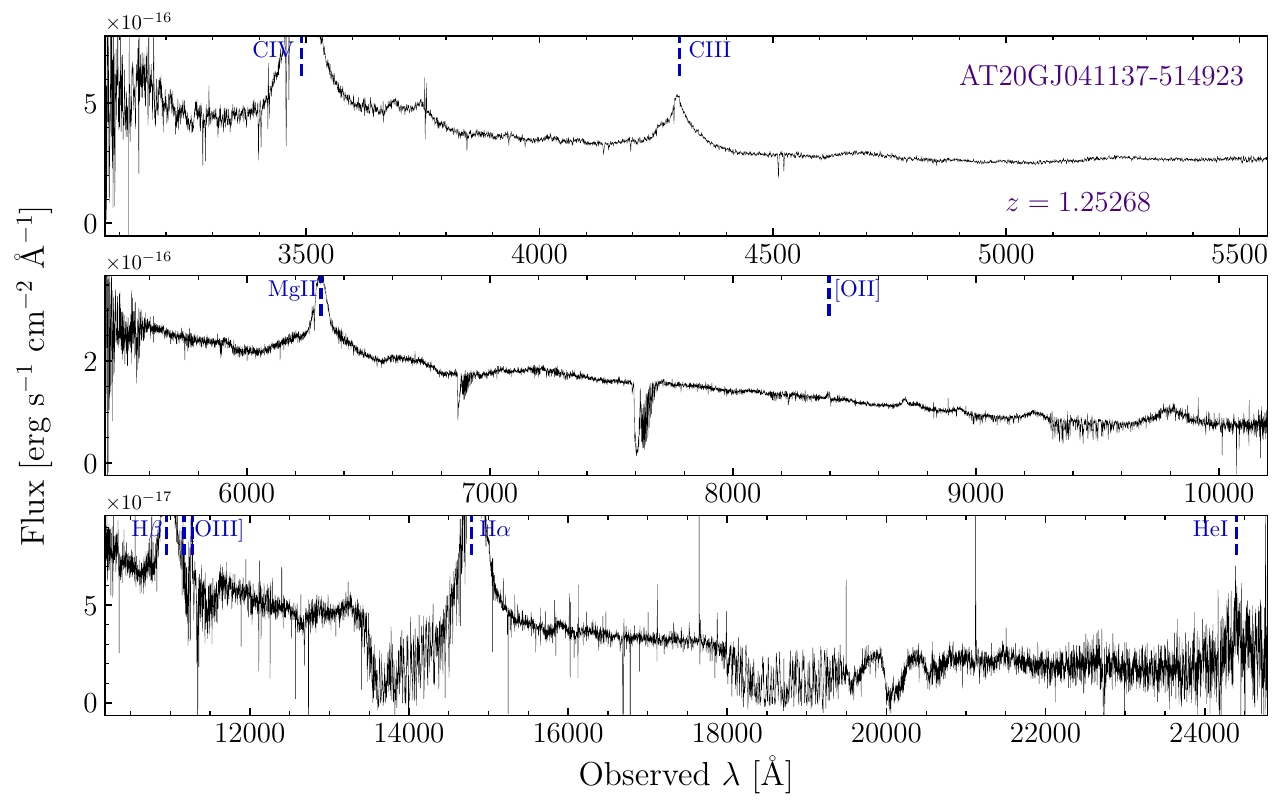}
    \caption{
    }
\end{figure*}

\begin{figure*}
    \ContinuedFloat
    \captionsetup{list=off,format=cont}
    \includegraphics[width=0.95\textwidth]{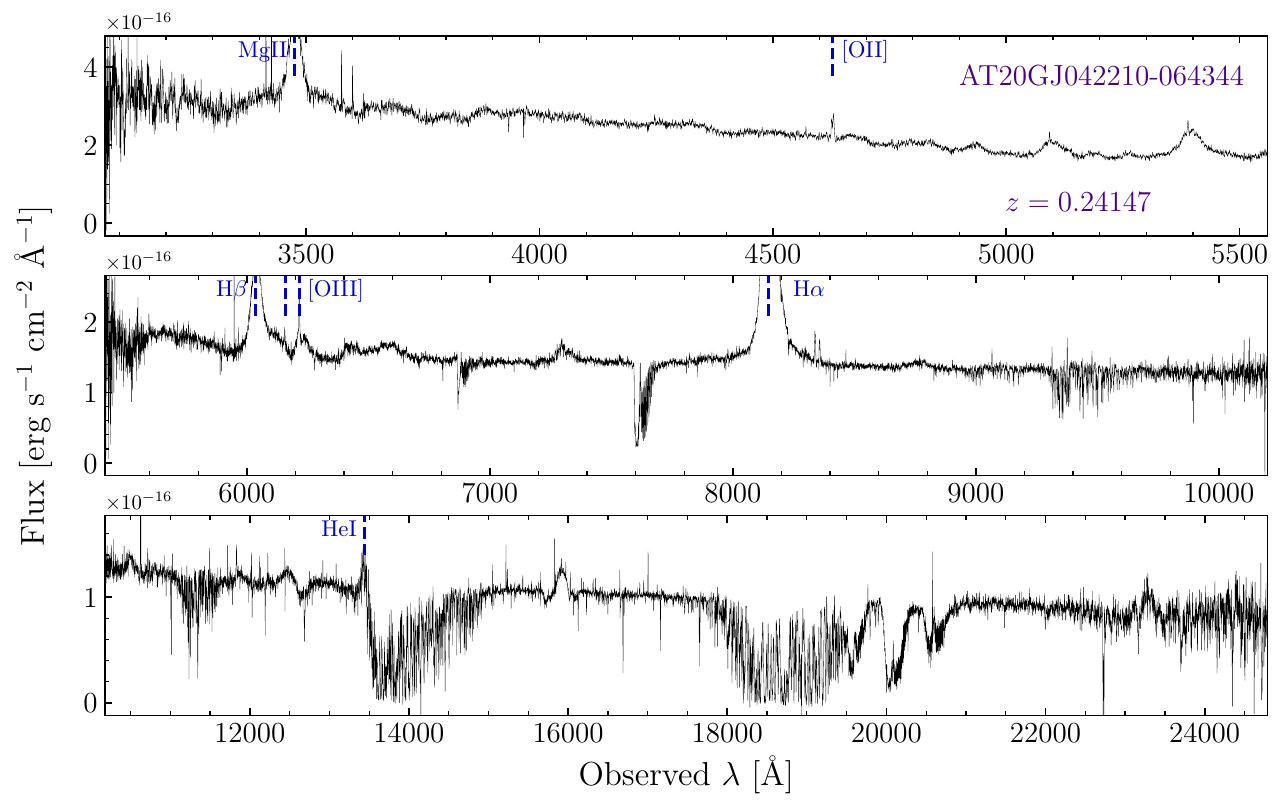}
    \caption{
    }
\end{figure*}

\begin{figure*}
    \ContinuedFloat
    \captionsetup{list=off,format=cont}
    \includegraphics[width=0.95\textwidth]{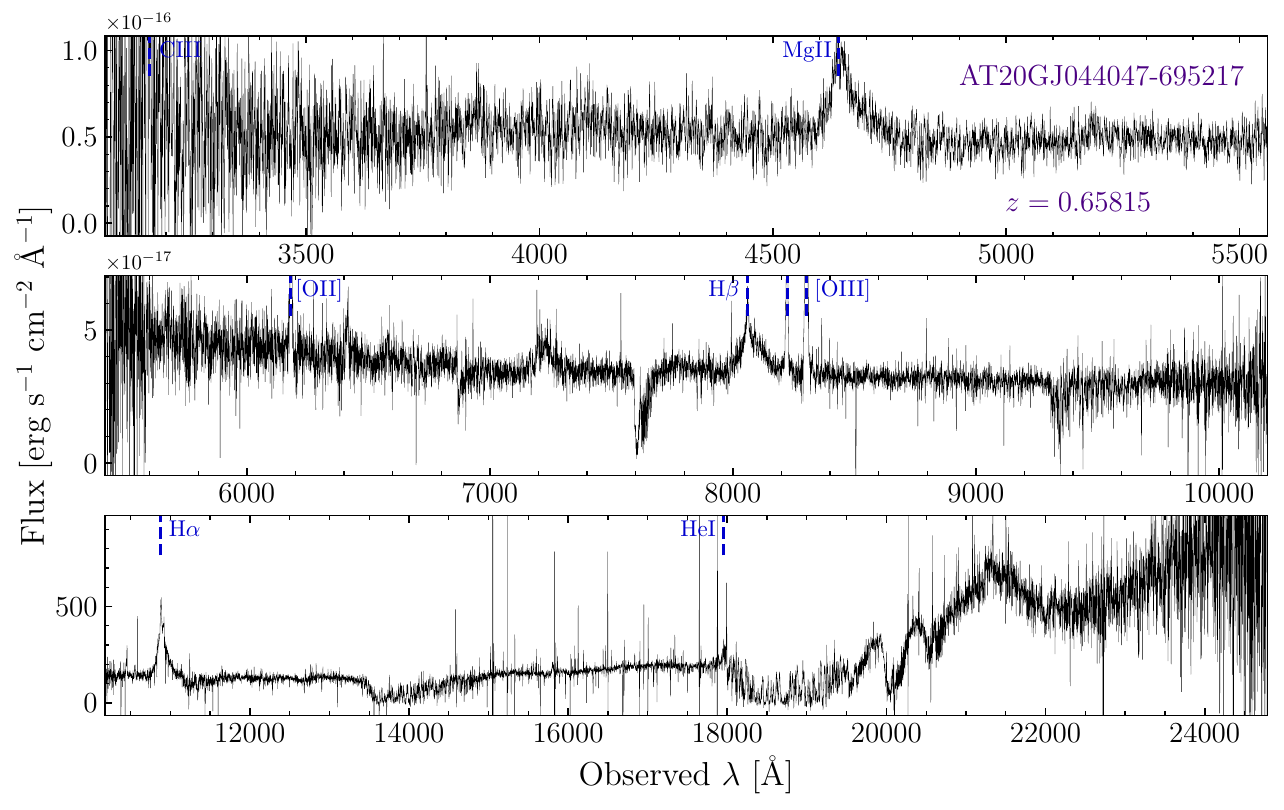}
    \caption{
    }
\end{figure*}

\begin{figure*}
    \ContinuedFloat
    \captionsetup{list=off,format=cont}
    \includegraphics[width=0.95\textwidth]{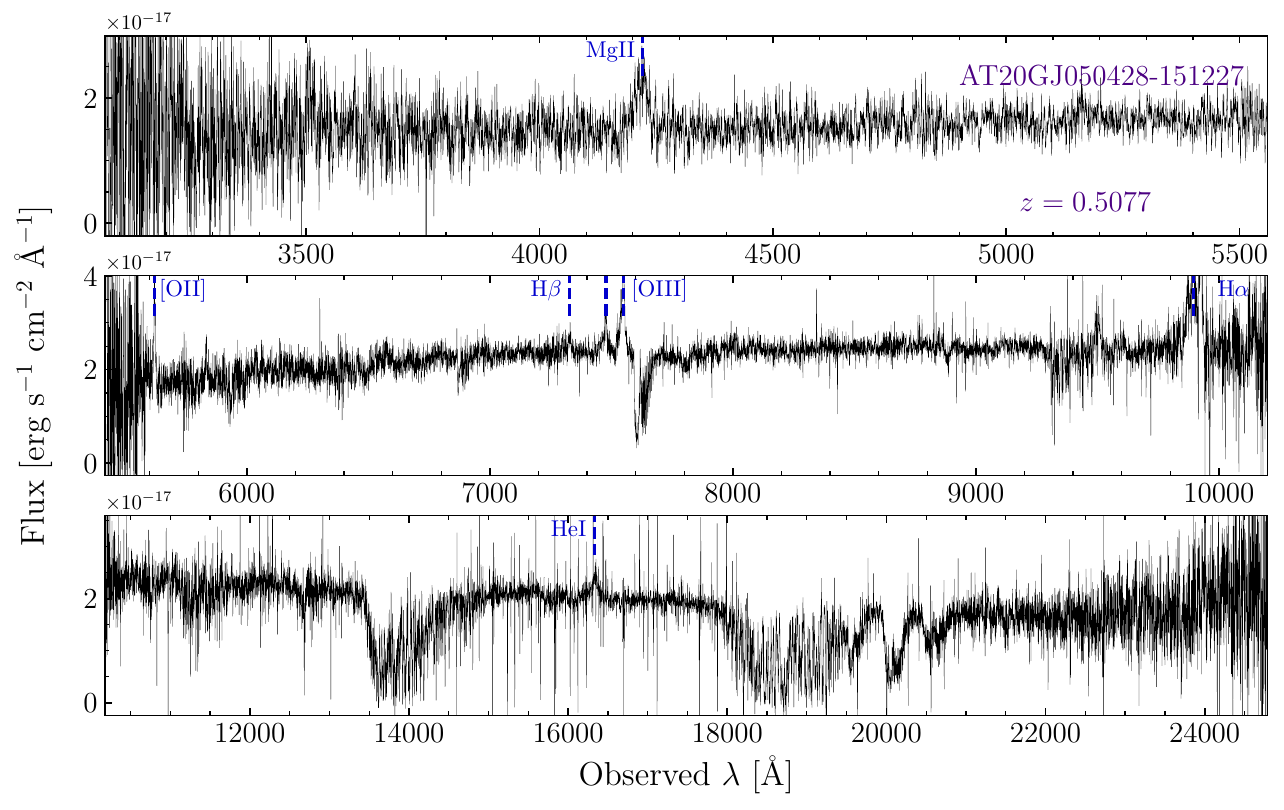}
    \caption{
    }
\end{figure*}

\begin{figure*}
    \ContinuedFloat
    \captionsetup{list=off,format=cont}
    \includegraphics[width=0.95\textwidth]{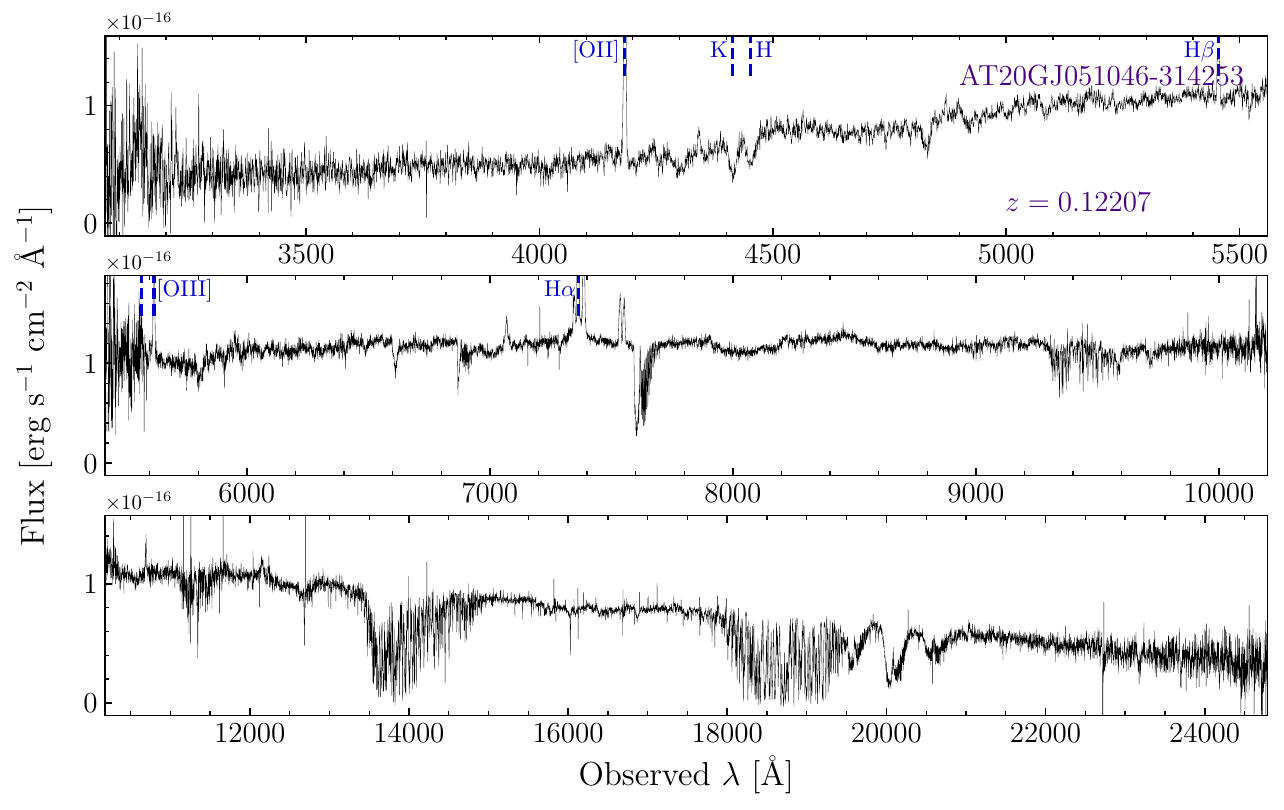}
    \caption{
    }
\end{figure*}

\begin{figure*}
    \ContinuedFloat
    \captionsetup{list=off,format=cont}
    \includegraphics[width=0.95\textwidth]{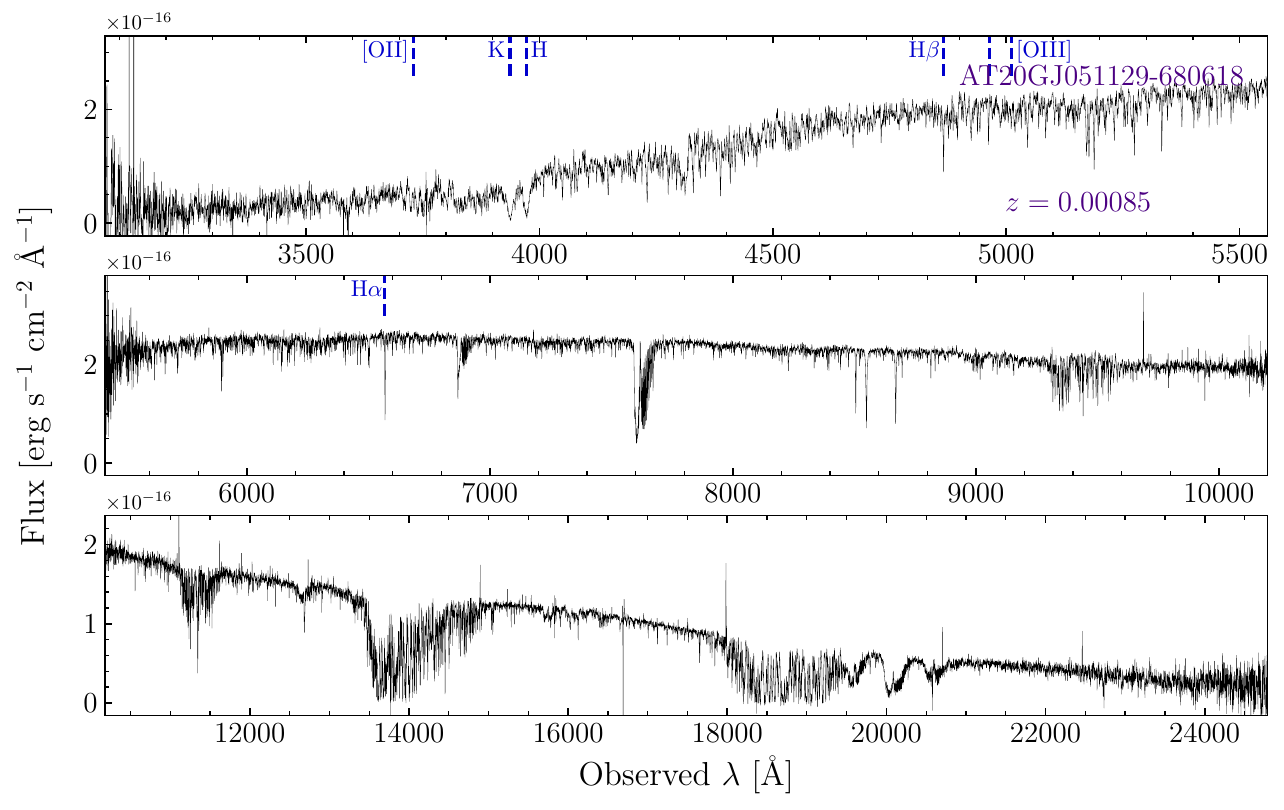}
    \caption{
    }
\end{figure*}

\begin{figure*}
    \ContinuedFloat
    \captionsetup{list=off,format=cont}
    \includegraphics[width=0.95\textwidth]{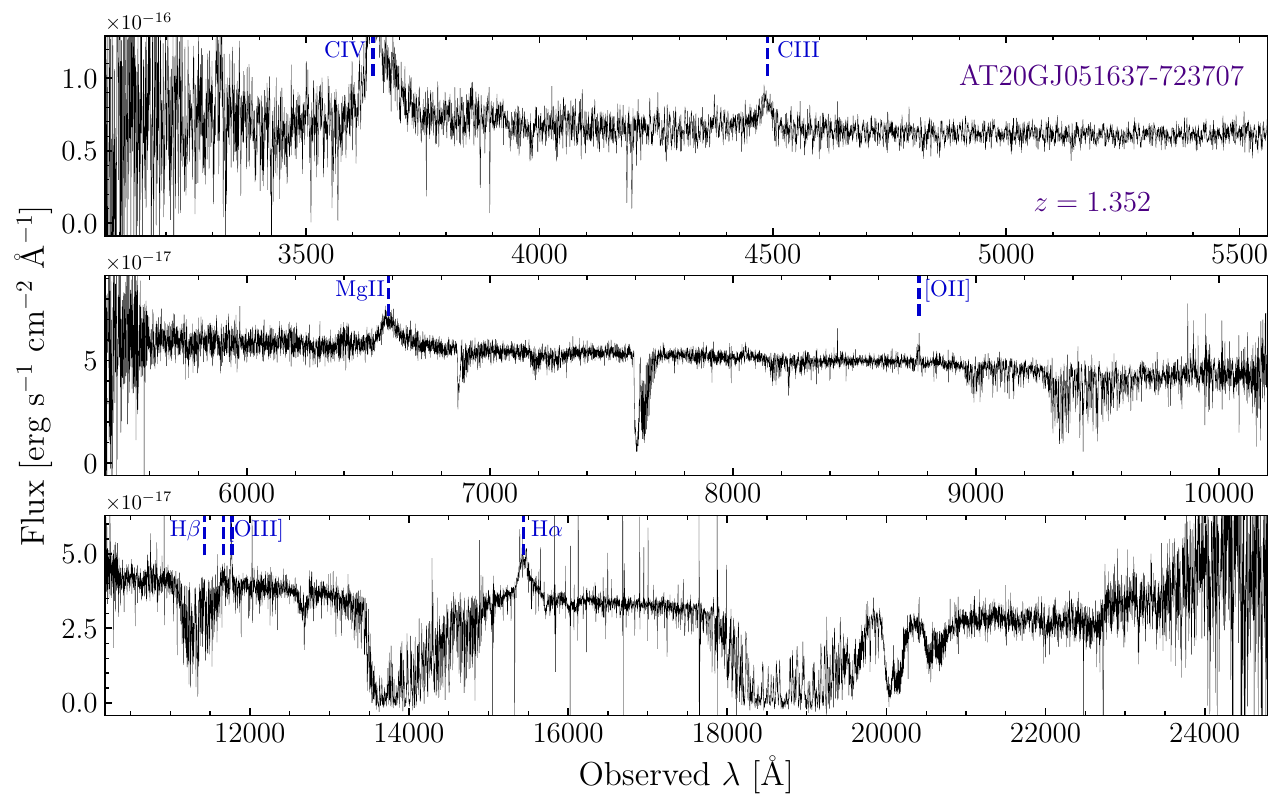}
    \caption{
    }
\end{figure*}

\begin{figure*}
    \ContinuedFloat
    \captionsetup{list=off,format=cont}
    \includegraphics[width=0.95\textwidth]{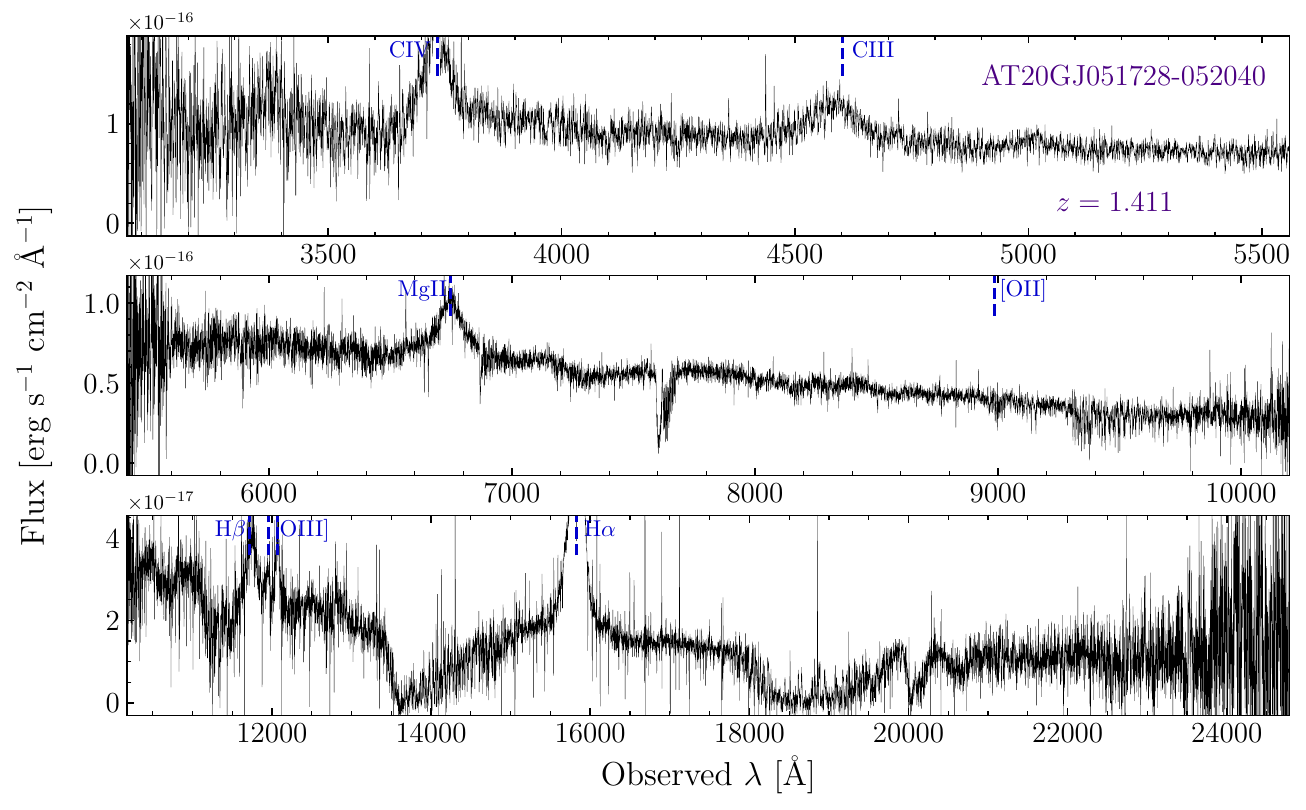}
    \caption{
    }
\end{figure*}

\begin{figure*}
    \ContinuedFloat
    \captionsetup{list=off,format=cont}
    \includegraphics[width=0.95\textwidth]{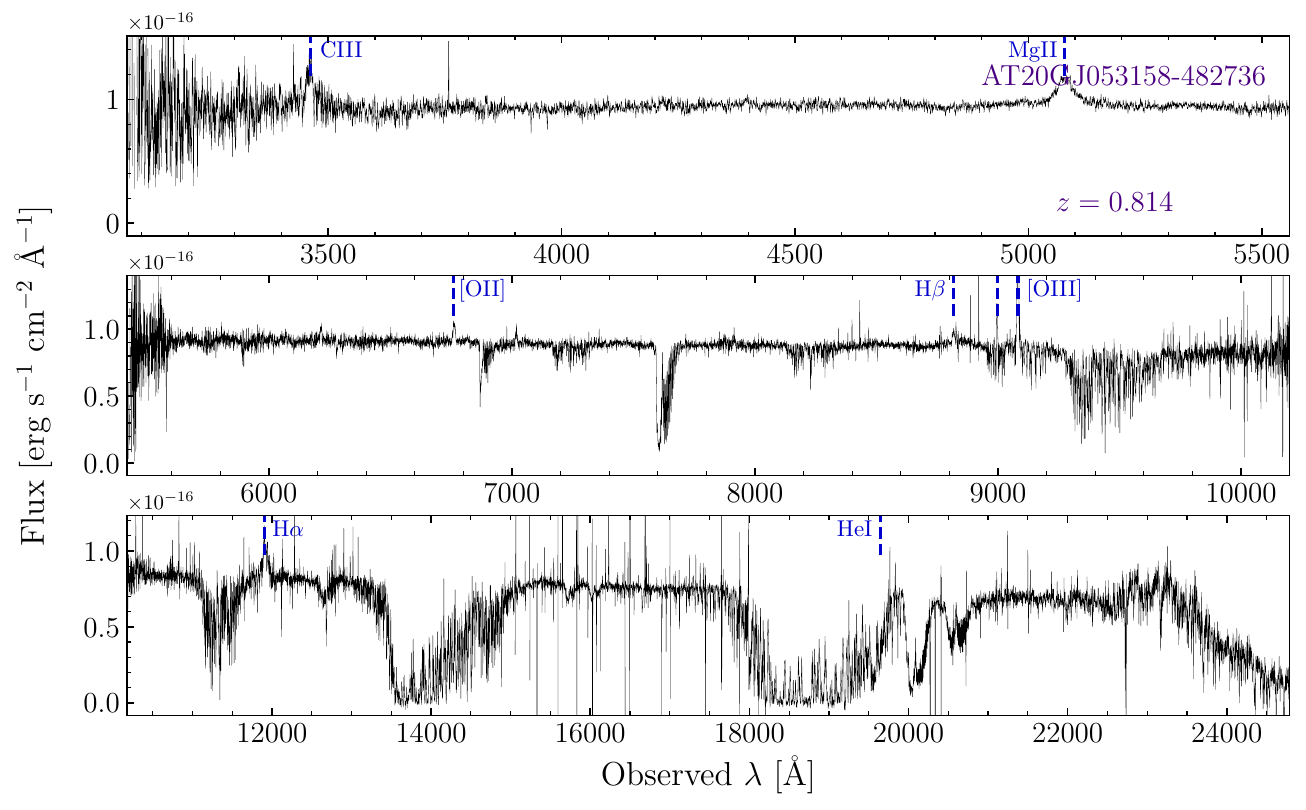}
    \caption{
    }
\end{figure*}

\begin{figure*}
    \ContinuedFloat
    \captionsetup{list=off,format=cont}
    \includegraphics[width=0.95\textwidth]{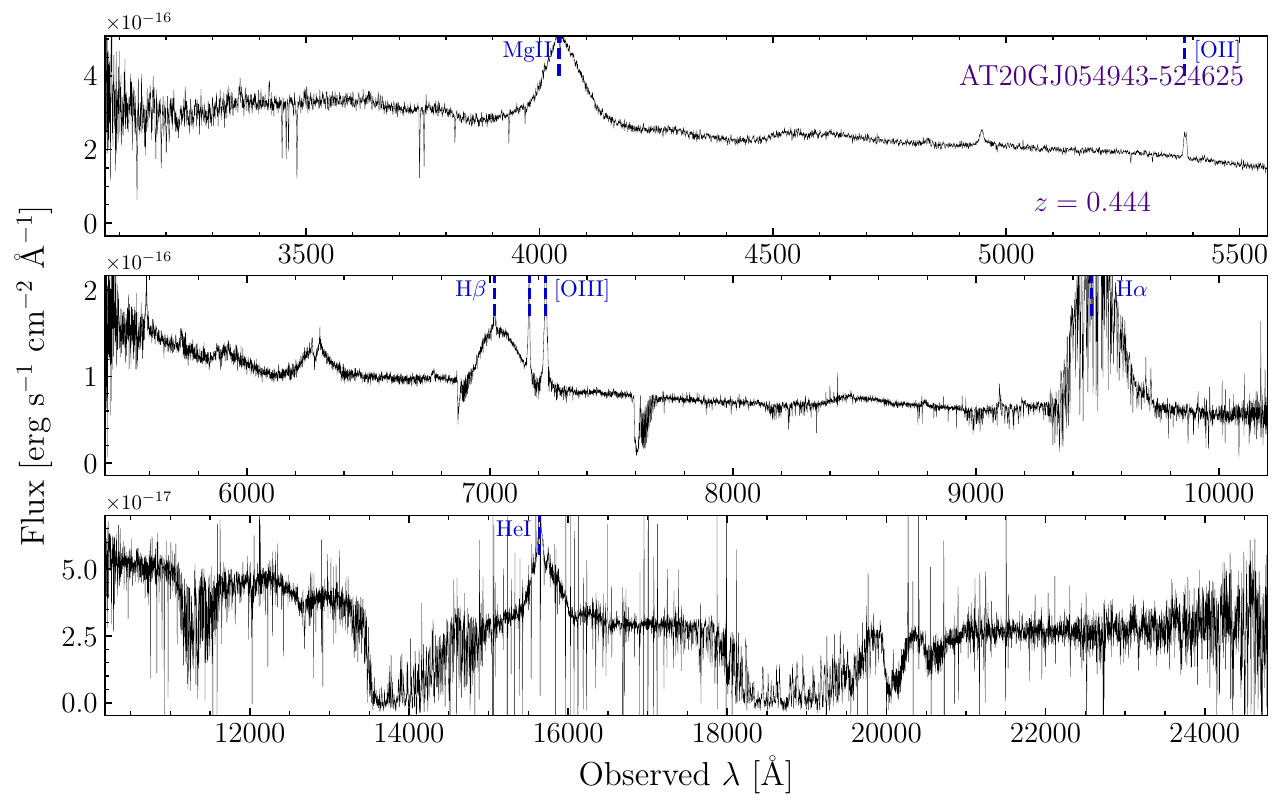}
    \caption{
    }
\end{figure*}

\begin{figure*}
    \ContinuedFloat
    \captionsetup{list=off,format=cont}
    \includegraphics[width=0.95\textwidth]{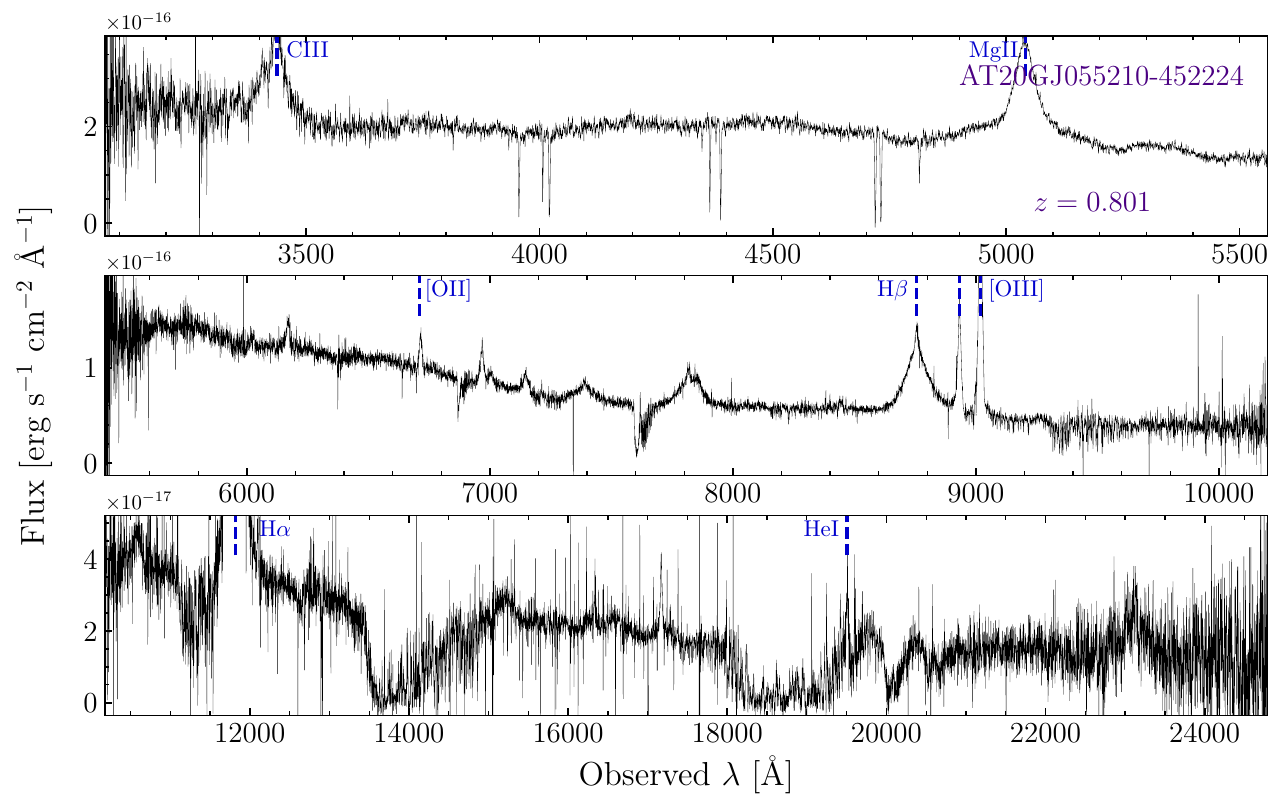}
    \caption{
    }
\end{figure*}

\begin{figure*}
    \ContinuedFloat
    \captionsetup{list=off,format=cont}
    \includegraphics[width=0.95\textwidth]{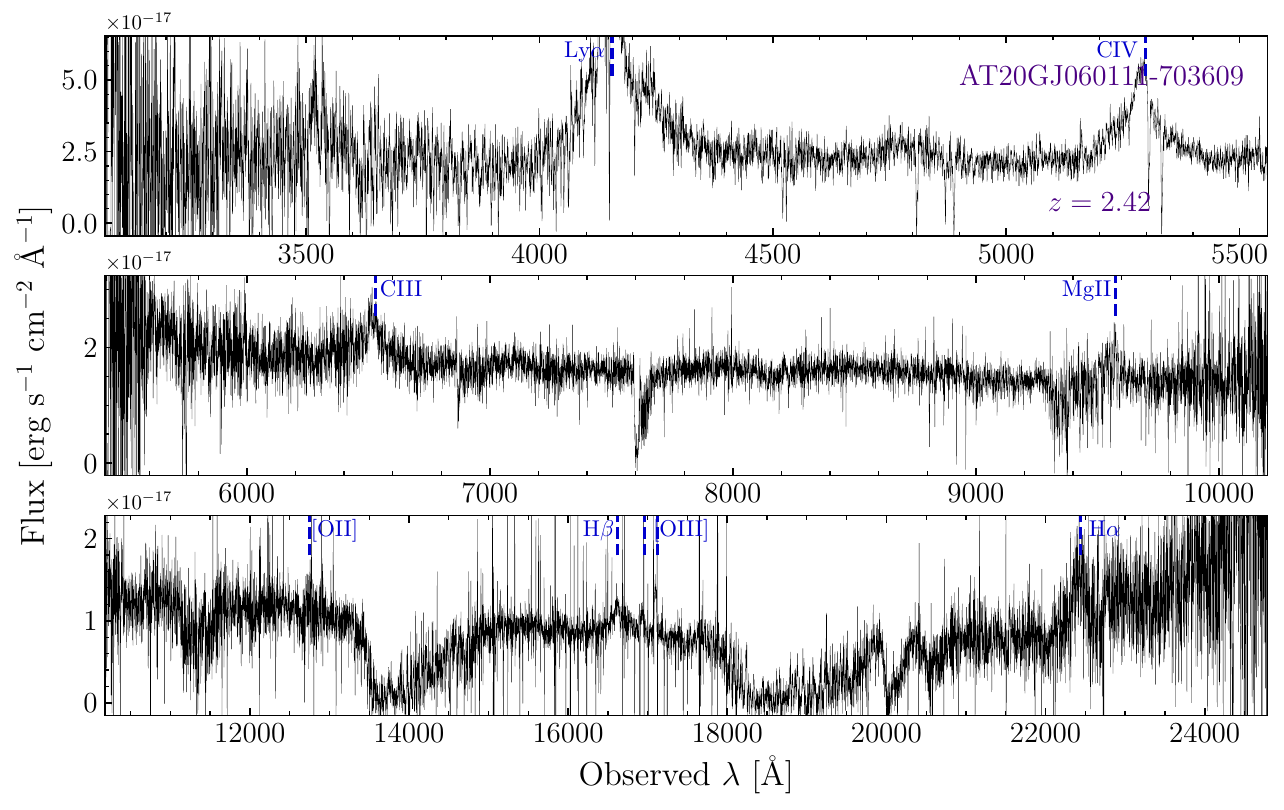}
    \caption{
    }
\end{figure*}

\begin{figure*}
    \ContinuedFloat
    \captionsetup{list=off,format=cont}
    \includegraphics[width=0.95\textwidth]{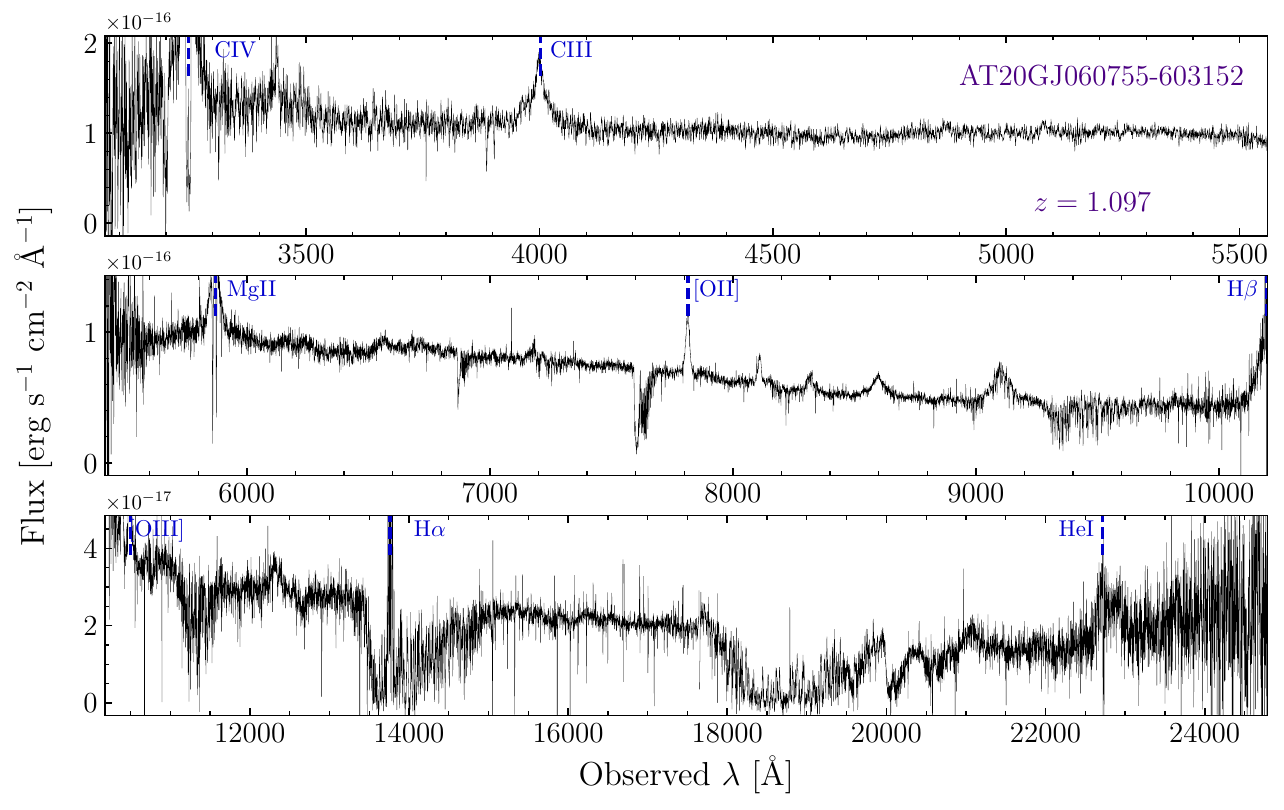}
    \caption{
    }
\end{figure*}

\begin{figure*}
    \ContinuedFloat
    \captionsetup{list=off,format=cont}
    \includegraphics[width=0.95\textwidth]{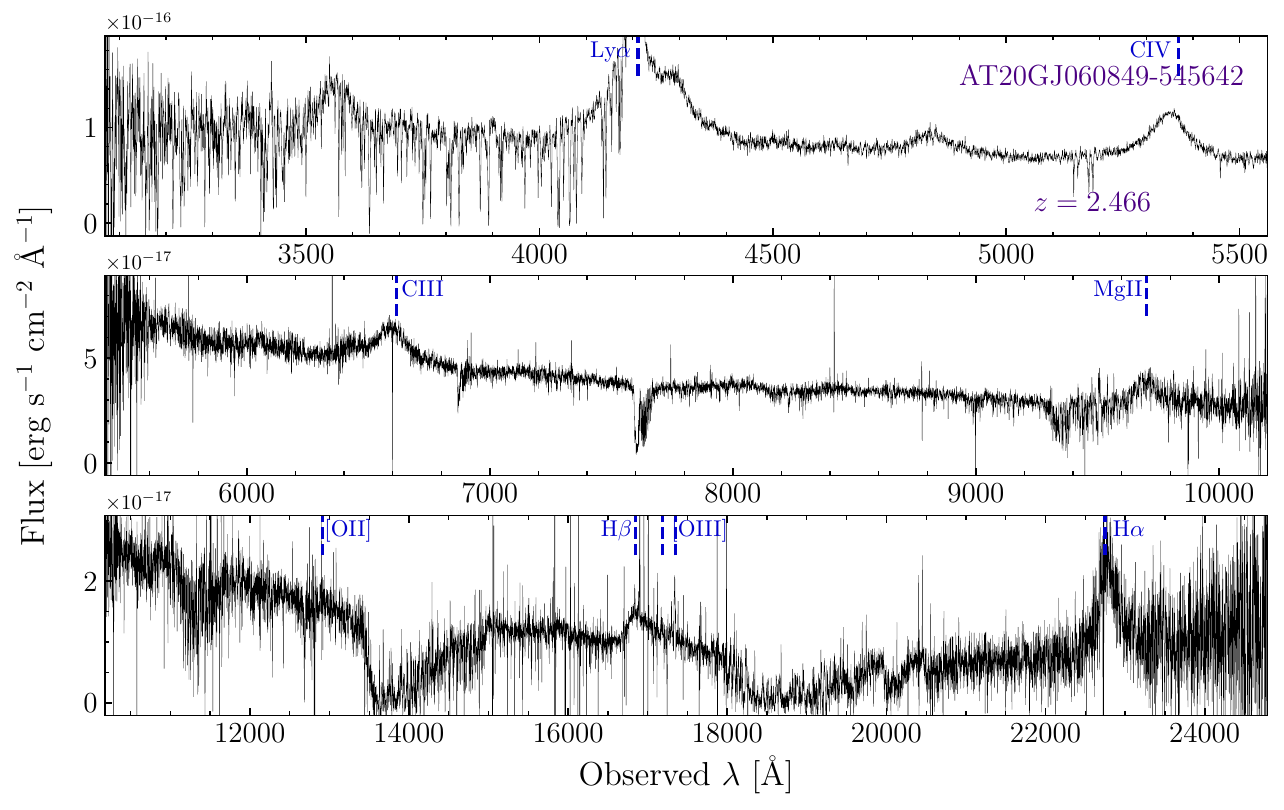}
    \caption{
    }
\end{figure*}

\begin{figure*}
    \ContinuedFloat
    \captionsetup{list=off,format=cont}
    \includegraphics[width=0.95\textwidth]{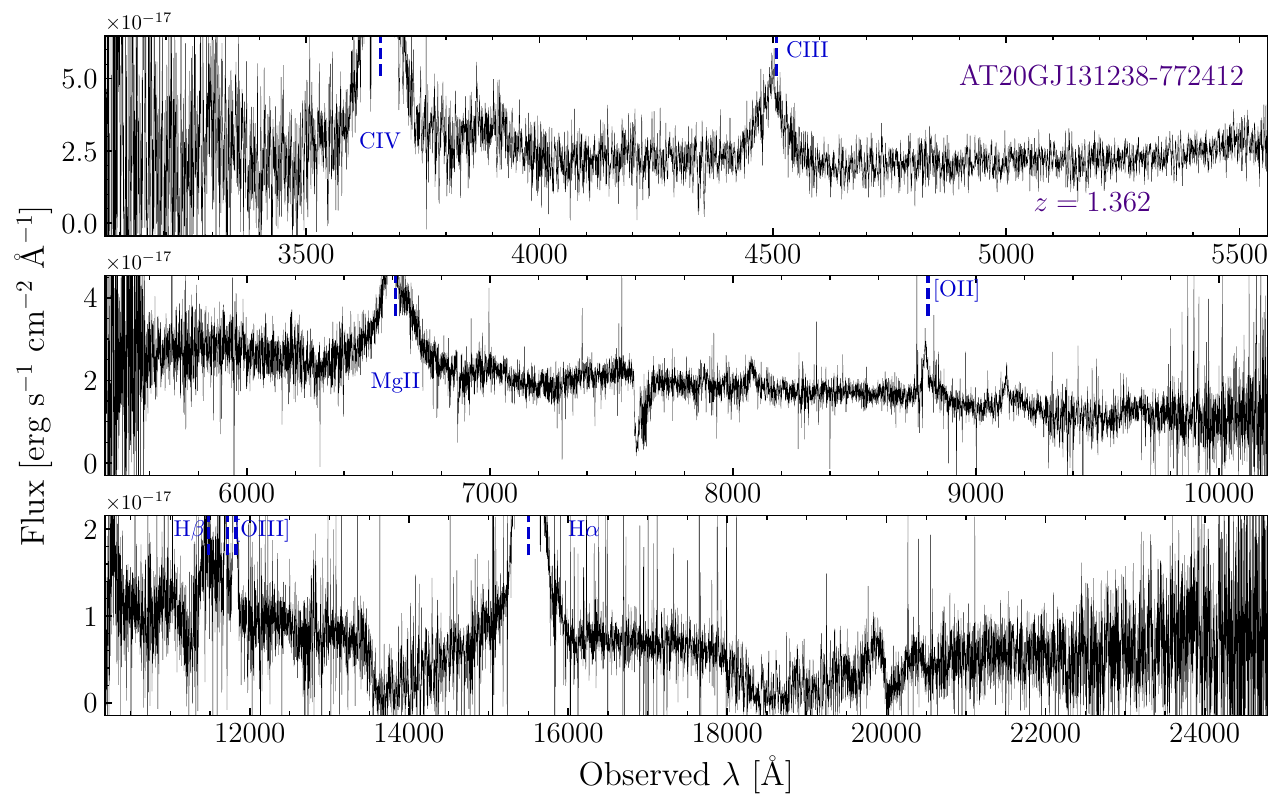}
    \caption{
    }
\end{figure*}

\begin{figure*}
    \ContinuedFloat
    \captionsetup{list=off,format=cont}
    \includegraphics[width=0.95\textwidth]{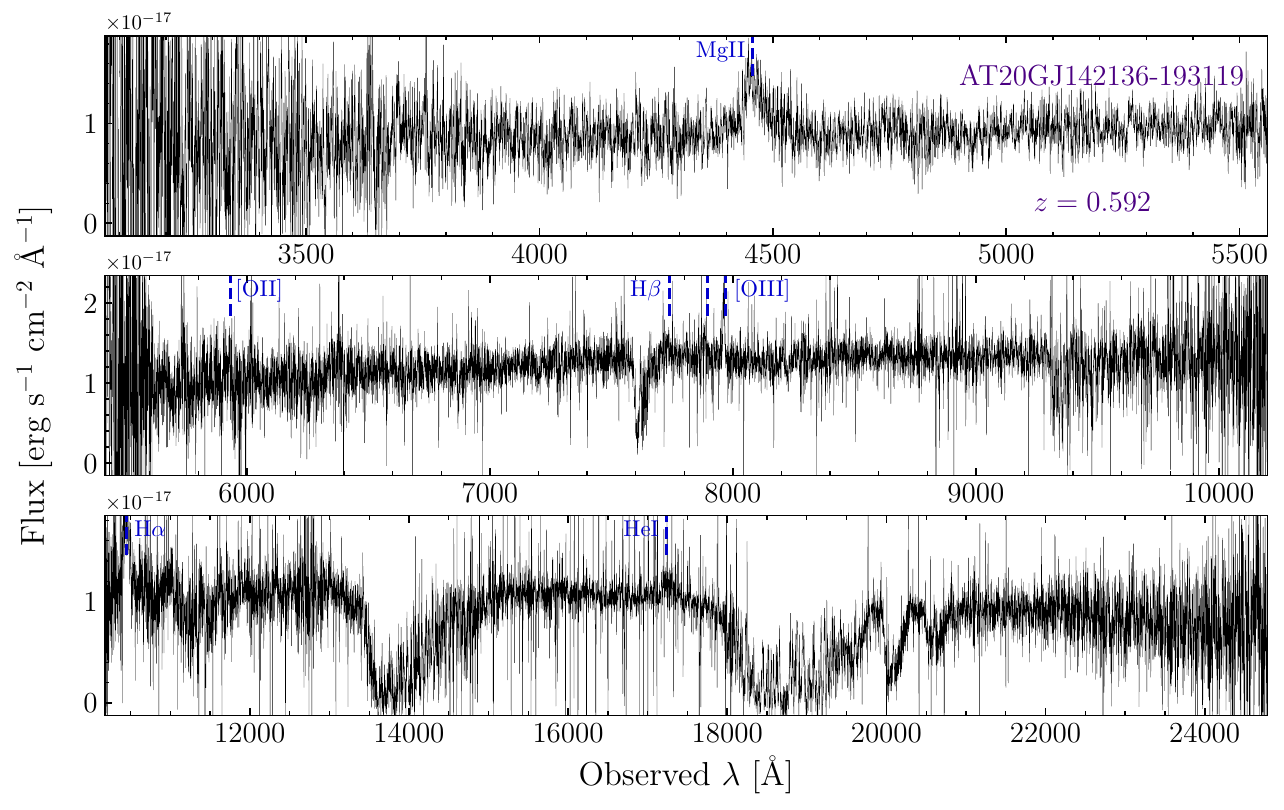}
    \caption{
    }
\end{figure*}

\begin{figure*}
    \ContinuedFloat
    \captionsetup{list=off,format=cont}
    \includegraphics[width=0.95\textwidth]{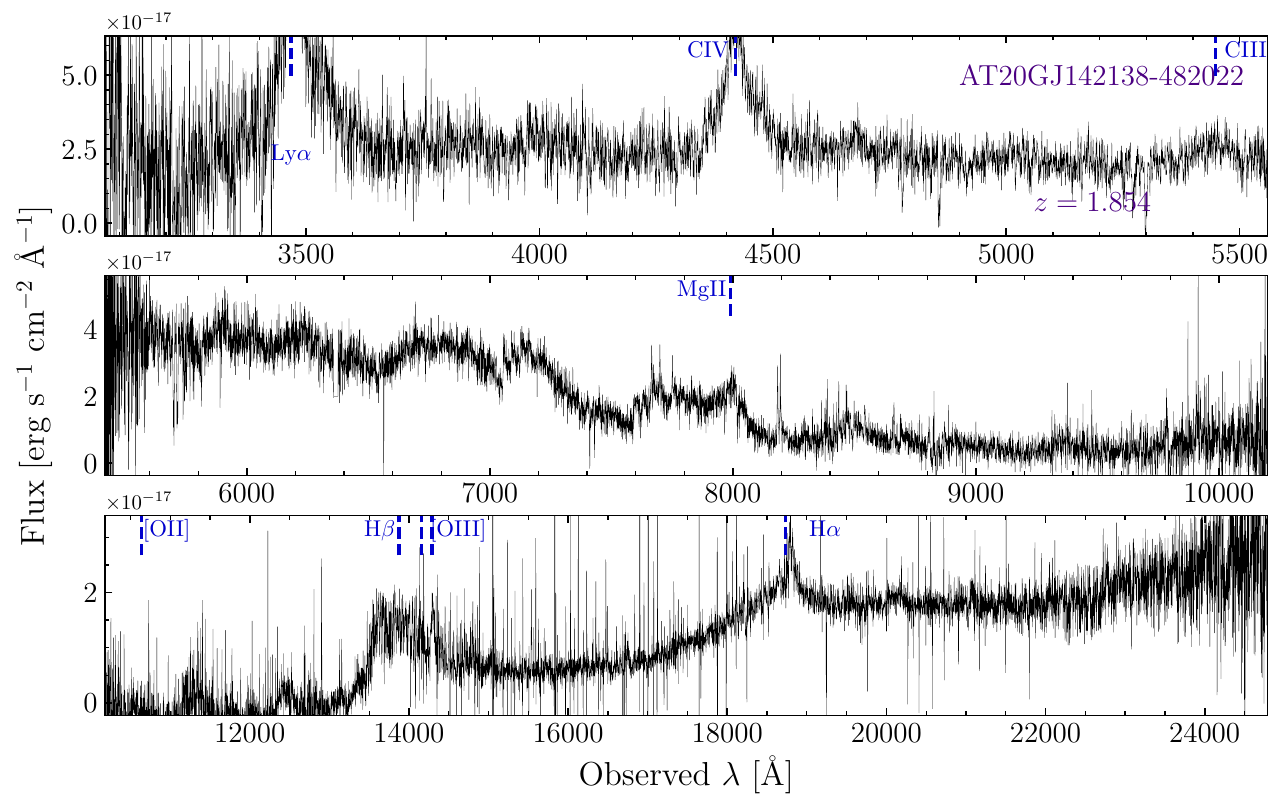}
    \caption{
    }
\end{figure*}

\begin{figure*}
    \ContinuedFloat
    \captionsetup{list=off,format=cont}
    \includegraphics[width=0.95\textwidth]{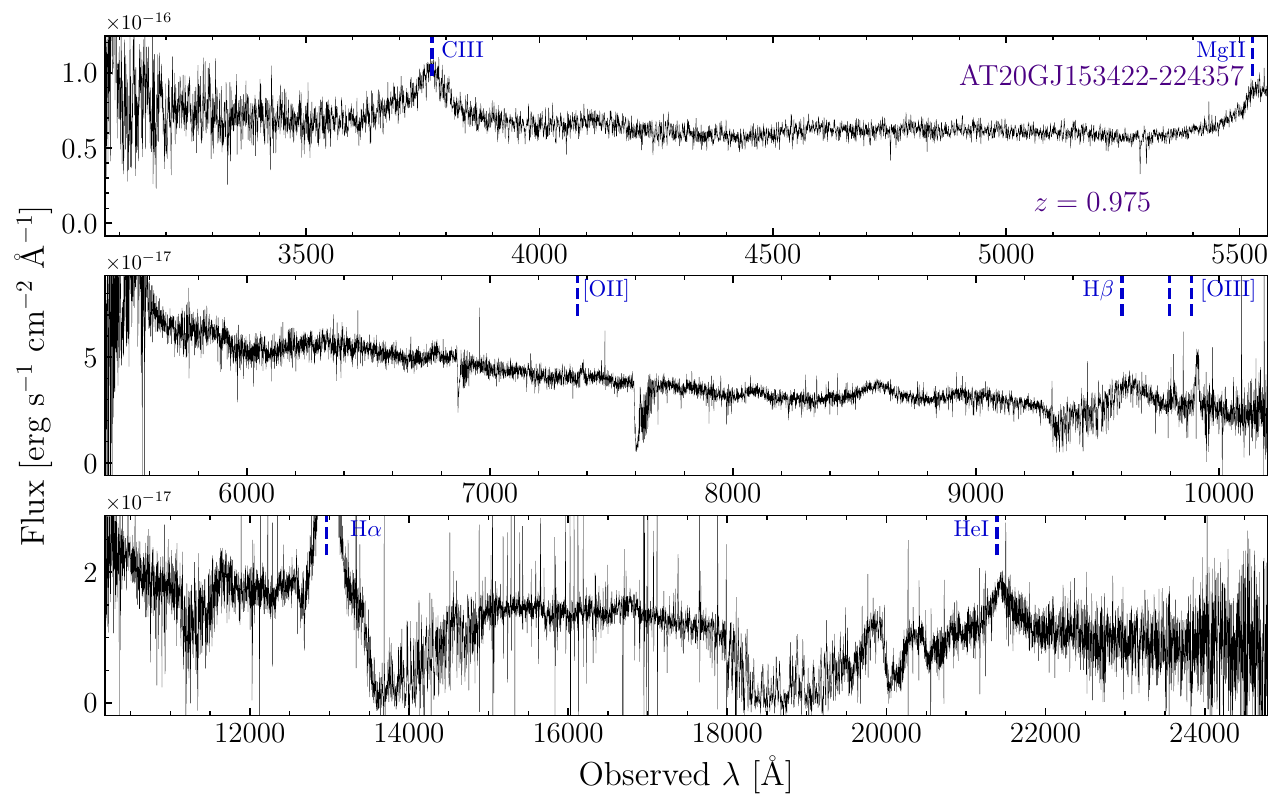}
    \caption{
    }
\end{figure*}

\begin{figure*}
    \ContinuedFloat
    \captionsetup{list=off,format=cont}
    \includegraphics[width=0.95\textwidth]{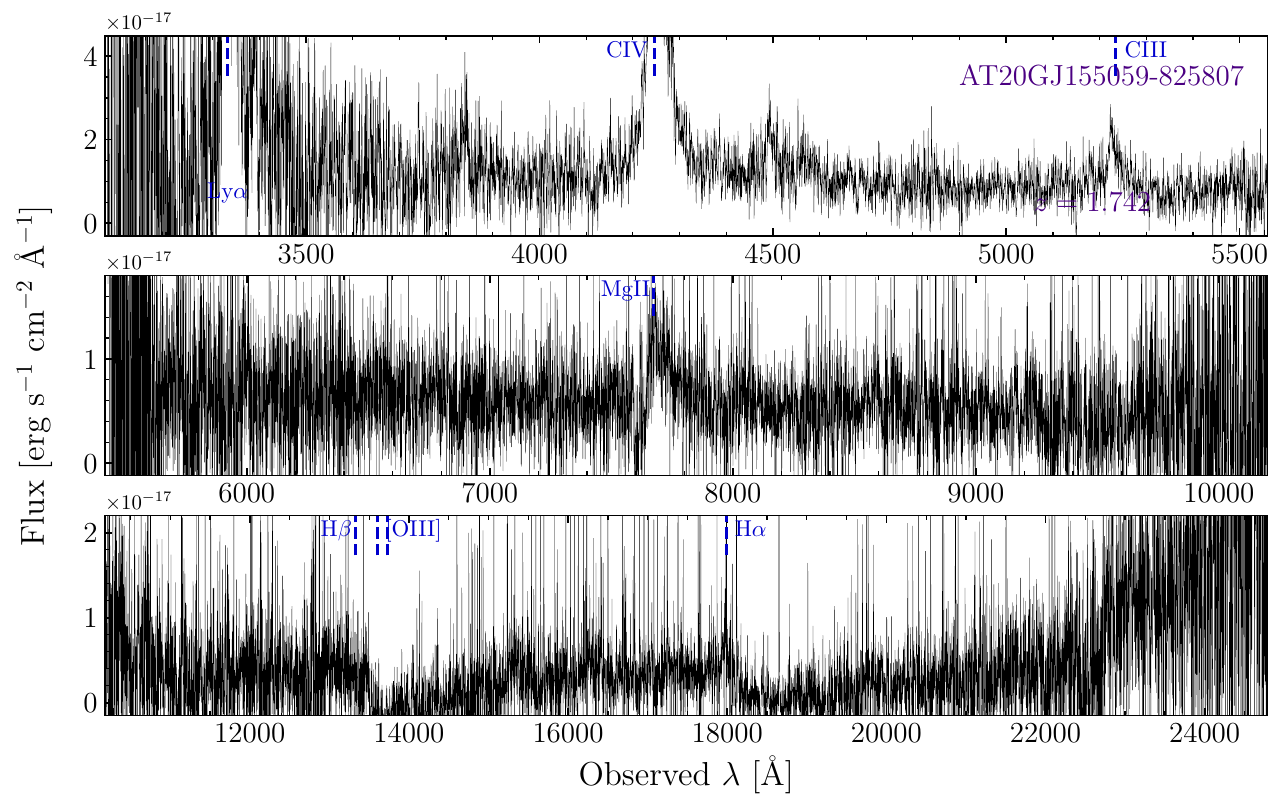}
    \caption{
    }
\end{figure*}

\begin{figure*}
    \ContinuedFloat
    \captionsetup{list=off,format=cont}
    \includegraphics[width=0.95\textwidth]{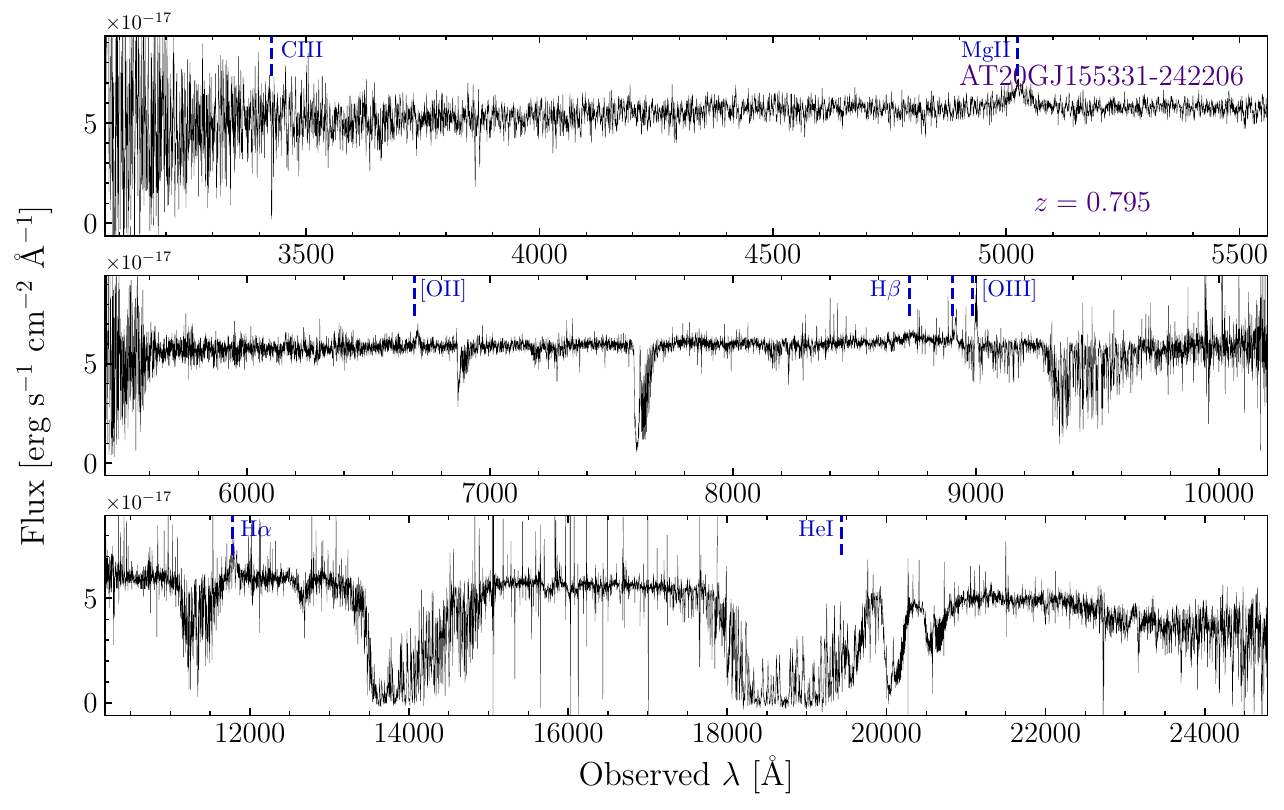}
    \caption{
    }
\end{figure*}

\begin{figure*}
    \ContinuedFloat
    \captionsetup{list=off,format=cont}
    \includegraphics[width=0.95\textwidth]{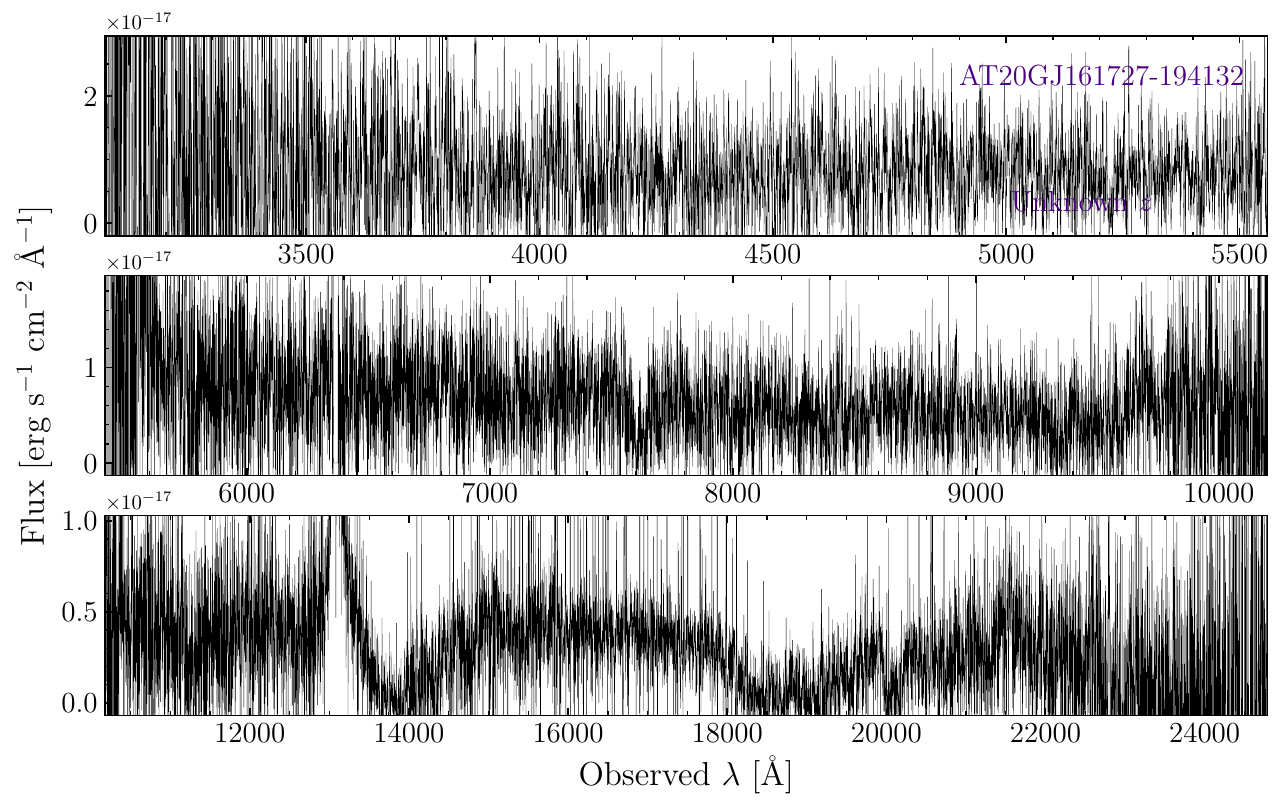}
    \caption{
    }
\end{figure*}

\begin{figure*}
    \ContinuedFloat
    \captionsetup{list=off,format=cont}
    \includegraphics[width=0.95\textwidth]{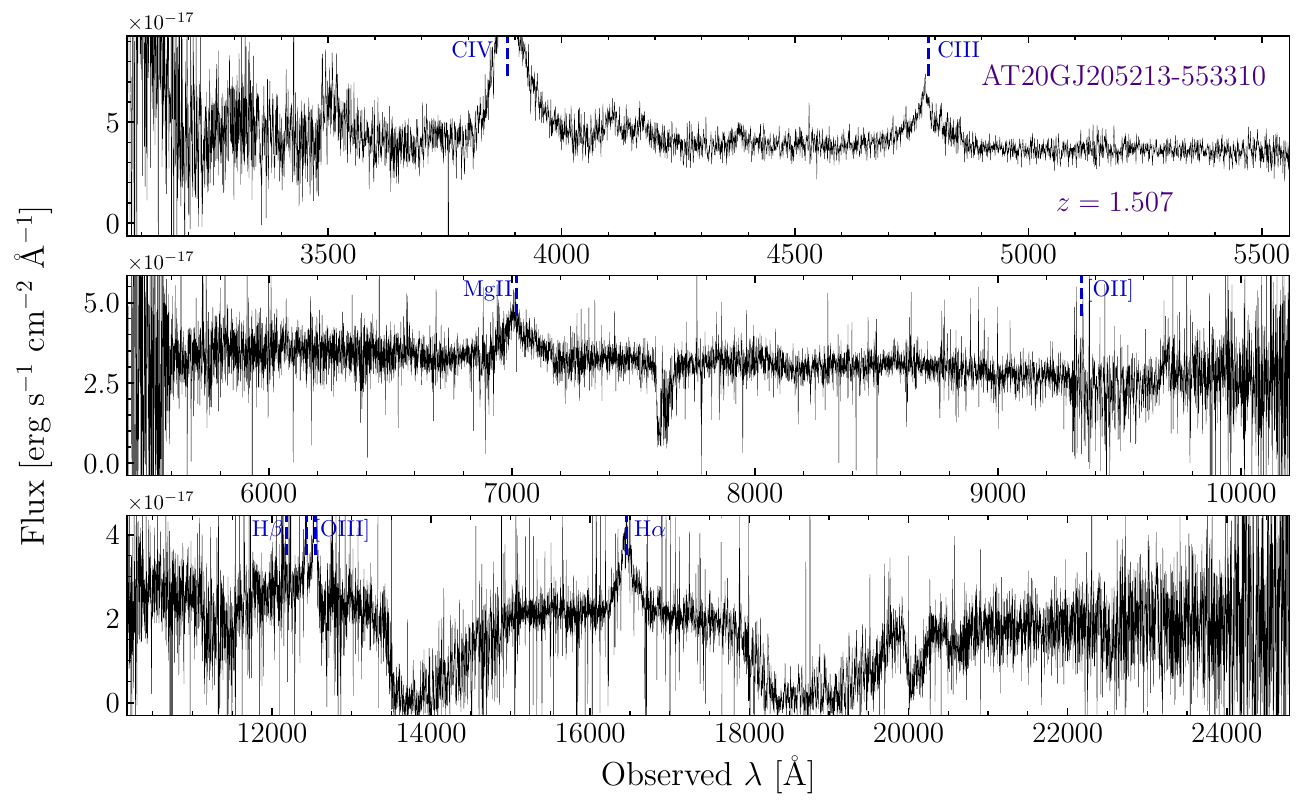}
    \caption{
    }
\end{figure*}

\begin{figure*}
    \ContinuedFloat
    \captionsetup{list=off,format=cont}
    \includegraphics[width=0.95\textwidth]{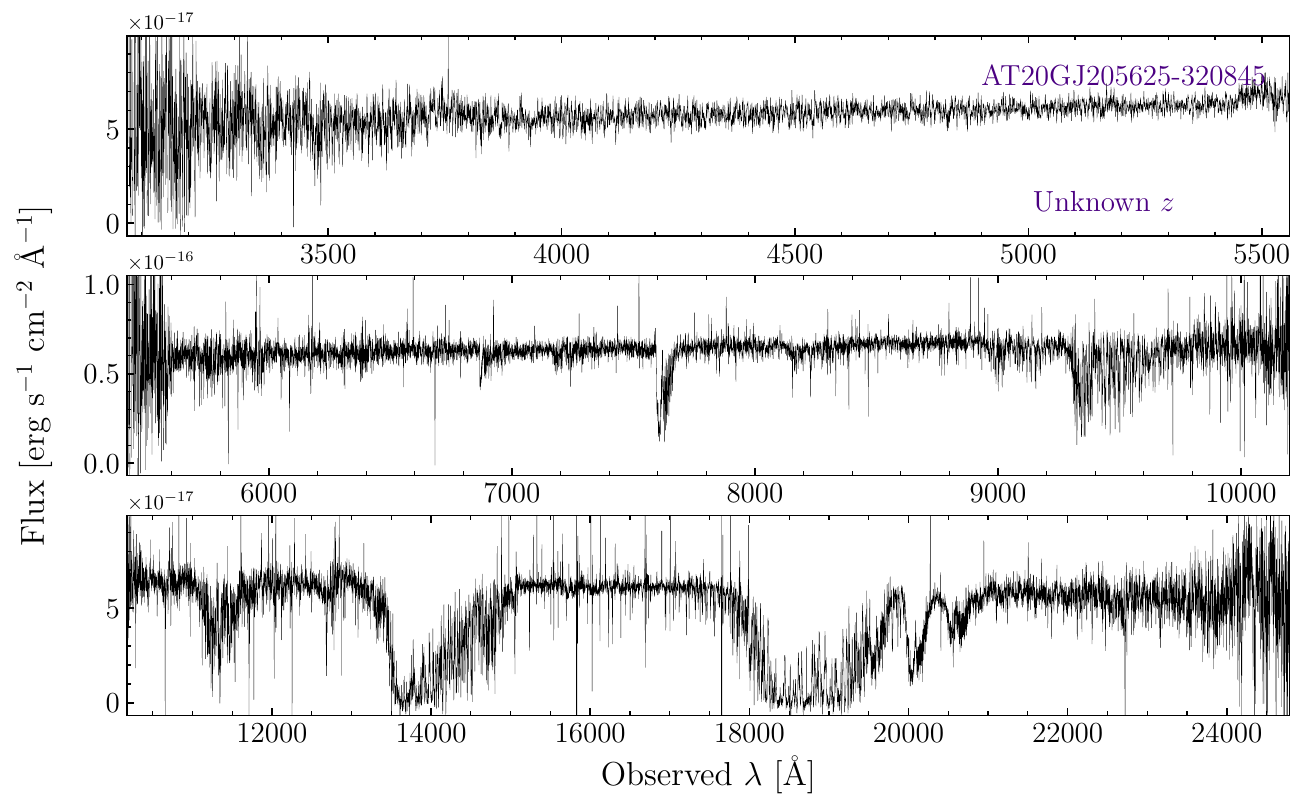}
    \caption{
    }
\end{figure*}

\begin{figure*}
    \ContinuedFloat
    \captionsetup{list=off,format=cont}
    \includegraphics[width=0.95\textwidth]{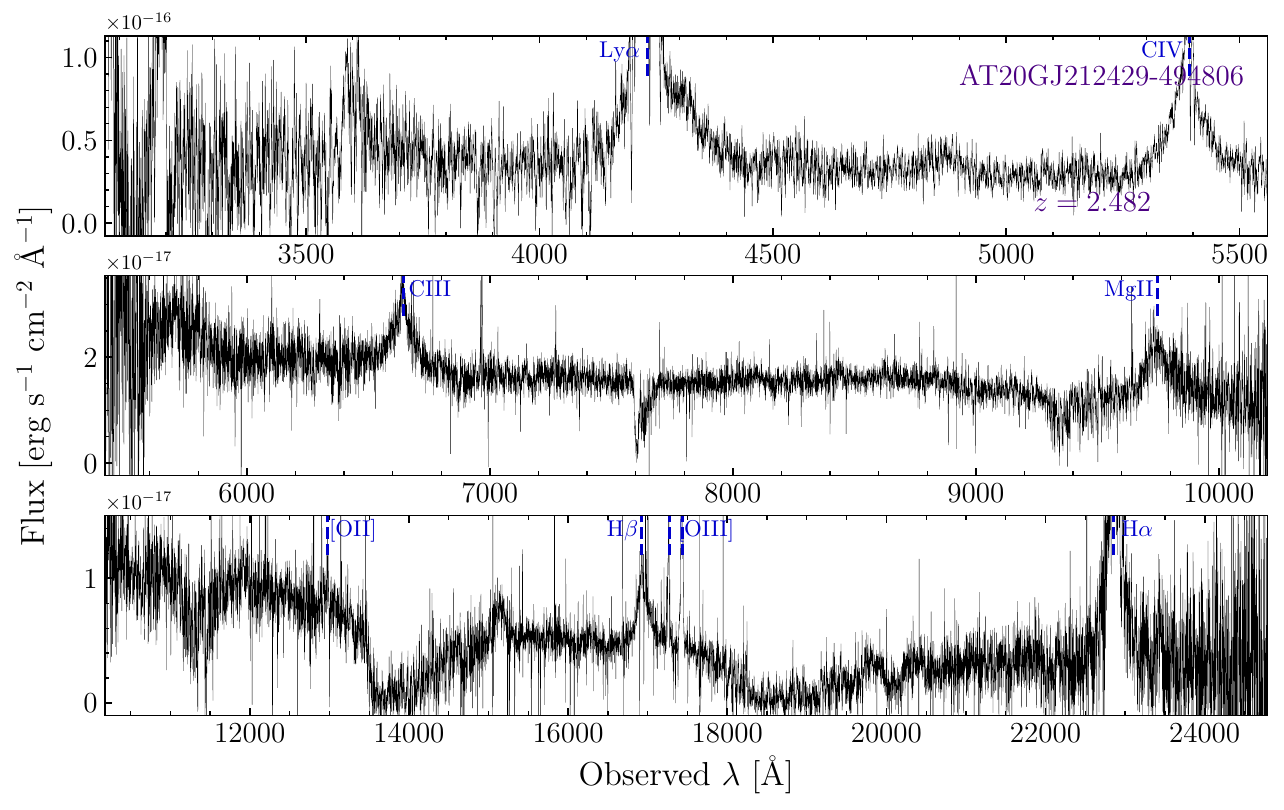}
    \caption{
    }
\end{figure*}

\begin{figure*}
    \ContinuedFloat
    \captionsetup{list=off,format=cont}
    \includegraphics[width=0.95\textwidth]{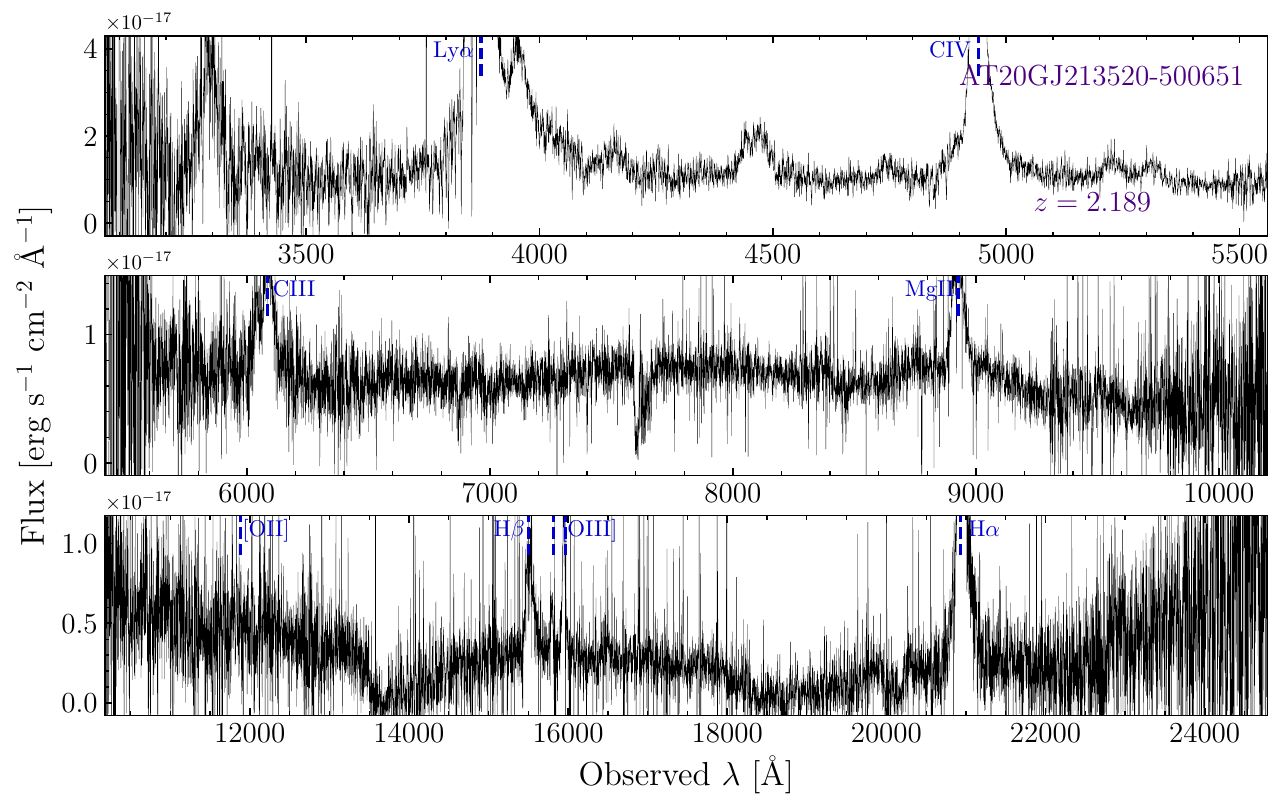}
    \caption{
    }
\end{figure*}

\begin{figure*}
    \ContinuedFloat
    \captionsetup{list=off,format=cont}
    \includegraphics[width=0.95\textwidth]{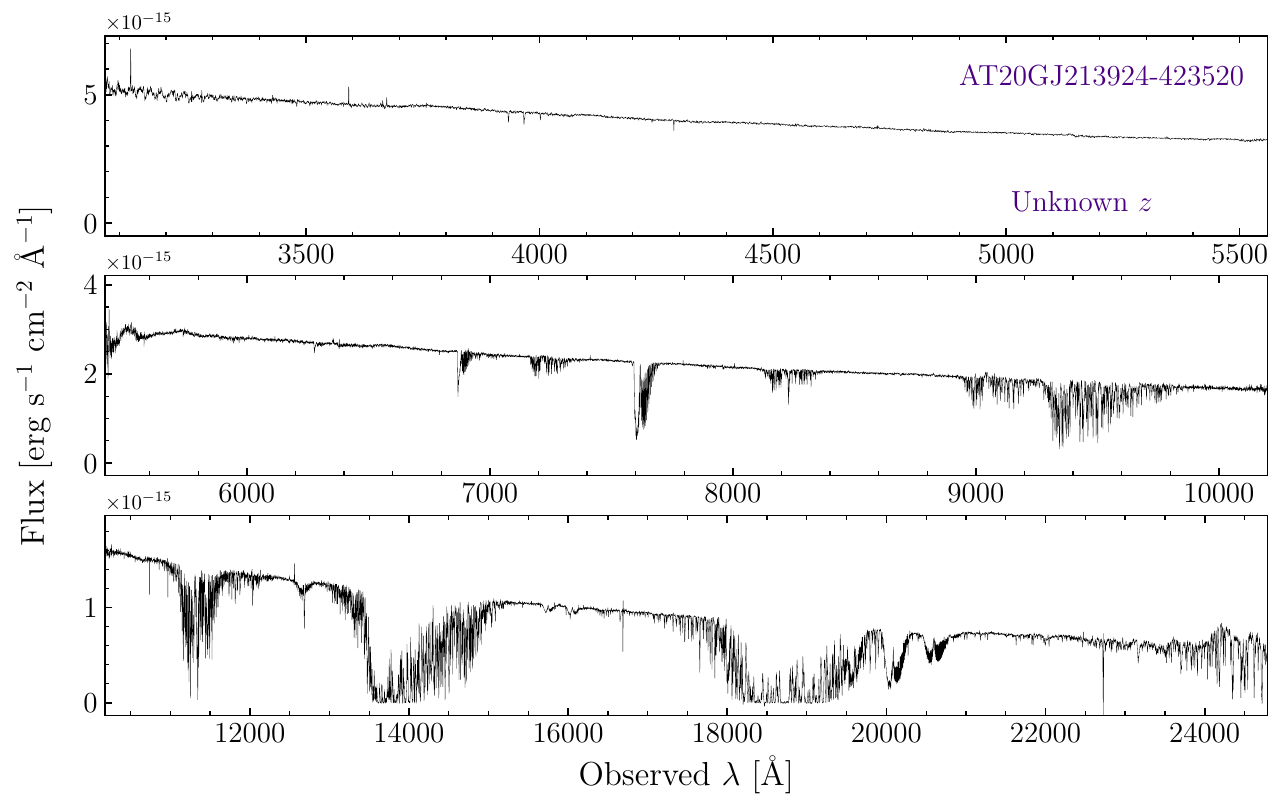}
    \caption{
    }
\end{figure*}

\begin{figure*}
    \ContinuedFloat
    \captionsetup{list=off,format=cont}
    \includegraphics[width=0.95\textwidth]{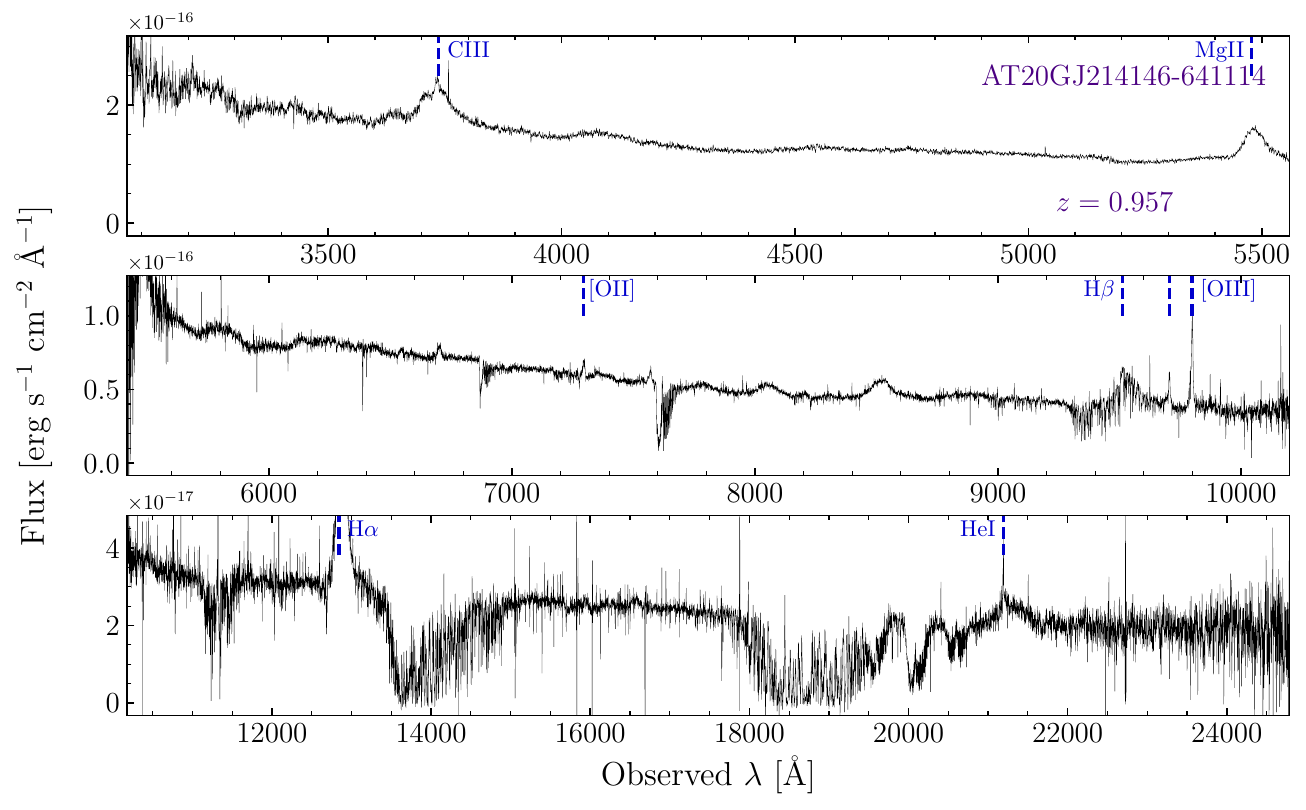}
    \caption{
    }
\end{figure*}

\begin{figure*}
    \ContinuedFloat
    \captionsetup{list=off,format=cont}
    \includegraphics[width=0.95\textwidth]{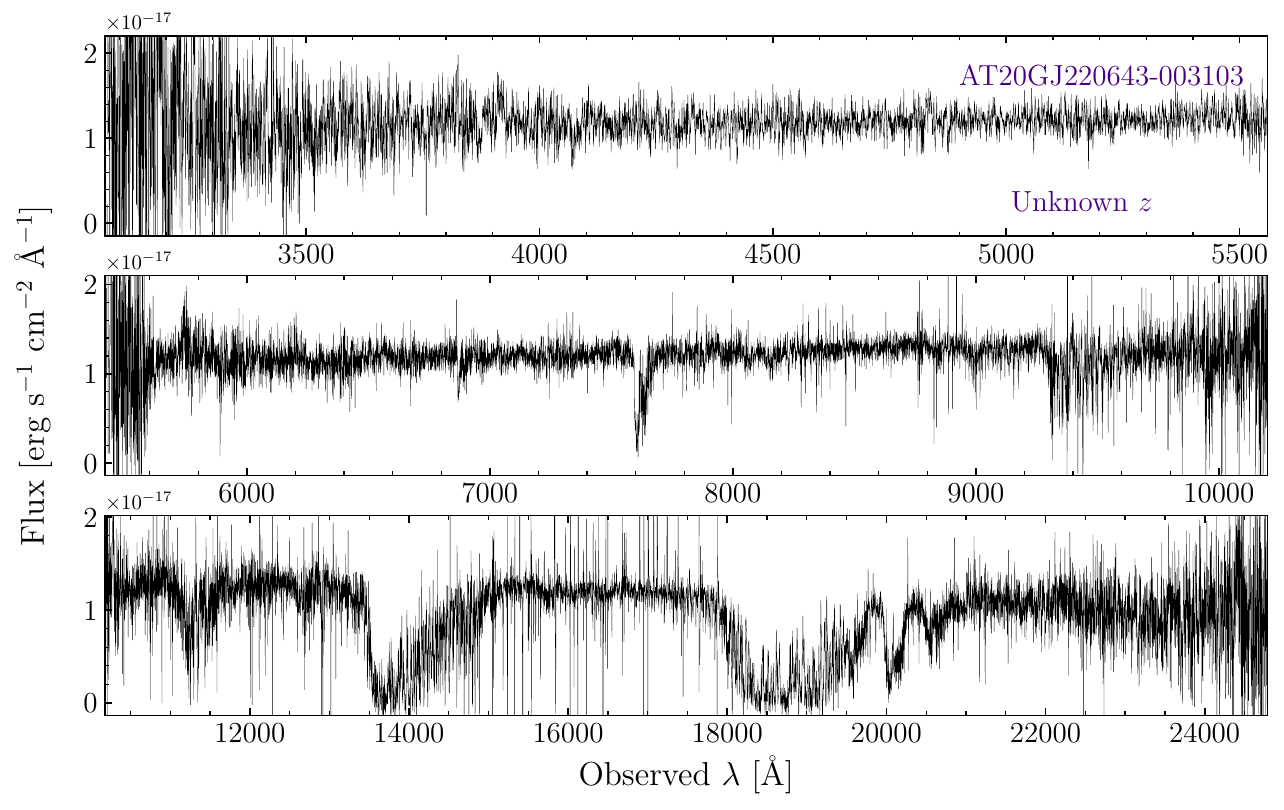}
    \caption{
    }
\end{figure*}

\begin{figure*}
    \ContinuedFloat
    \captionsetup{list=off,format=cont}
    \includegraphics[width=0.95\textwidth]{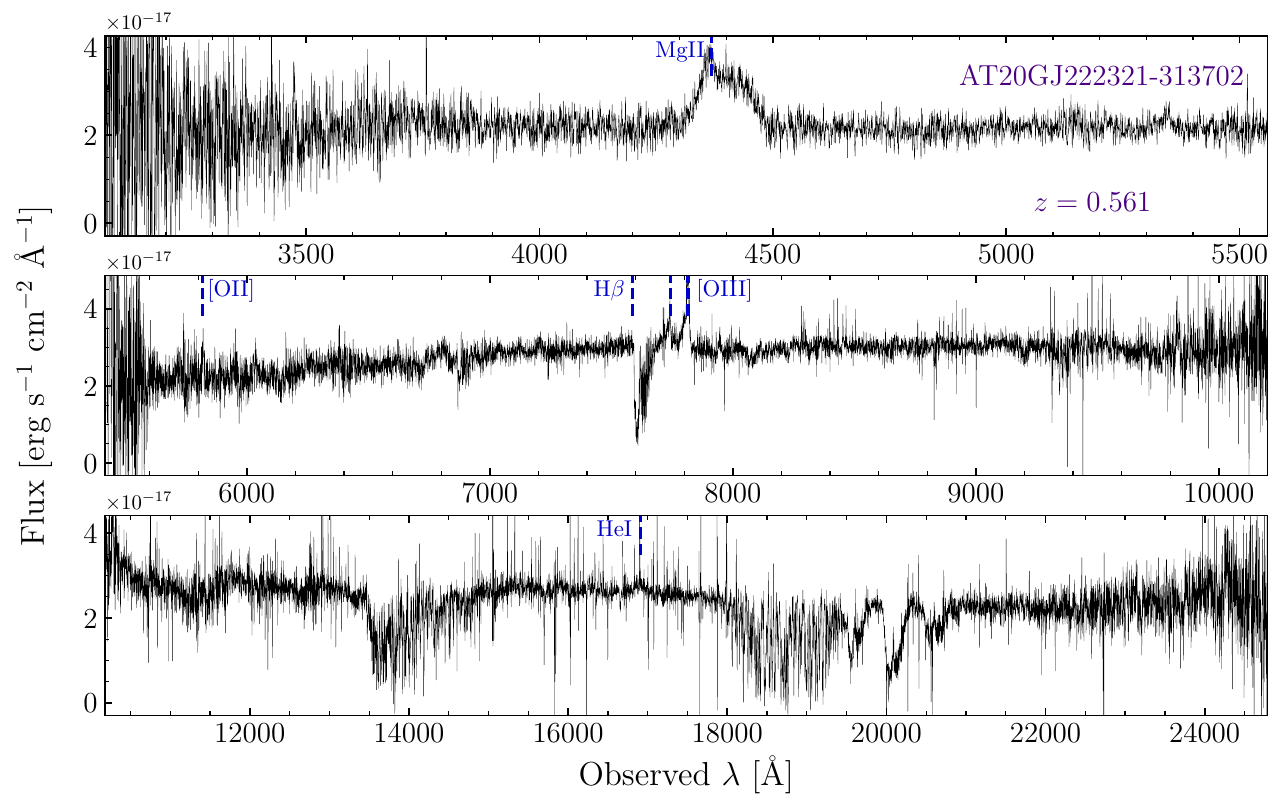}
    \caption{
    }
\end{figure*}

\begin{figure*}
    \ContinuedFloat
    \captionsetup{list=off,format=cont}
    \includegraphics[width=0.95\textwidth]{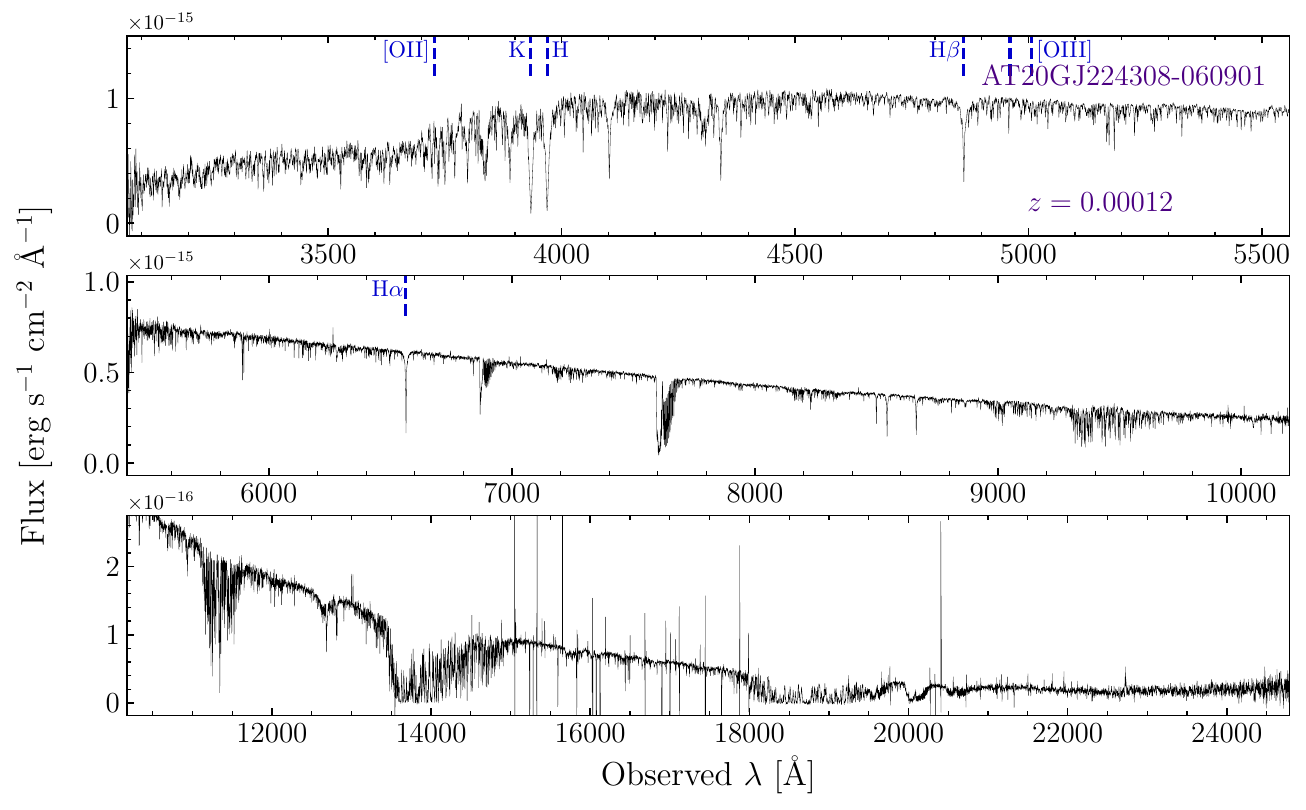}
    \caption{
    }
\end{figure*}

\begin{figure*}
    \ContinuedFloat
    \captionsetup{list=off,format=cont}
    \includegraphics[width=0.95\textwidth]{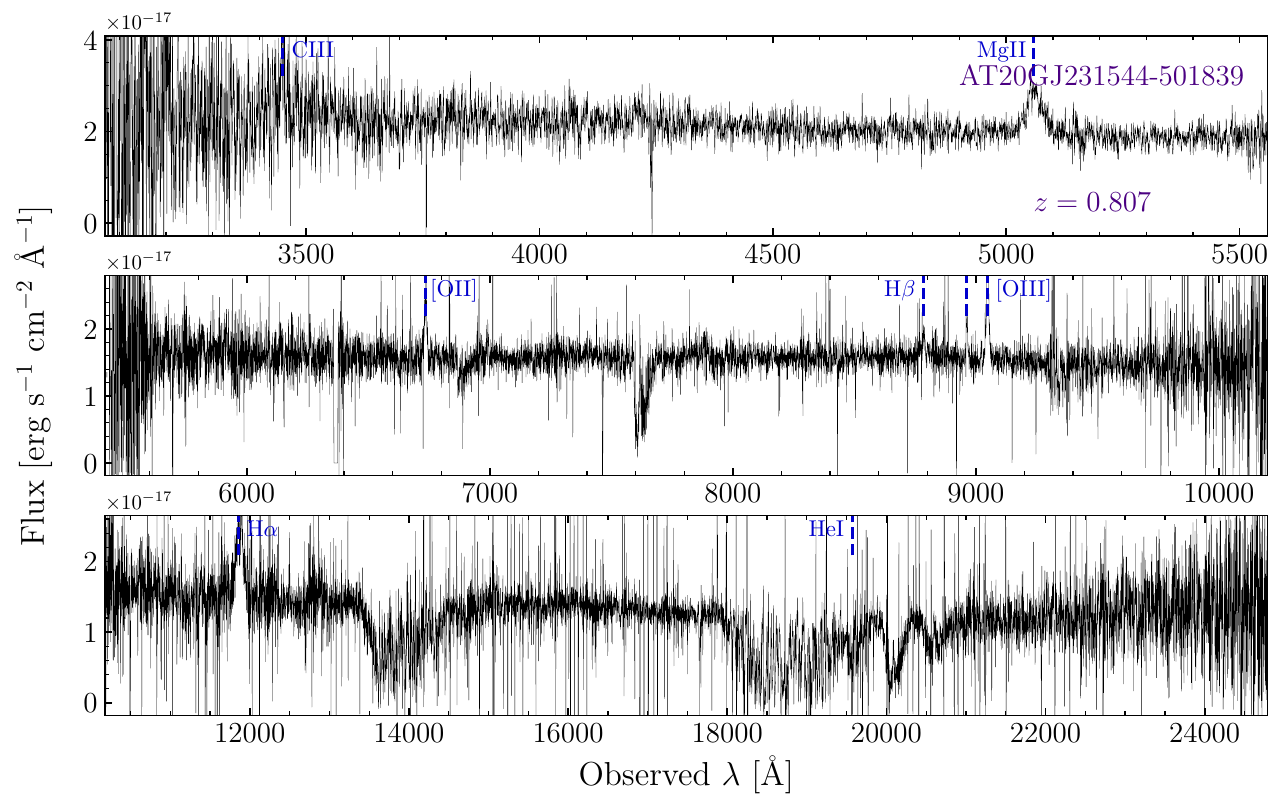}
    \caption{
    }
\end{figure*}

\begin{figure*}
    \ContinuedFloat
    \captionsetup{list=off,format=cont}
    \includegraphics[width=0.95\textwidth]{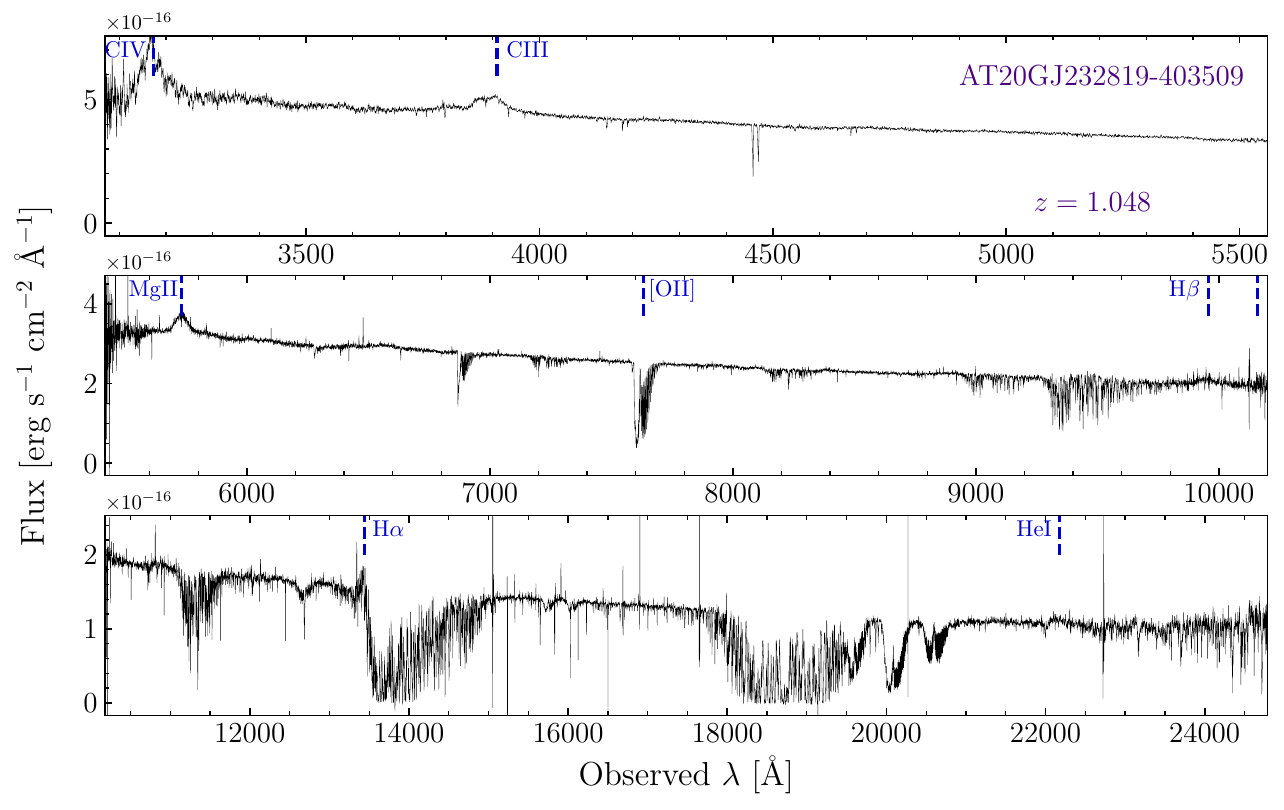}
    \caption{
    }
\end{figure*}

\begin{figure*}
    \ContinuedFloat
    \captionsetup{list=off,format=cont}
    \includegraphics[width=0.95\textwidth]{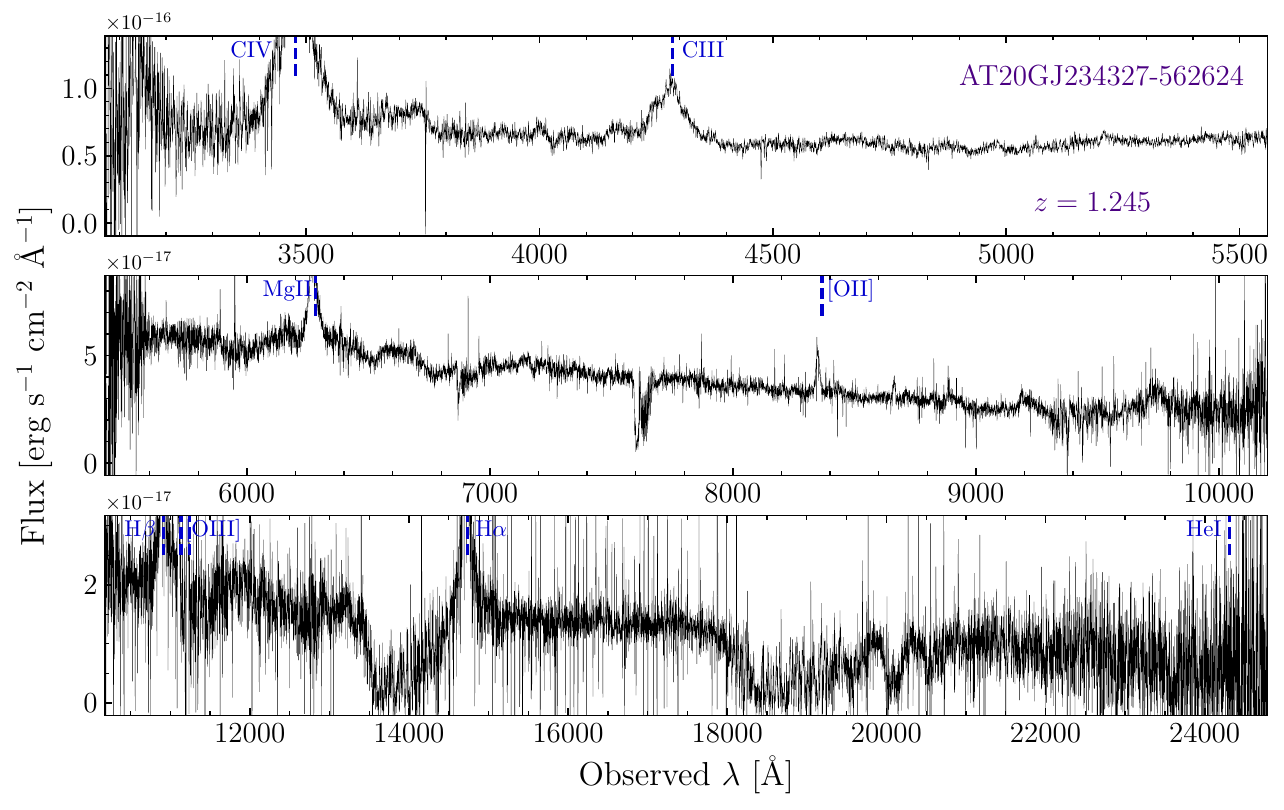}
    \caption{
    }
\end{figure*}
\begin{figure*}
    \ContinuedFloat
    \captionsetup{list=off,format=cont}
    \includegraphics[width=0.95\textwidth]{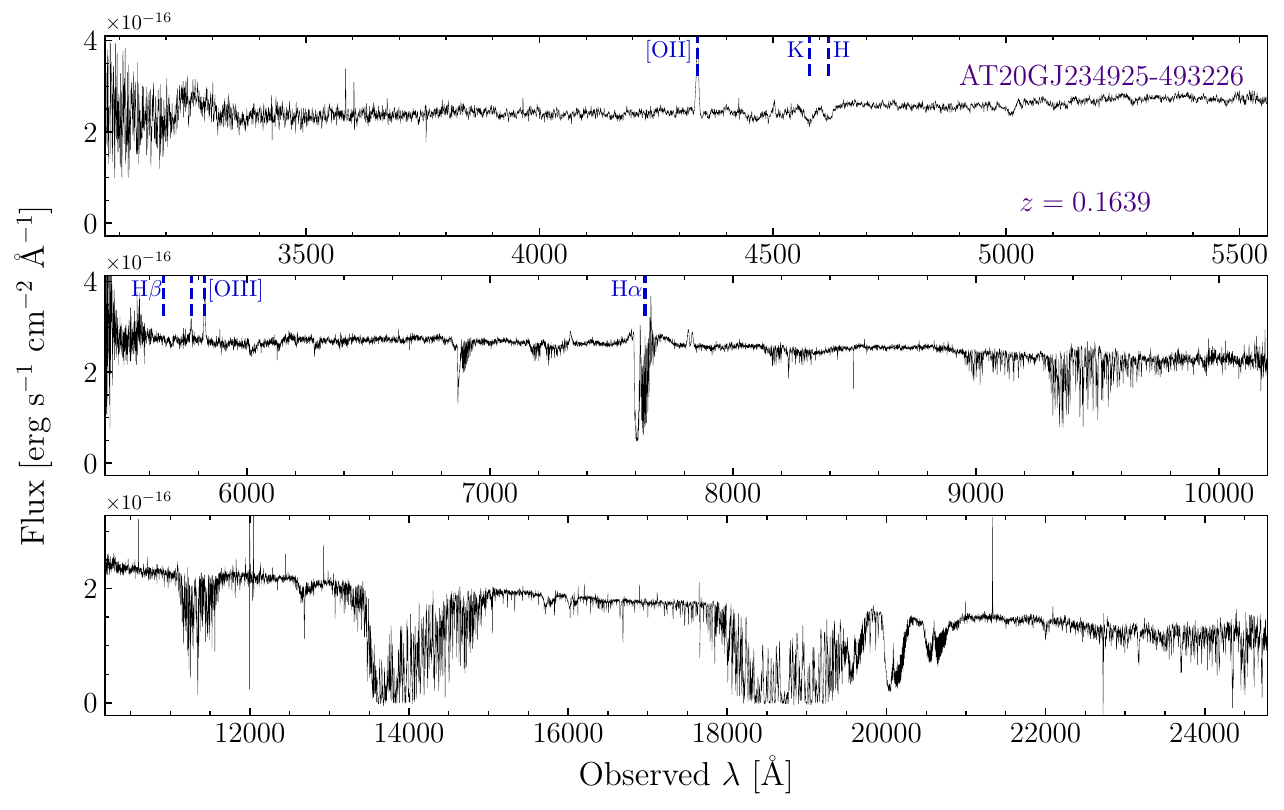}
    \caption{
    }
\end{figure*}
\clearpage

\begin{figure*}
    \includegraphics[width=0.95\textwidth]{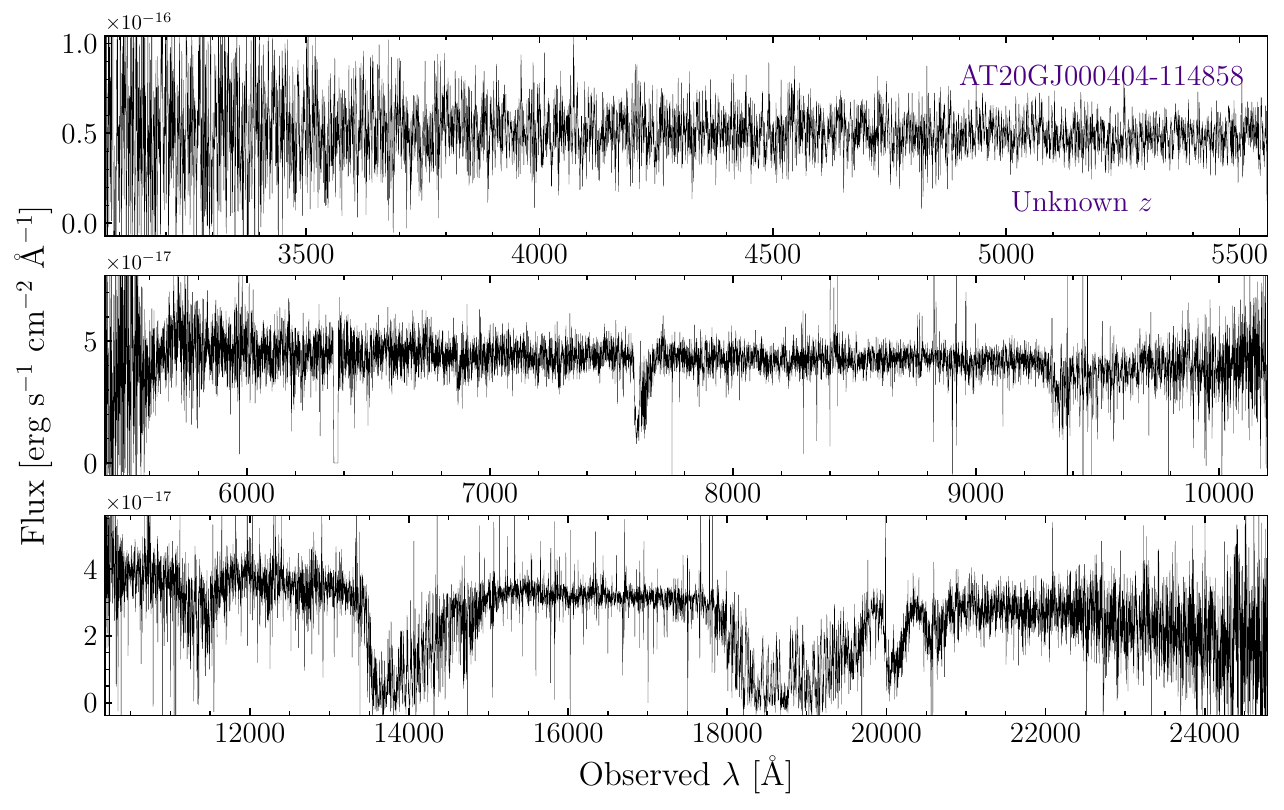}
    \caption{29 X-Shooter spectra from programme 111.253L.001
    }
    \label{fig:XSHOOP111}
\end{figure*}

\begin{figure*}
    \ContinuedFloat
    \captionsetup{list=off,format=cont}
    \includegraphics[width=0.95\textwidth]{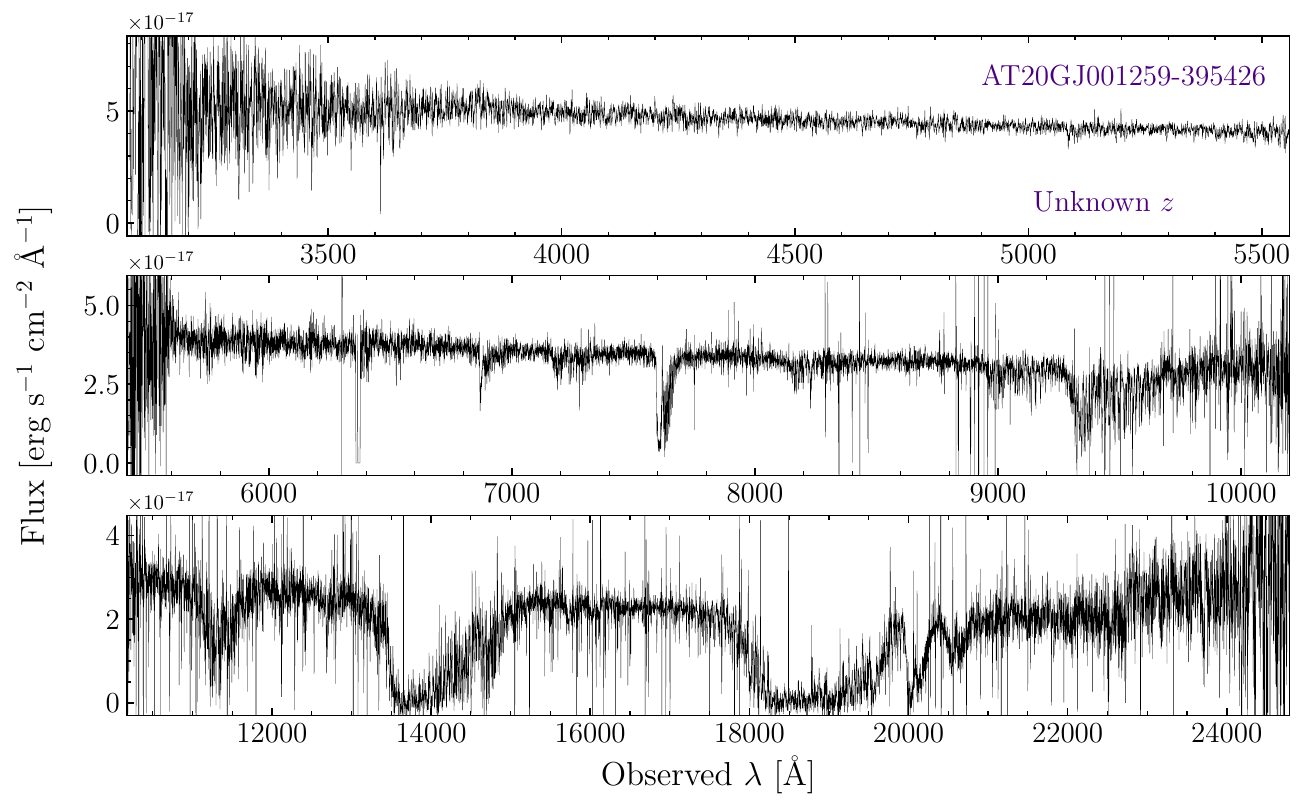}
    \caption{}
\end{figure*}

\begin{figure*}
    \ContinuedFloat
    \captionsetup{list=off,format=cont}
    \includegraphics[width=0.95\textwidth]{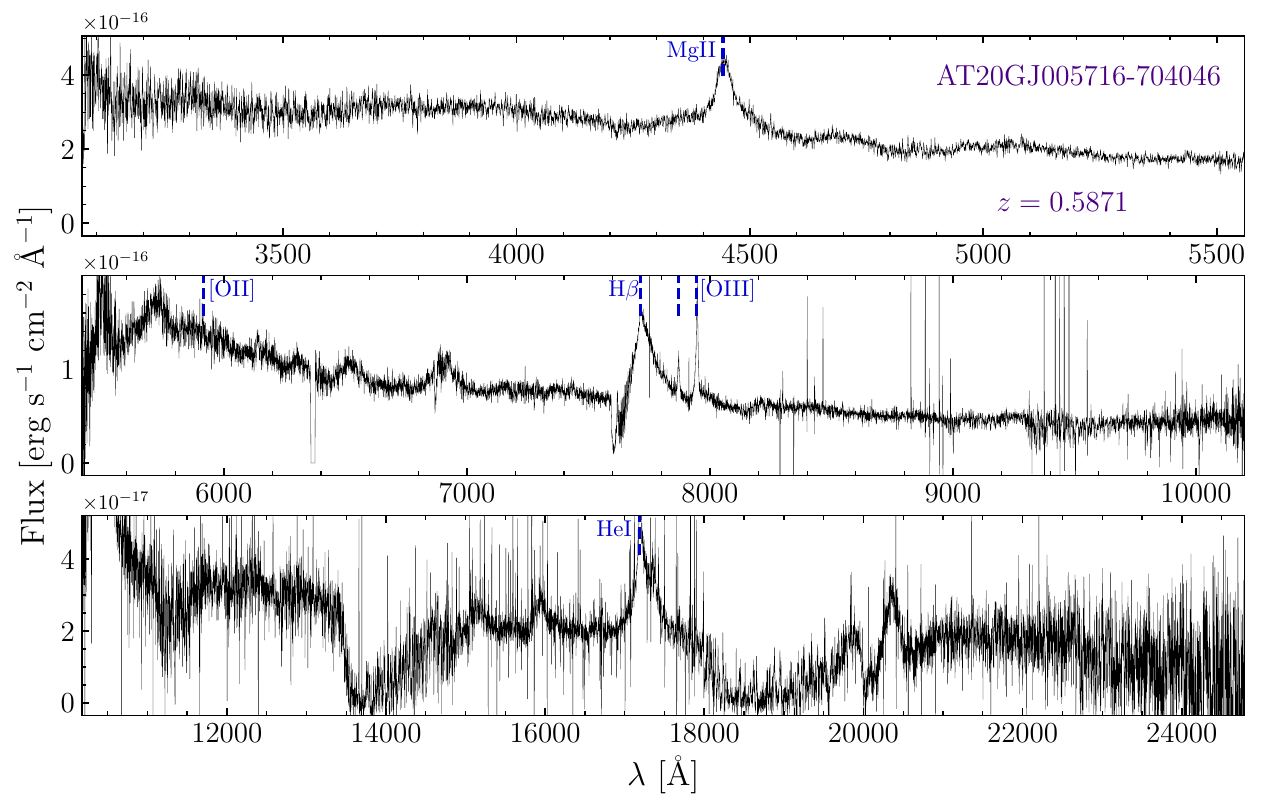}
    \caption{}
\end{figure*}

\begin{figure*}
    \ContinuedFloat
    \captionsetup{list=off,format=cont}
    \includegraphics[width=0.95\textwidth]{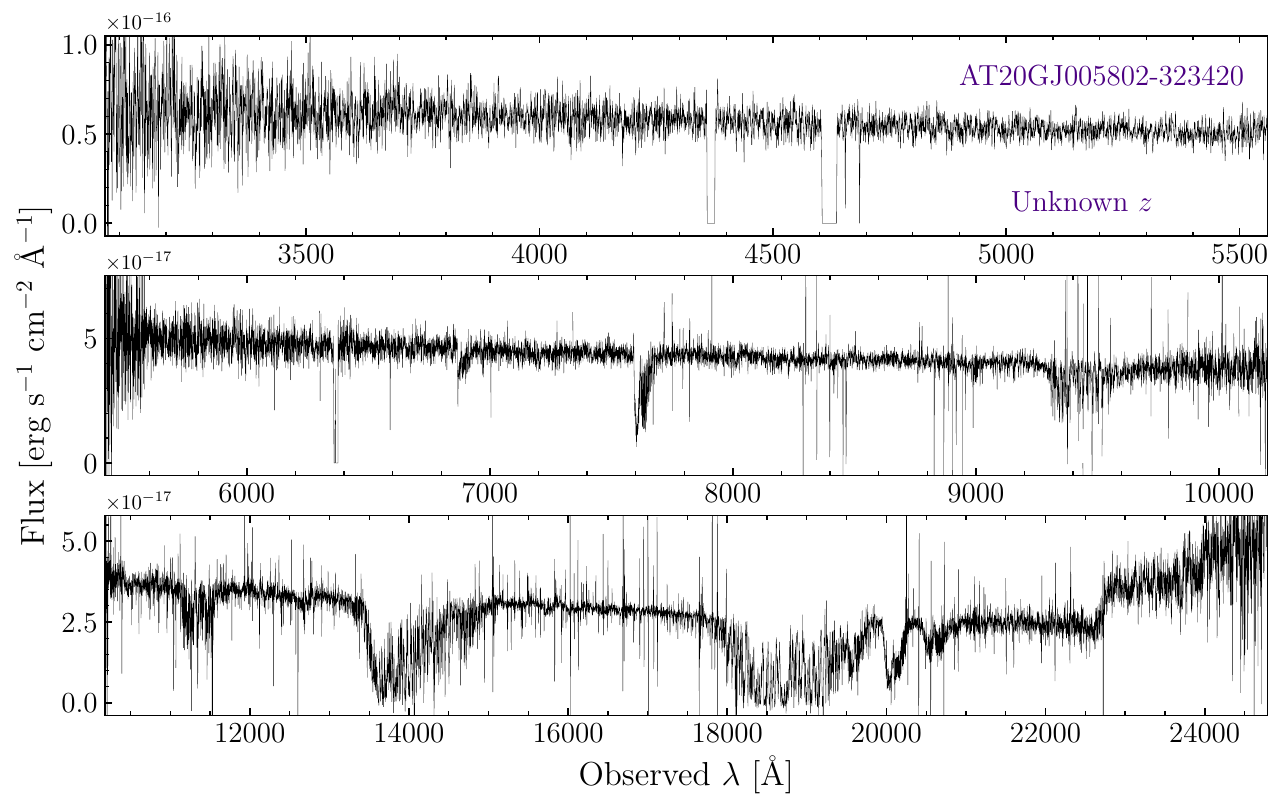}
    \caption{}

\end{figure*}

\begin{figure*}
    \ContinuedFloat
    \captionsetup{list=off,format=cont}
    \includegraphics[width=0.95\textwidth]{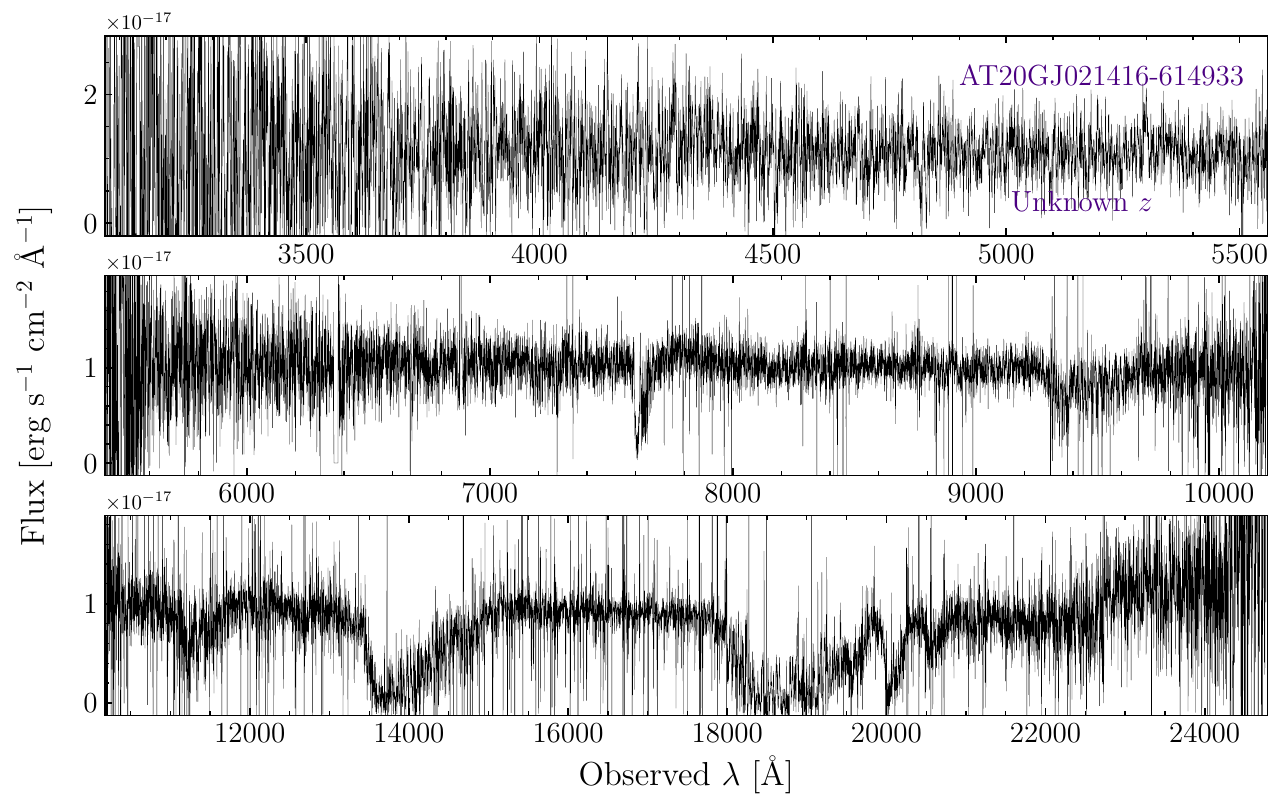}
    \caption{}

\end{figure*}

\begin{figure*}
    \ContinuedFloat
    \captionsetup{list=off,format=cont}
    \includegraphics[width=0.95\textwidth]{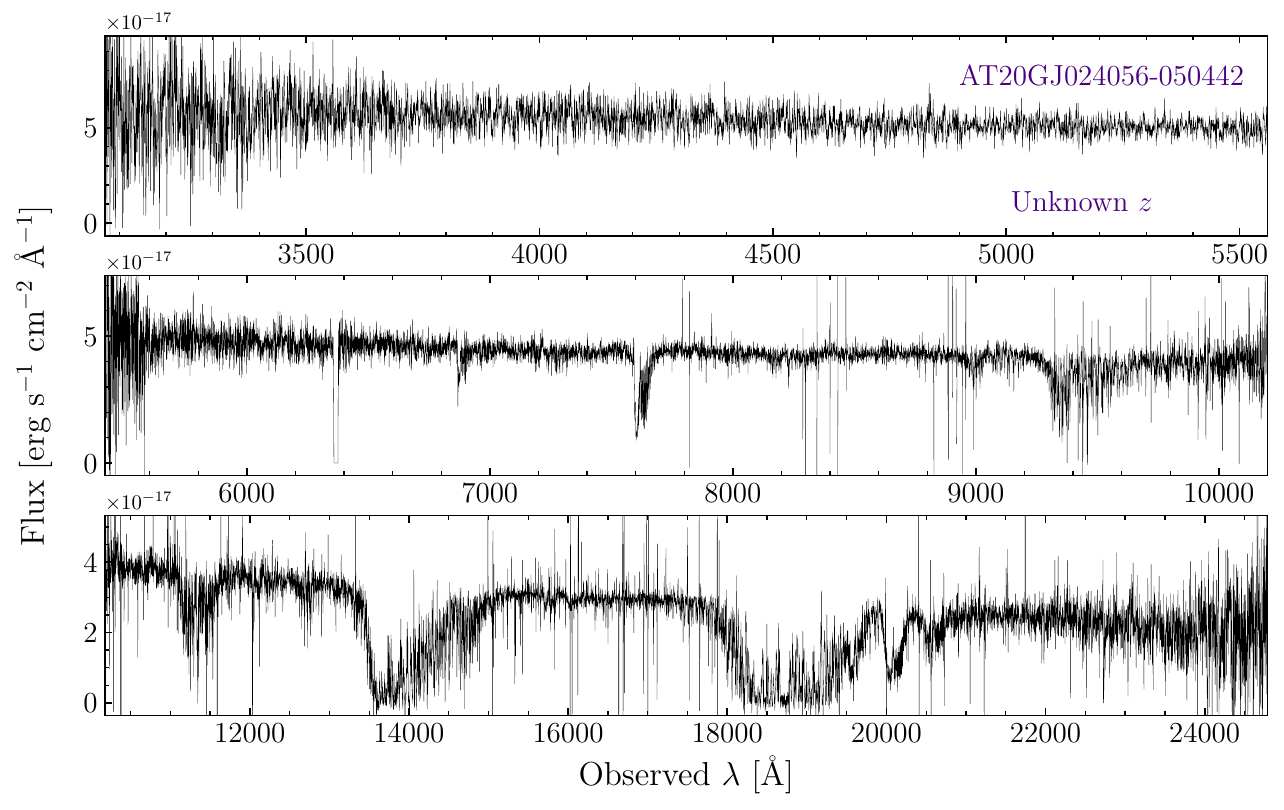}
    \caption{}

\end{figure*}

\begin{figure*}
    \ContinuedFloat
    \captionsetup{list=off,format=cont}
    \includegraphics[width=0.95\textwidth]{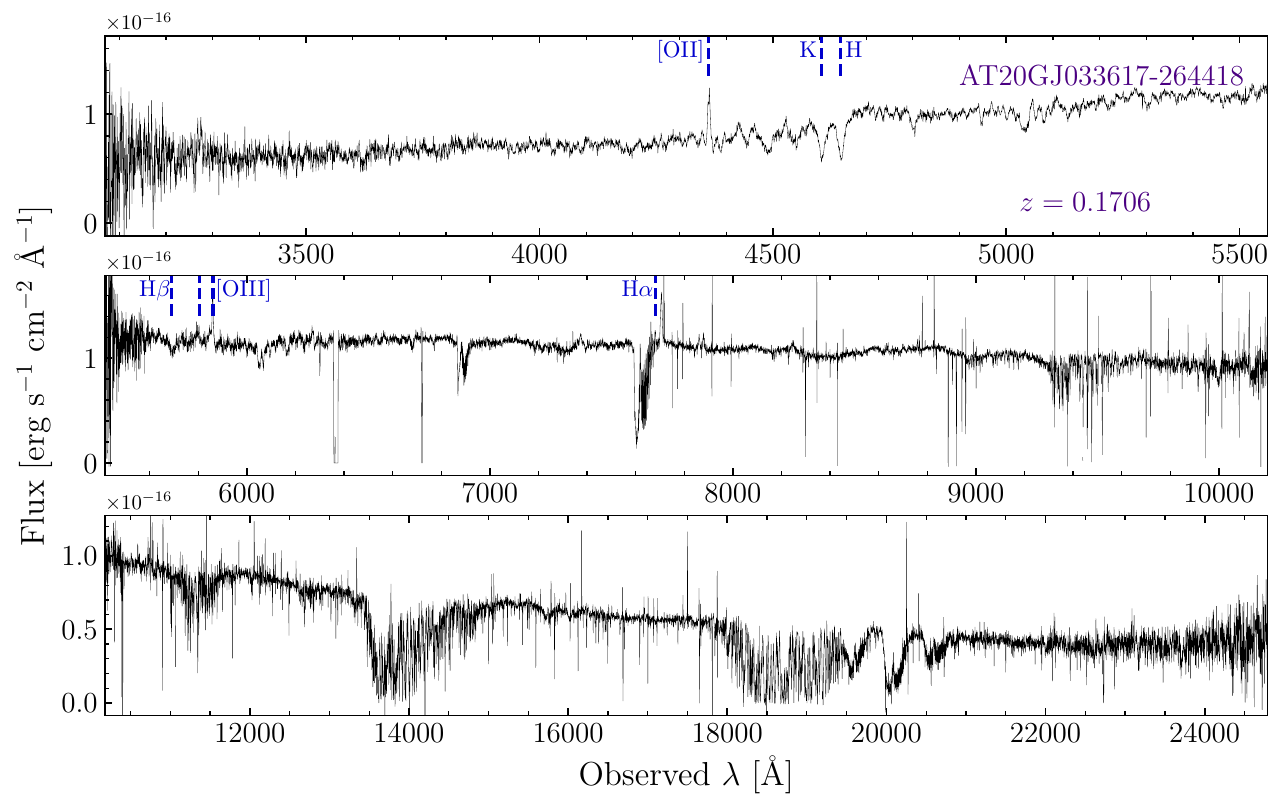}
    \caption{}

\end{figure*}

\begin{figure*}
    \ContinuedFloat
    \captionsetup{list=off,format=cont}
    \includegraphics[width=0.95\textwidth]{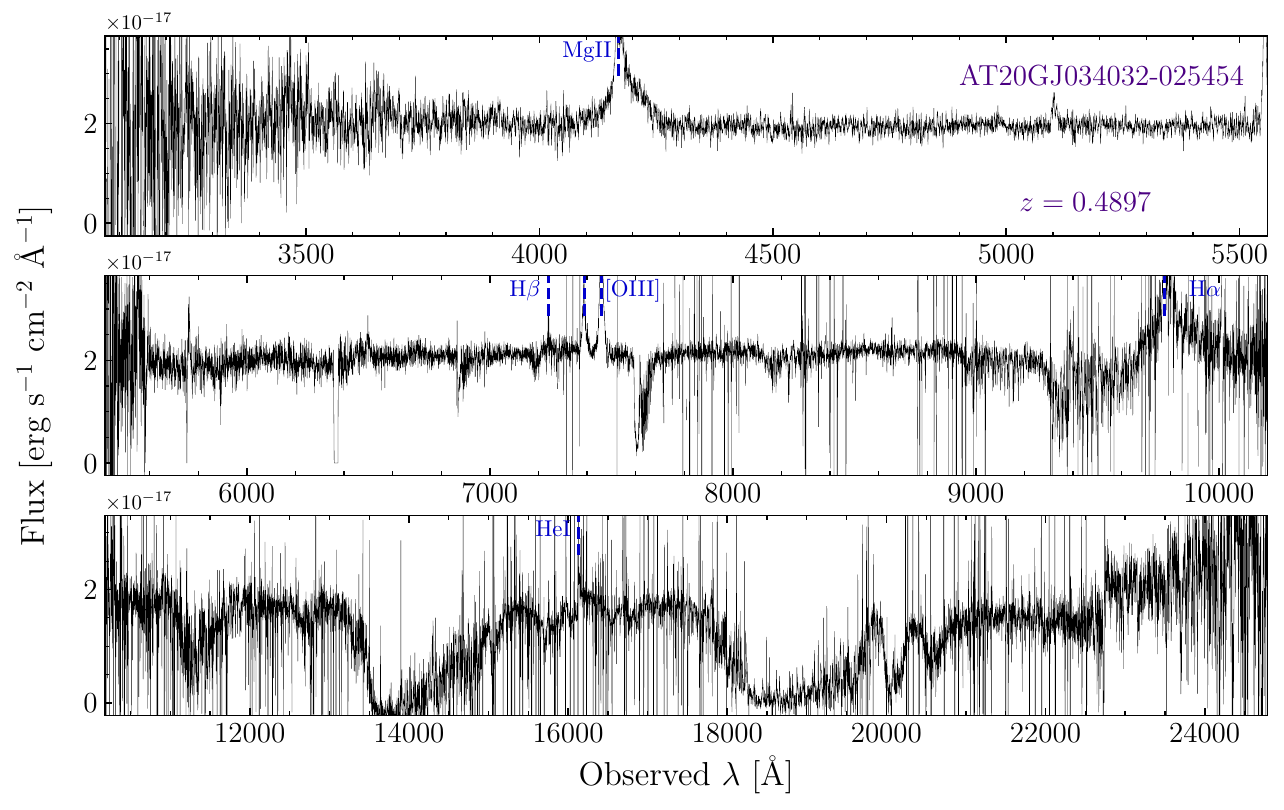}
    \caption{}

\end{figure*}

\begin{figure*}
    \ContinuedFloat
    \captionsetup{list=off,format=cont}
    \includegraphics[width=0.95\textwidth]{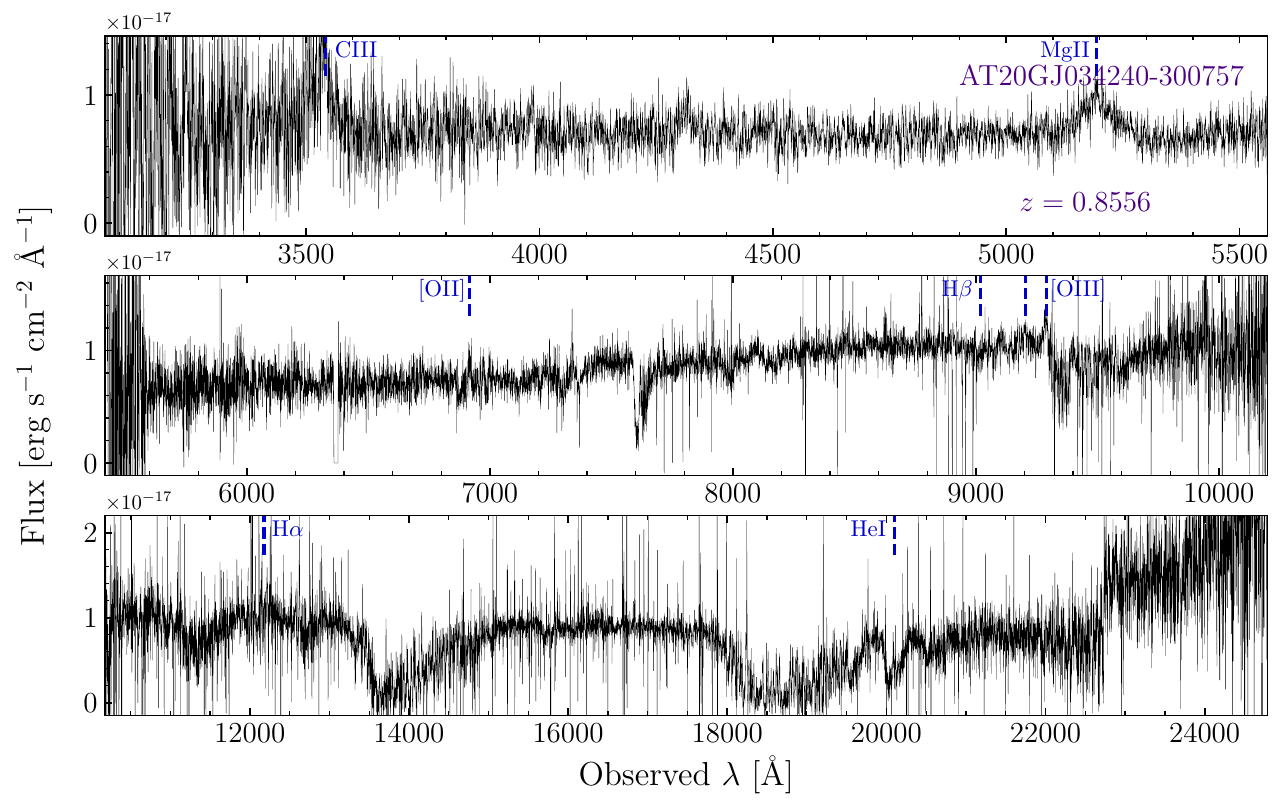}
    \caption{}

\end{figure*}

\begin{figure*}
    \ContinuedFloat
    \captionsetup{list=off,format=cont}
    \includegraphics[width=0.95\textwidth]{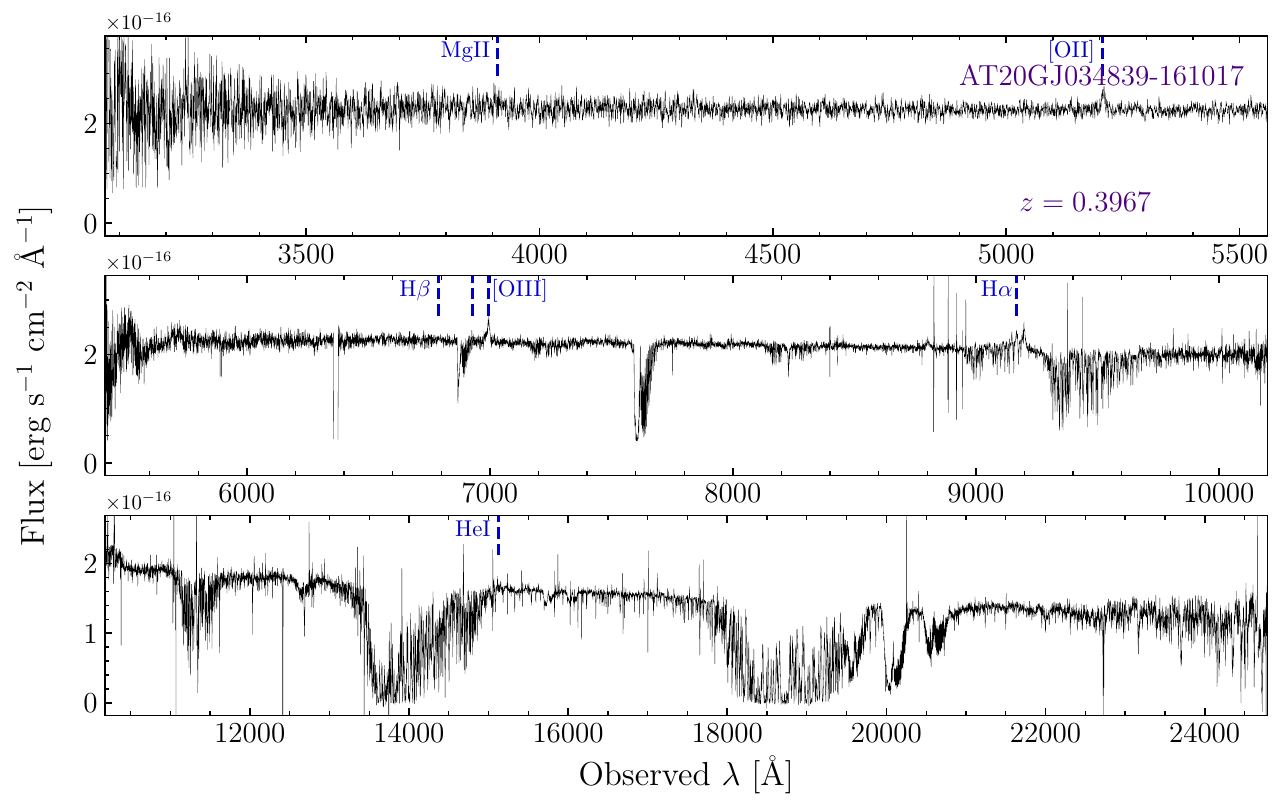}
    \caption{}

\end{figure*}

\begin{figure*}
    \ContinuedFloat
    \captionsetup{list=off,format=cont}
    \includegraphics[width=0.95\textwidth]{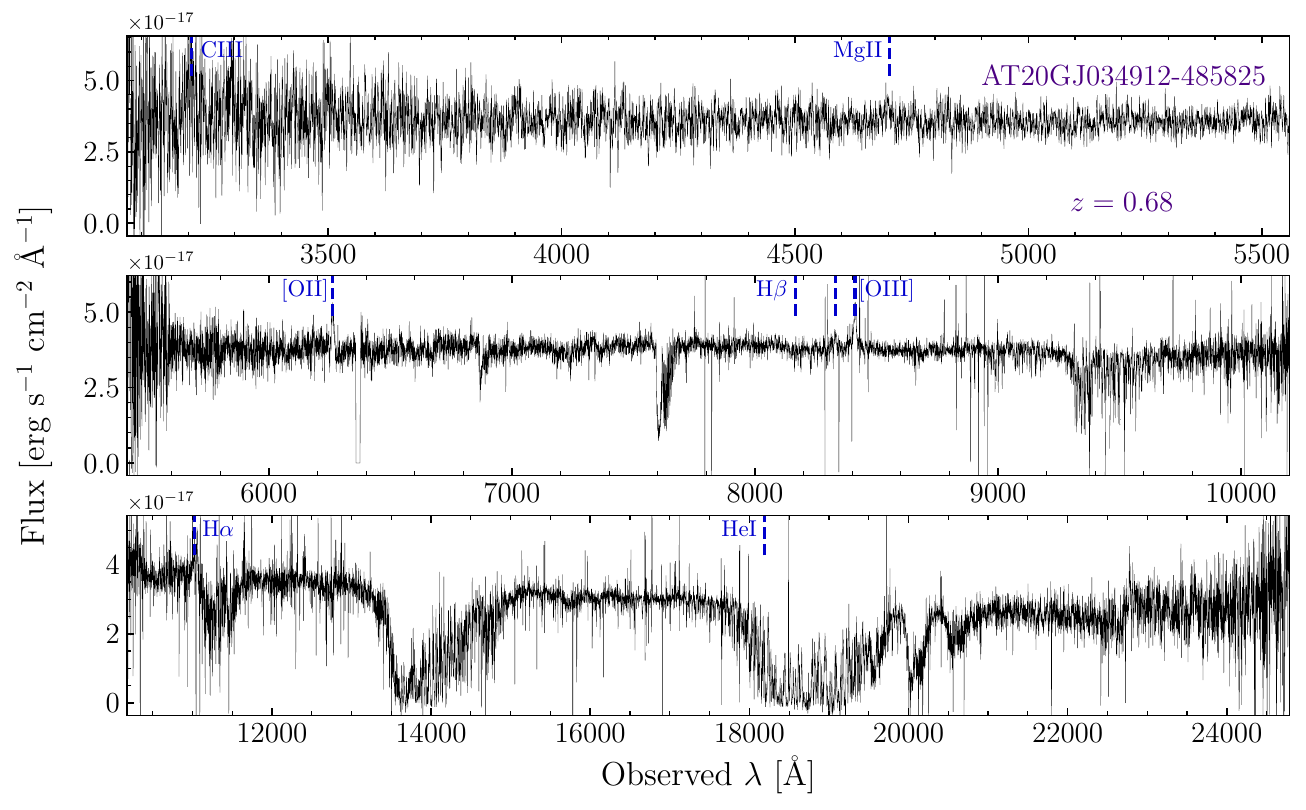}
    \caption{}

\end{figure*}

\begin{figure*}
    \ContinuedFloat
    \captionsetup{list=off,format=cont}
    \includegraphics[width=0.95\textwidth]{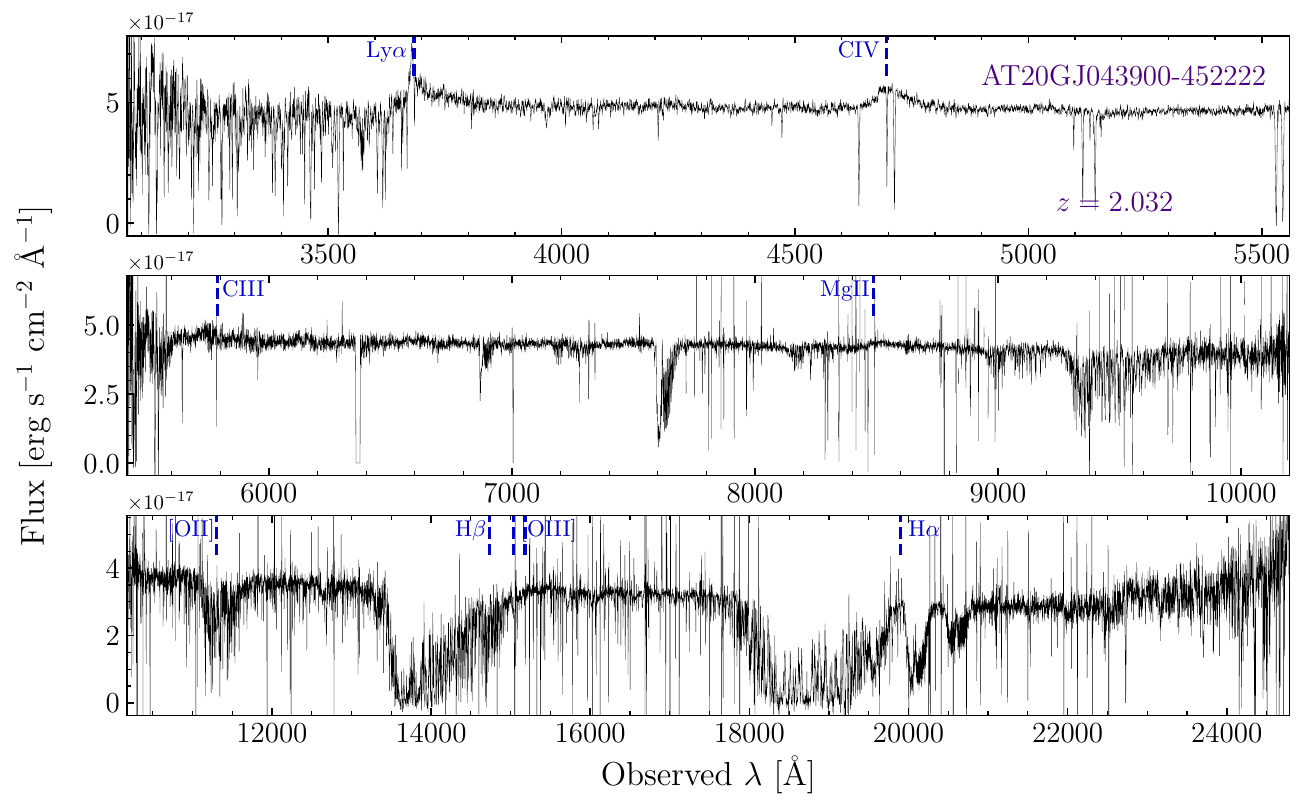}
    \caption{}

\end{figure*}

\begin{figure*}
    \ContinuedFloat
    \captionsetup{list=off,format=cont}
    \includegraphics[width=0.95\textwidth]{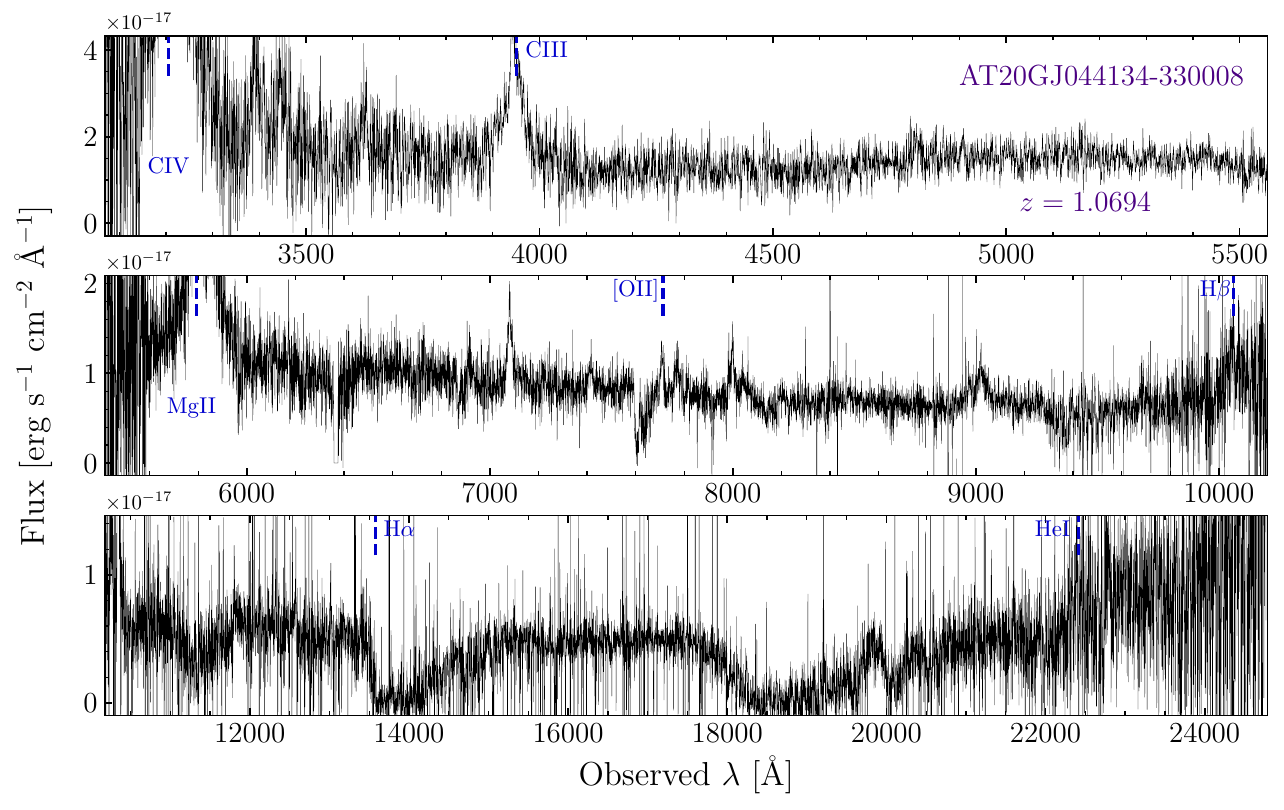}
    \caption{}

\end{figure*}

\begin{figure*}
    \ContinuedFloat
    \captionsetup{list=off,format=cont}
    \includegraphics[width=0.95\textwidth]{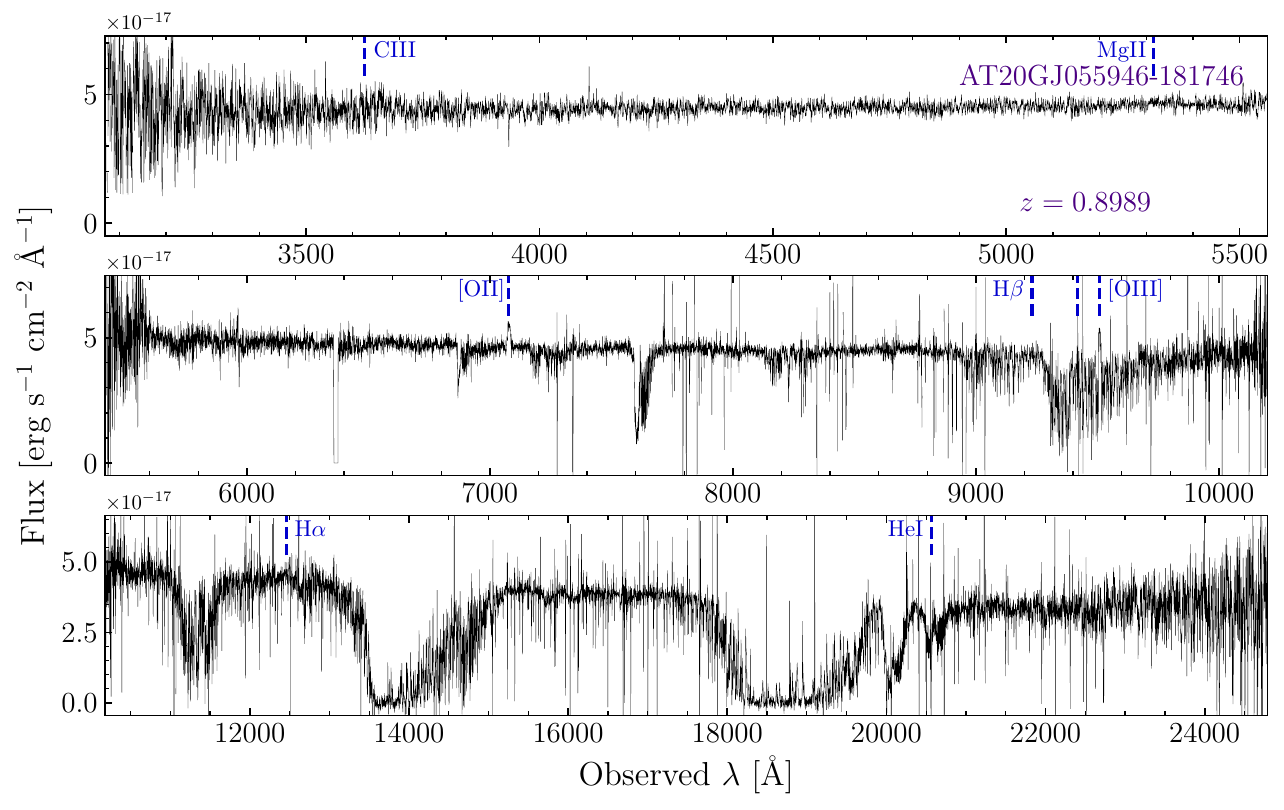}
    \caption{}

\end{figure*}

\begin{figure*}
    \ContinuedFloat
    \captionsetup{list=off,format=cont}
    \includegraphics[width=0.95\textwidth]{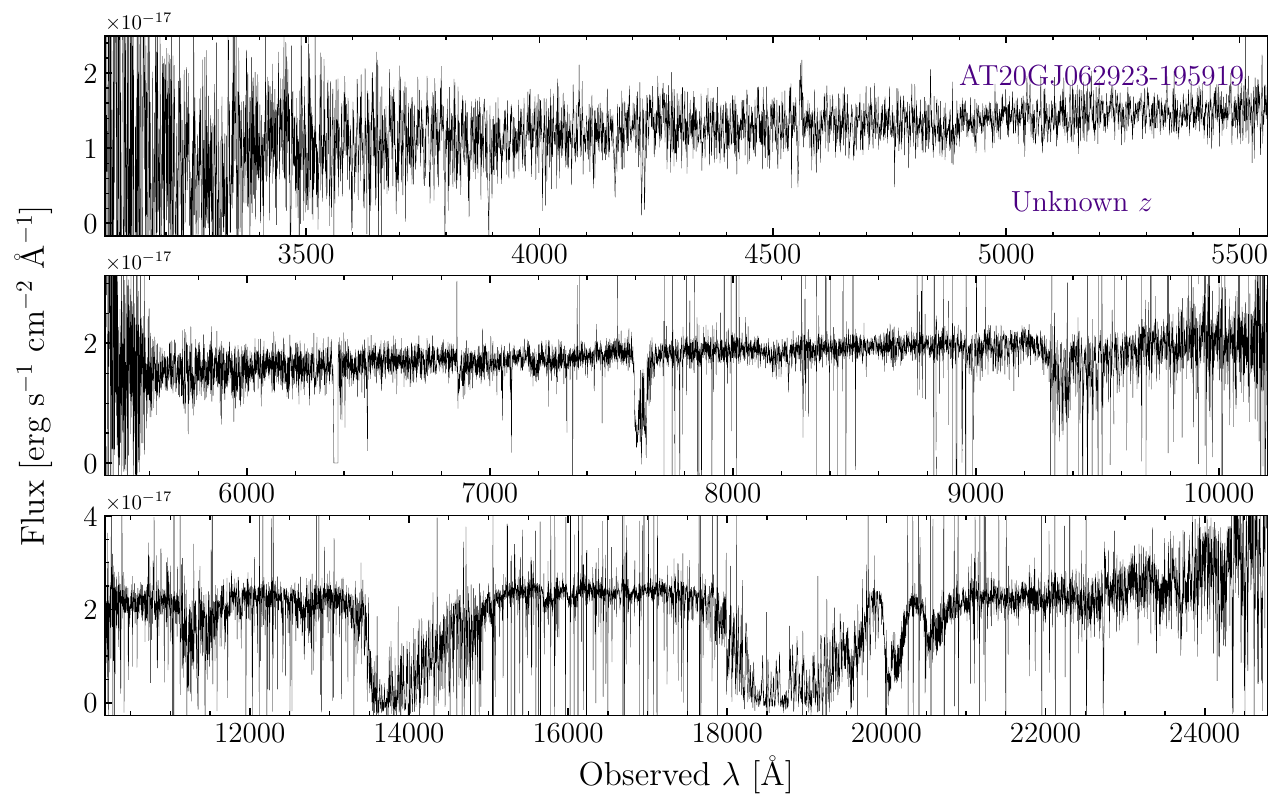}
    \caption{}

\end{figure*}

\begin{figure*}
    \ContinuedFloat
    \captionsetup{list=off,format=cont}
    \includegraphics[width=0.95\textwidth]{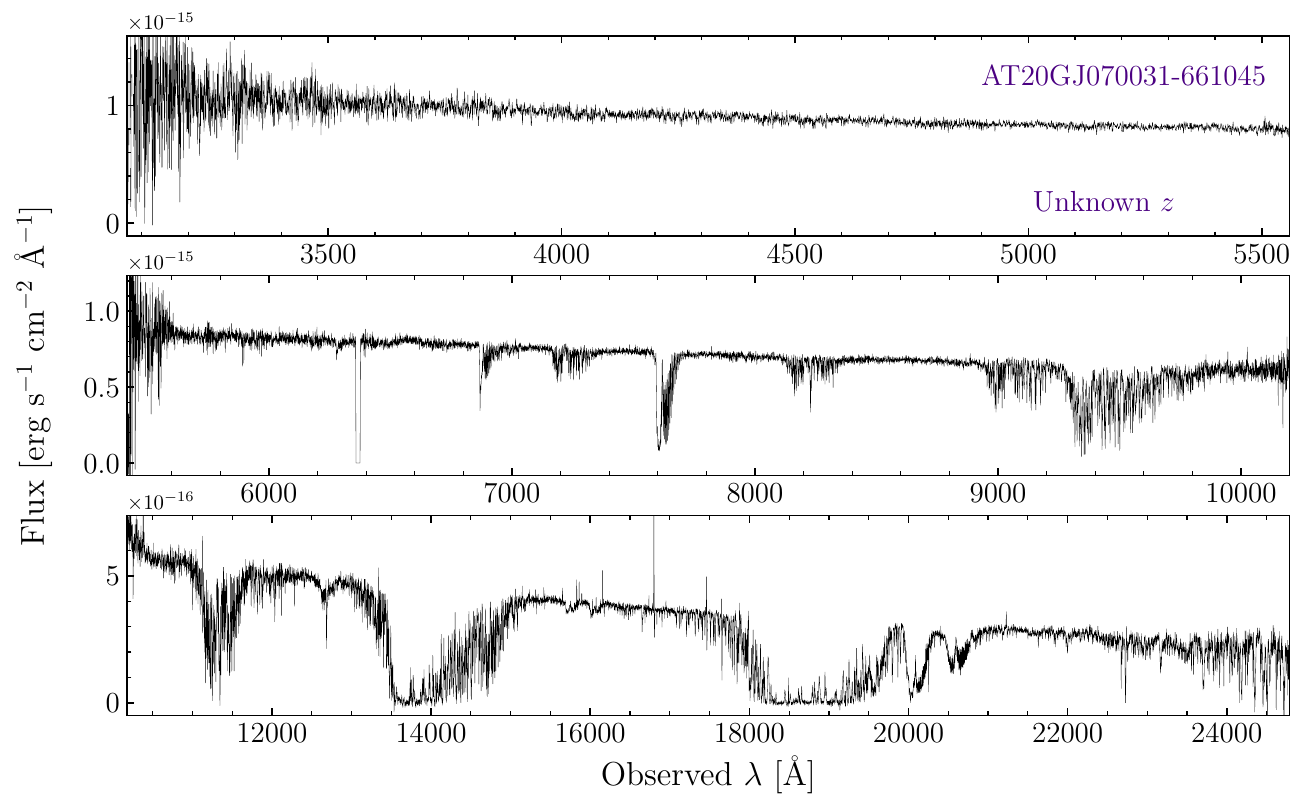}
    \caption{}

\end{figure*}

\begin{figure*}
    \ContinuedFloat
    \captionsetup{list=off,format=cont}
    \includegraphics[width=0.95\textwidth]{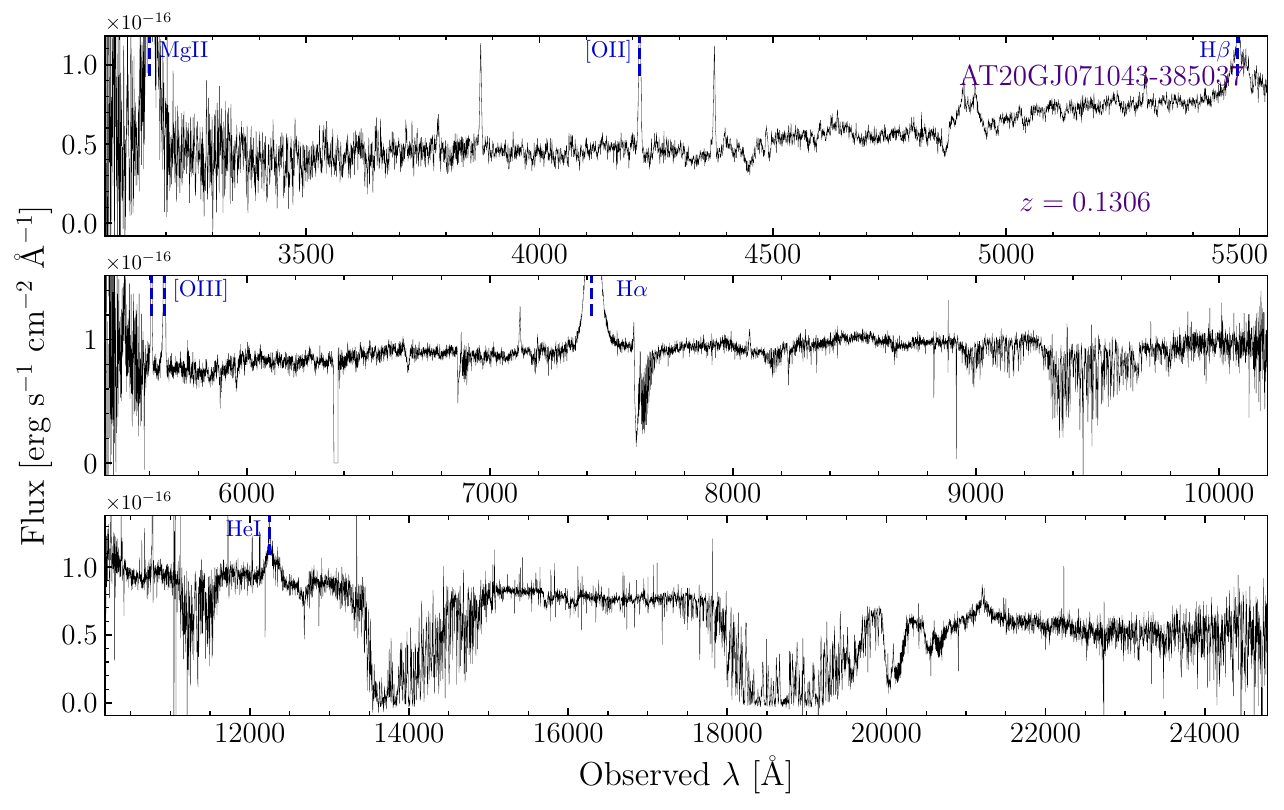}
    \caption{}

\end{figure*}

\begin{figure*}
    \ContinuedFloat
    \captionsetup{list=off,format=cont}
    \includegraphics[width=0.95\textwidth]{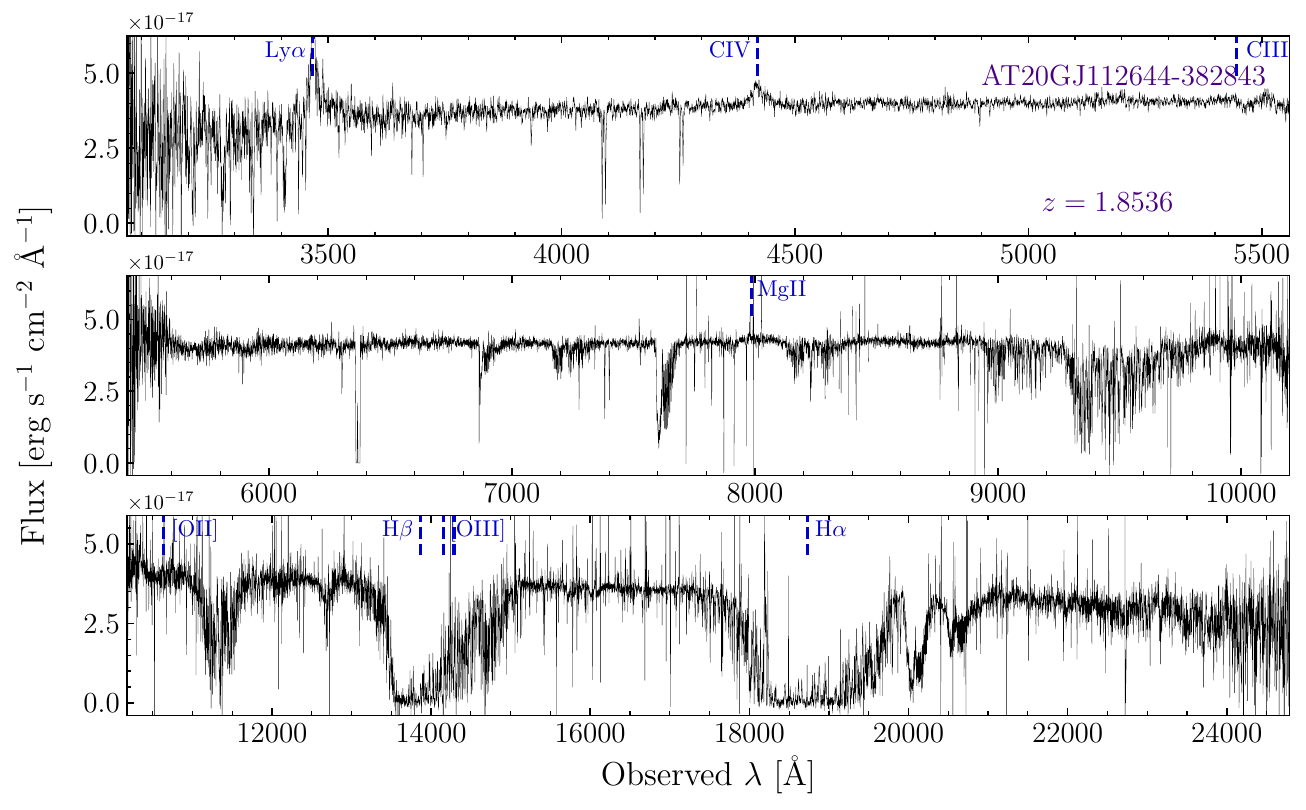}
    \caption{}

\end{figure*}

\begin{figure*}
    \ContinuedFloat
    \captionsetup{list=off,format=cont}
    \includegraphics[width=0.95\textwidth]{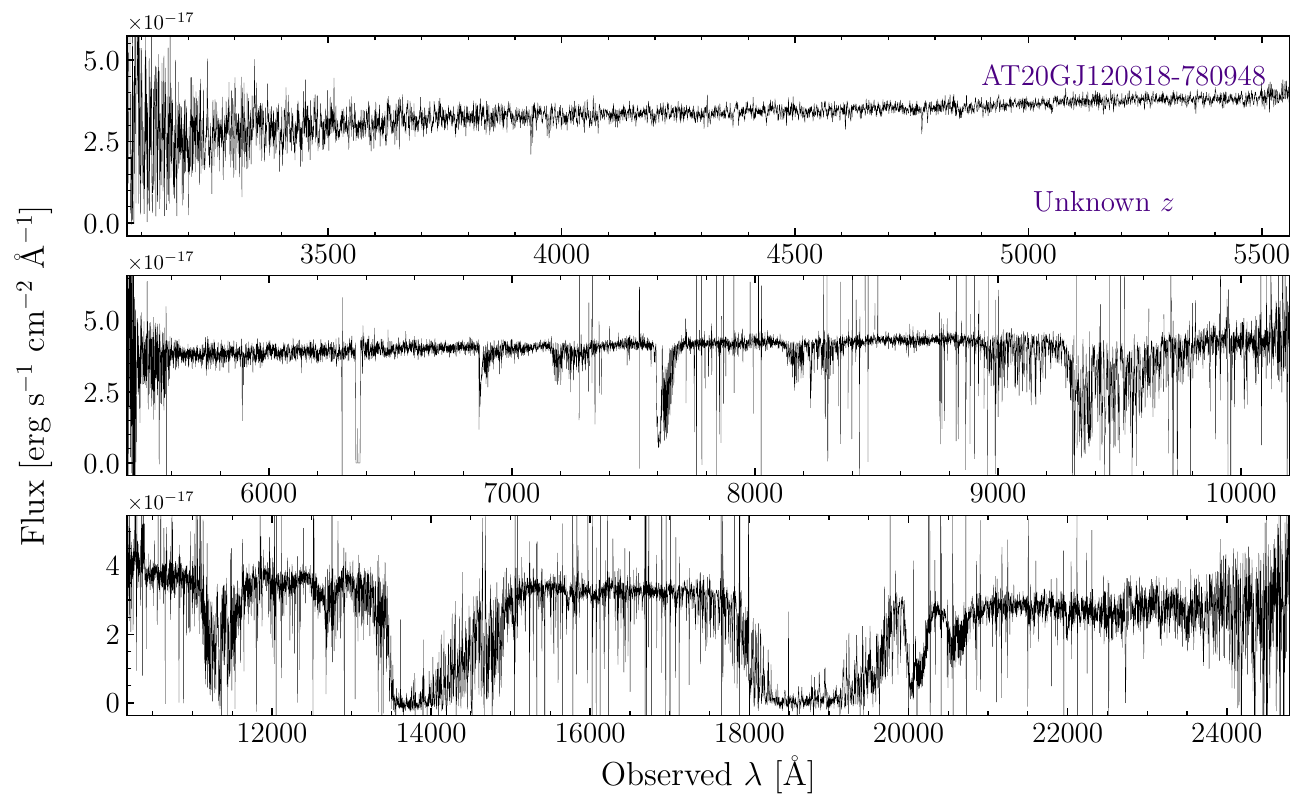}
    \caption{}

\end{figure*}

\begin{figure*}
    \ContinuedFloat
    \captionsetup{list=off,format=cont}
    \includegraphics[width=0.95\textwidth]{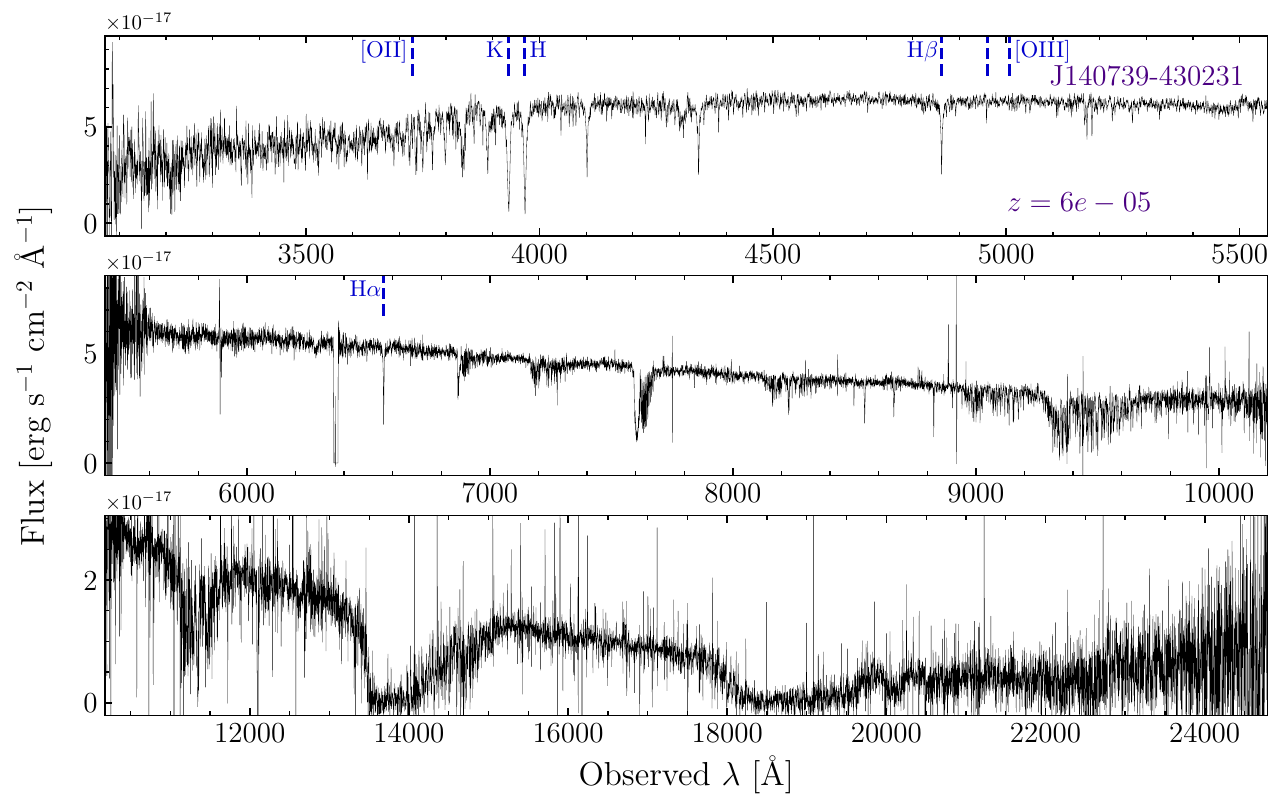}
    \caption{}

\end{figure*}

\begin{figure*}
    \ContinuedFloat
    \captionsetup{list=off,format=cont}
    \includegraphics[width=0.95\textwidth]{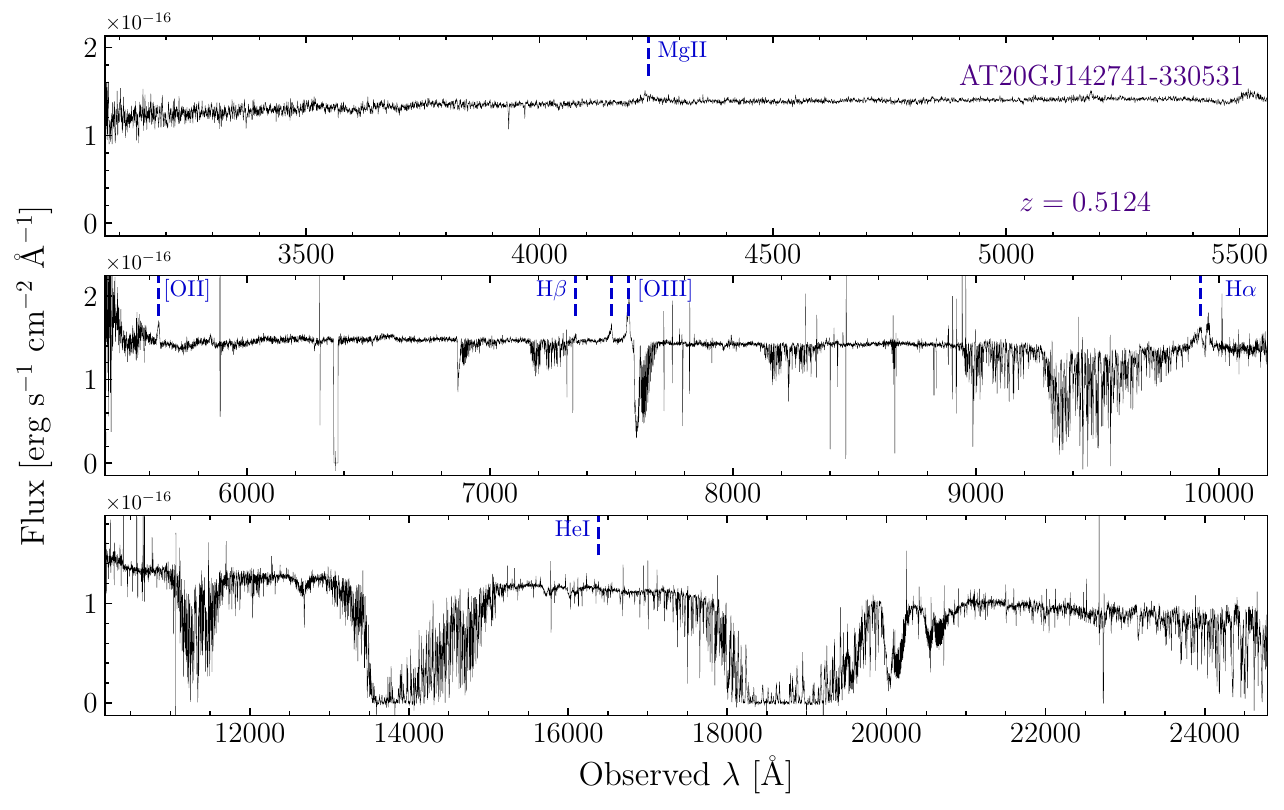}
    \caption{}

\end{figure*}

\begin{figure*}
    \ContinuedFloat
    \captionsetup{list=off,format=cont}
    \includegraphics[width=0.95\textwidth]{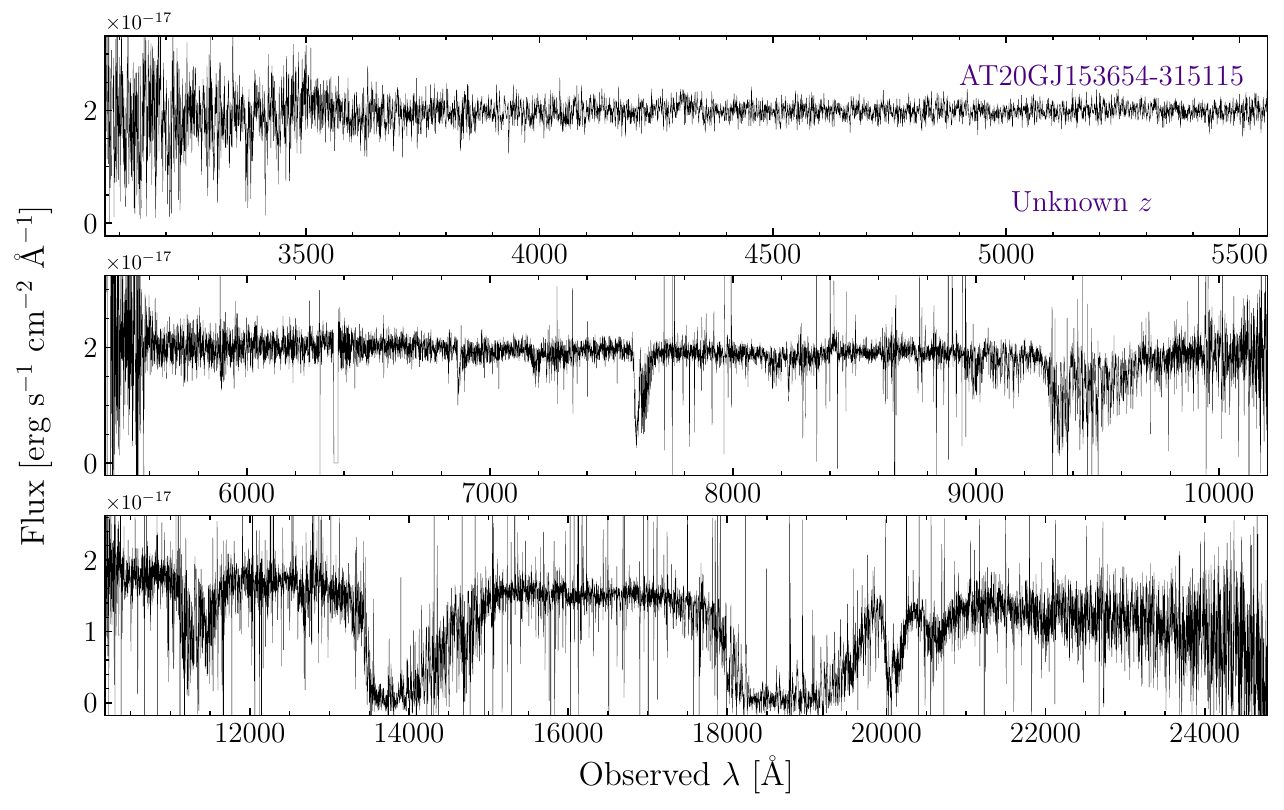}
    \caption{}

\end{figure*}

\begin{figure*}
    \ContinuedFloat
    \captionsetup{list=off,format=cont}
    \includegraphics[width=0.95\textwidth]{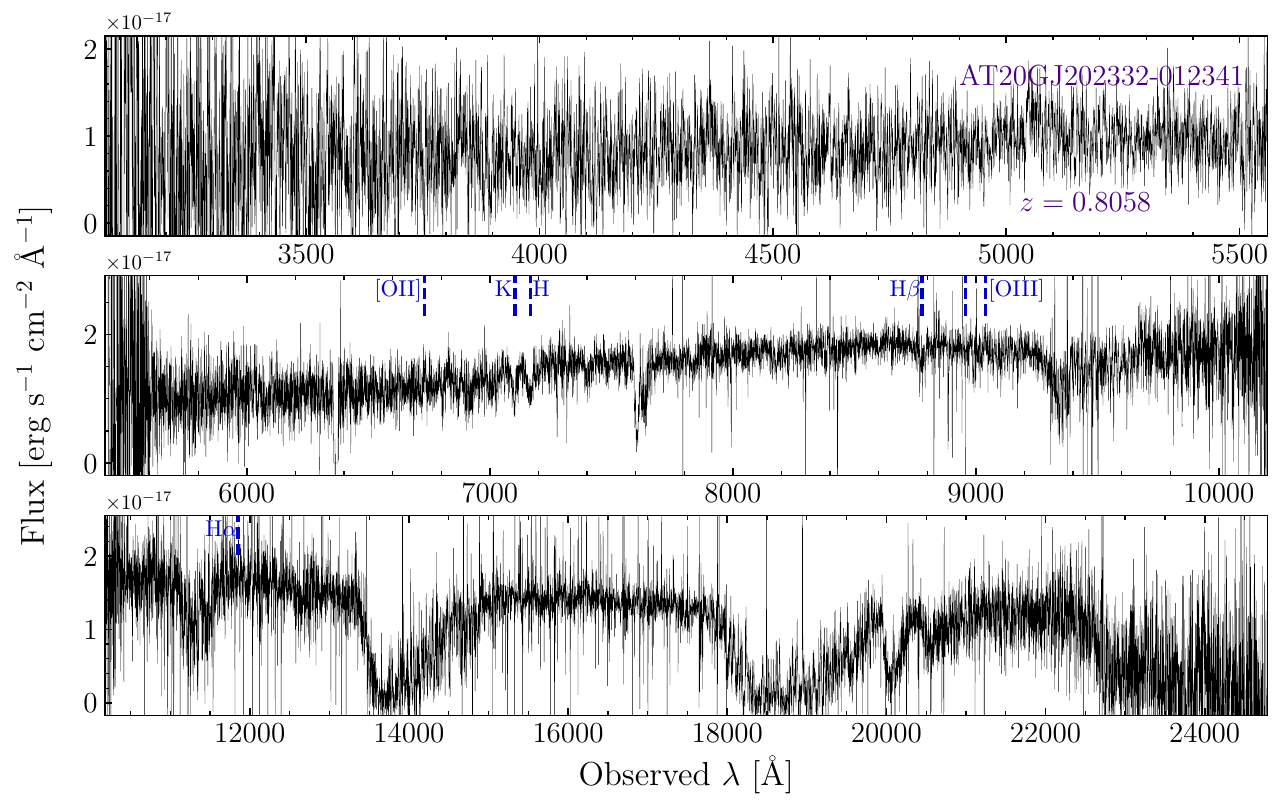}
    \caption{}

\end{figure*}

\begin{figure*}
    \ContinuedFloat
    \captionsetup{list=off,format=cont}
    \includegraphics[width=0.95\textwidth]{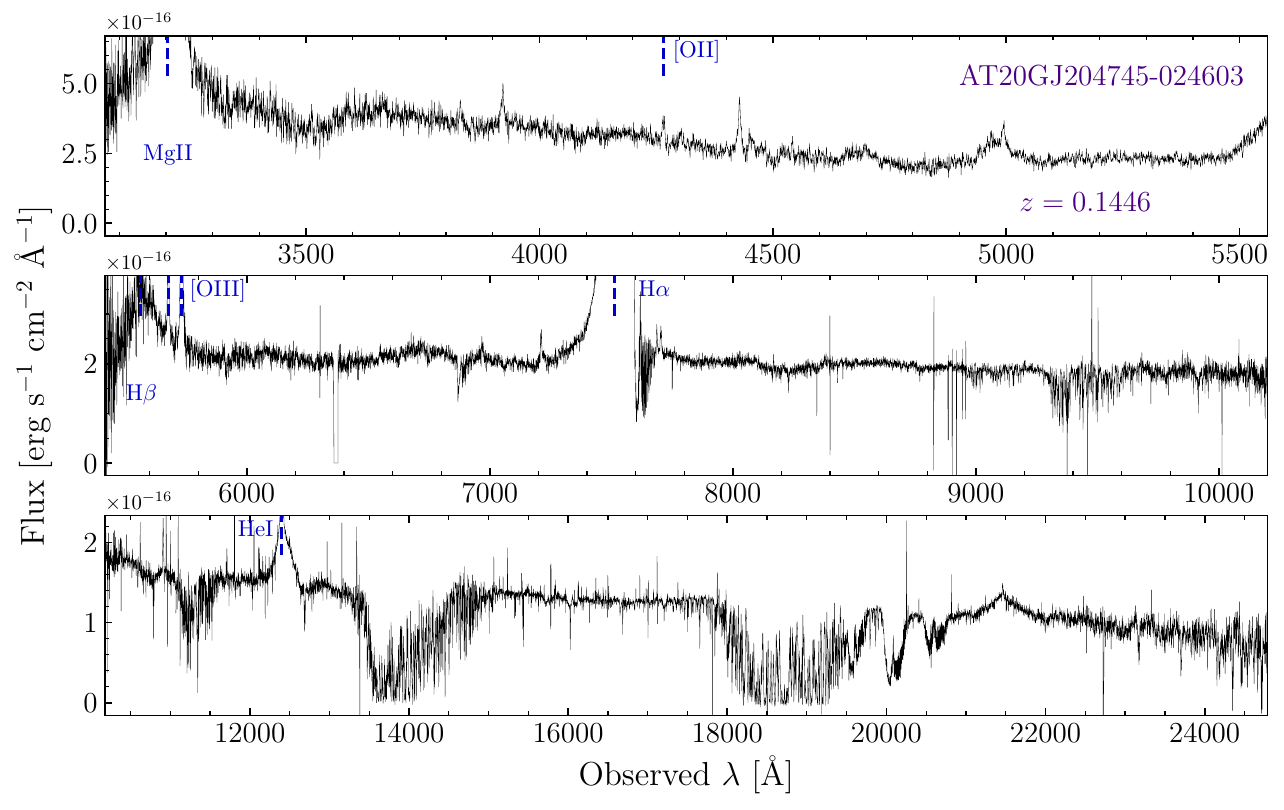}
    \caption{}

\end{figure*}

\begin{figure*}
    \ContinuedFloat
    \captionsetup{list=off,format=cont}
    \includegraphics[width=0.95\textwidth]{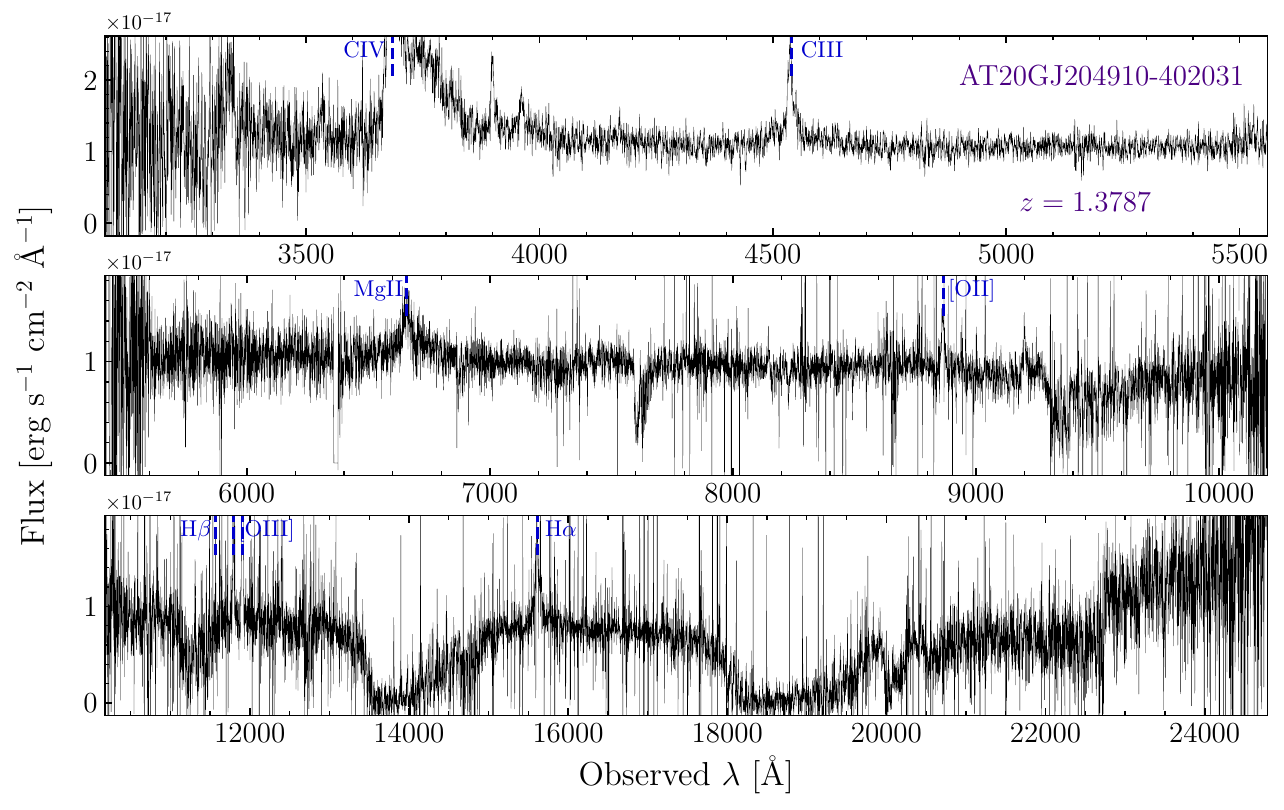}
    \caption{}

\end{figure*}

\begin{figure*}
    \ContinuedFloat
    \captionsetup{list=off,format=cont}
    \includegraphics[width=0.95\textwidth]{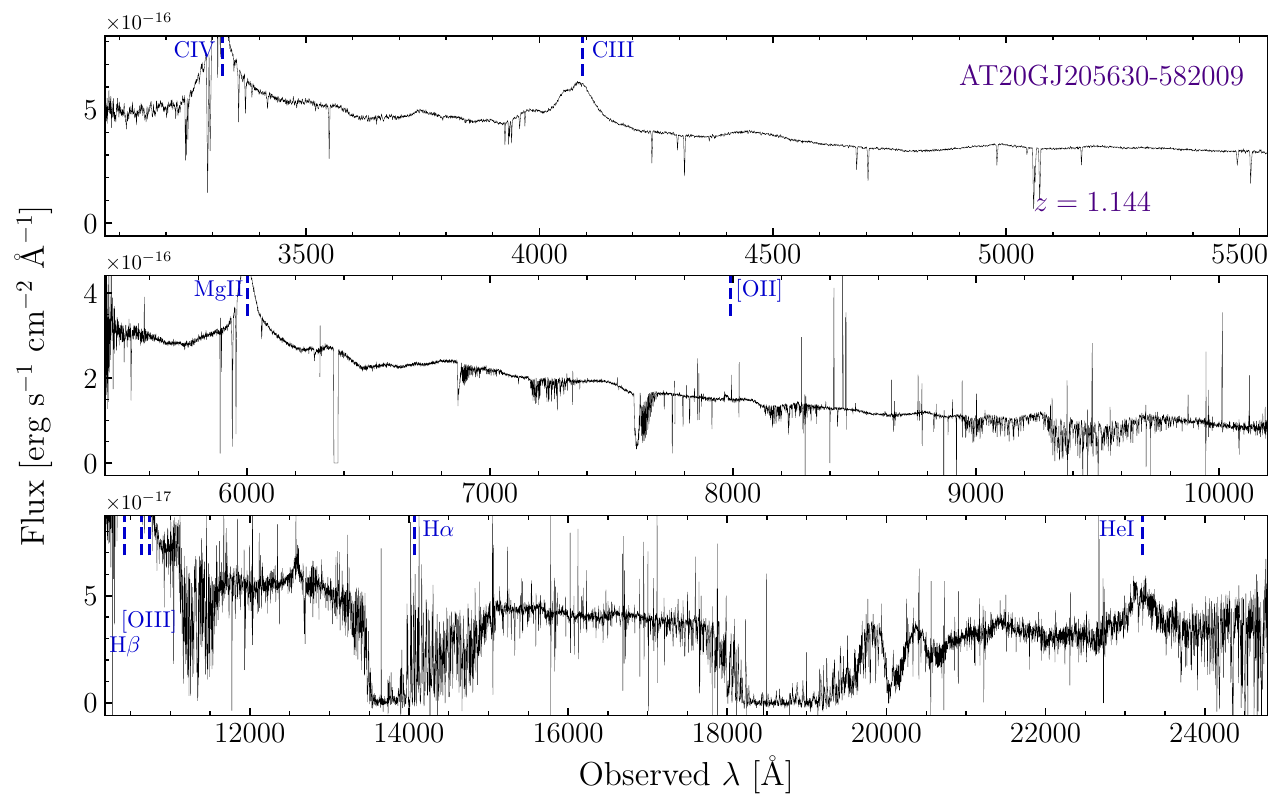}
    \caption{}

\end{figure*}

\begin{figure*}
    \ContinuedFloat
    \captionsetup{list=off,format=cont}
    \includegraphics[width=0.95\textwidth]{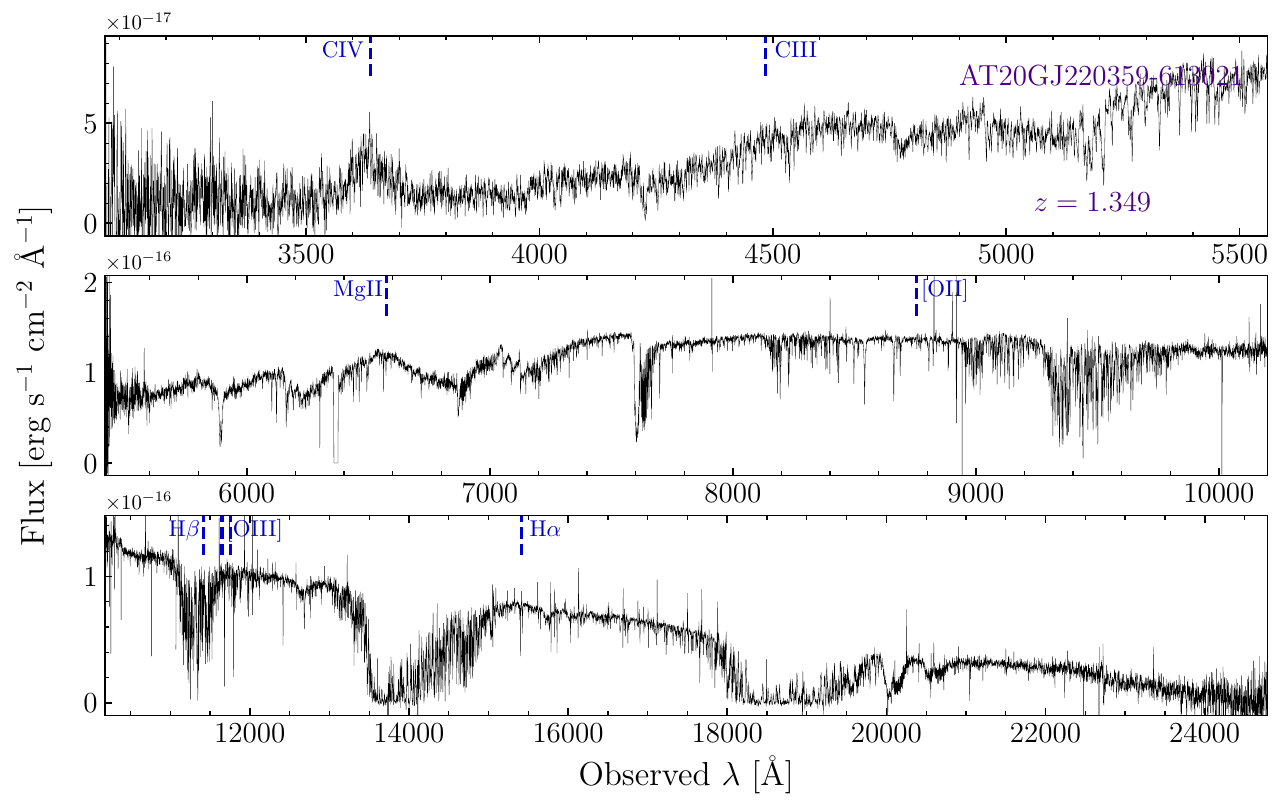}
    \caption{}

\end{figure*}

\begin{figure*}
    \ContinuedFloat
    \captionsetup{list=off,format=cont}
    \includegraphics[width=0.95\textwidth]{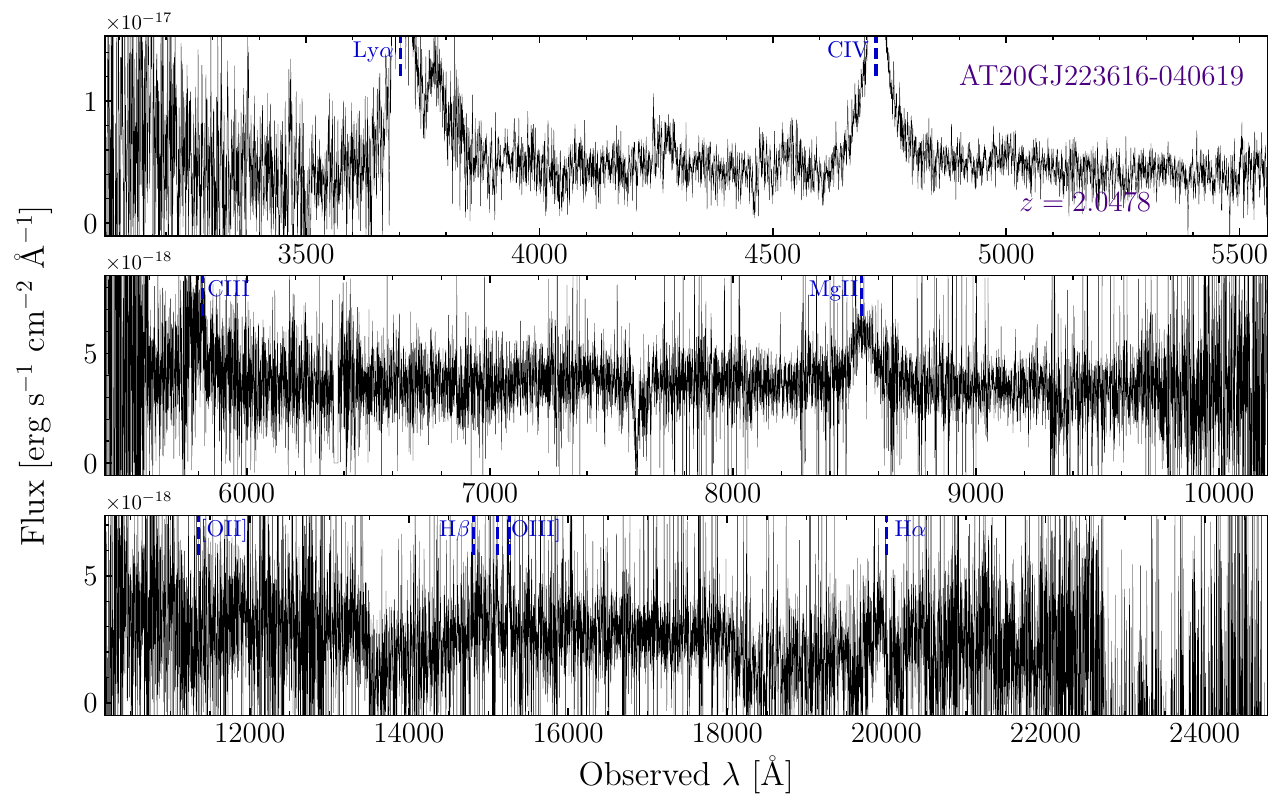}
    \caption{}

\end{figure*}

\begin{figure*}
    \ContinuedFloat
    \captionsetup{list=off,format=cont}
    \includegraphics[width=0.95\textwidth]{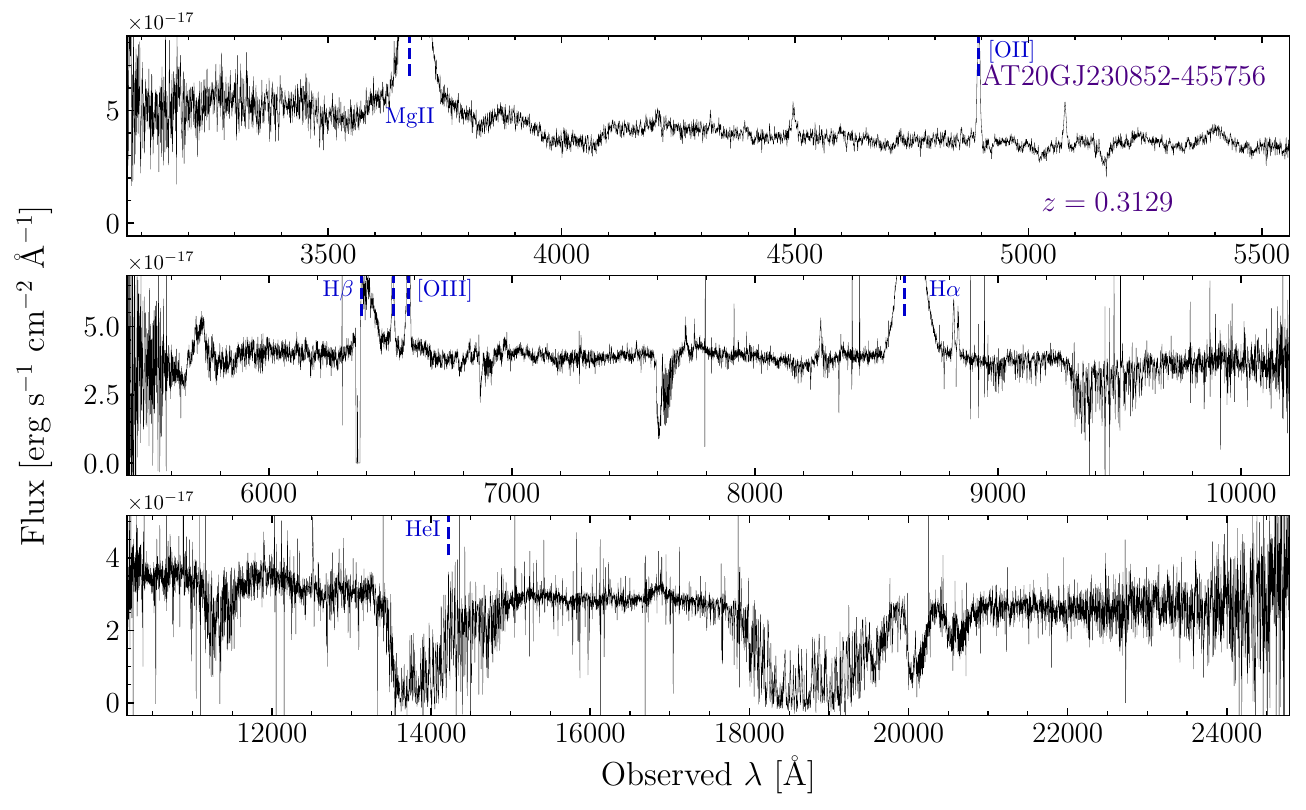}
    \caption{}
\end{figure*}

\bsp	
\label{lastpage}
\end{document}